\pdfoutput=1
\documentclass[twocolumn,10pt]{article}
\usepackage[margin=0.5in]{geometry}
\usepackage{amsmath}
\usepackage{amsfonts}
\usepackage[mathscr]{euscript}
\usepackage{dsfont}
\usepackage{amssymb}
\usepackage{graphicx}
\usepackage{url}
\usepackage{rotating}
\usepackage{verbatim}
\usepackage{color}
\usepackage{svg}
\usepackage{phonetic}
\usepackage{cite}
\usepackage{hyperref}
\usepackage{bm}
\usepackage{upgreek}
\usepackage{adjustbox}
\usepackage[ruled,vlined]{algorithm2e}

\usepackage[runin]{abstract}
\newtheorem{assumption}{Assumption}

\DeclareFontFamily{OT1}{pzc}{}
\DeclareFontShape{OT1}{pzc}{m}{it}{<-> s * [1.10] pzcmi7t}{}
\DeclareMathAlphabet{\mathpzc}{OT1}{pzc}{m}{it}

\newcommand{\angstrom}{\text{\normalfont \AA}}

\newcommand{\com}[1]{}

\newcommand{\blue}[1]{\textcolor{blue}{#1}}

\title{\textsf{\textbf{Protofold II}}:
\\
Enhanced Model and Implementation for Kinetostatic Protein Folding%
    \footnote{This article was submitted on 09/08/2015 and published on 03/22/2016 in the ASME JNEM. For citation, please use:
    \protect\\
    \protect\\
        \blue{Tavousi, Pouya and Behandish, Morad and Ilie\c{s}, Horea T. and Kazerounian, Kazem, 2016. ``Protofold II: Enhanced Model and Implementation for Kinetostatic Protein Folding.'' Journal of Nanotechnology in Engineering and Medicine, 6(3), p.034601.}
    \protect\\
    \protect\\
    This article is based on two shorter conference papers presented at the ASME IDETC/CIE'2013 \cite{Tavousi2013,Behandish2013}.}
}

\author{Pouya Tavousi$^1$, Morad Behandish$^2$, Horea T. Ilie\c{s}$^{1,2}$, and Kazem Kazerounian$^1$
    \\
        {\small $^1$Kinematics Design Laboratory and $^2$Computational Design Laboratory,}
    \\
        {\small Department of Mechanical Engineering, University of Connecticut, USA}
    }

\date{\small Technical Report No. CDL-TR-16-03, March 22, 2016}

\begin{document}

\maketitle

\noindent \hrule \vspace{5pt}
\begin{abstract}
A reliable prediction of 3D protein structures from sequence data remains a big challenge due to both theoretical and computational difficulties. We have previously shown that our kinetostatic compliance method (KCM) implemented into the {\sf Protofold} package can overcome some of the key difficulties faced by other {\it de novo} structure prediction methods, such as the very small time steps required by the molecular
dynamics (MD) approaches or the very large number of samples needed by the Monte Carlo (MC) sampling techniques.
In this article, we improve the free energy formulation used in {\sf Protofold} by including the typically underrated entropic effects, imparted due to differences in hydrophobicity of the chemical groups, which dominate the folding of most water-soluble proteins. In addition to the model enhancement, we revisit the numerical implementation by redesigning the algorithms and introducing efficient data structures that reduce the expected  complexity from quadratic to linear. Moreover, we develop and optimize parallel implementations of the algorithms on both central and graphics processing units (CPU/GPU) achieving speed-ups up to two orders of magnitude on the GPU.
Our simulations are consistent with the general behavior observed in the folding process in aqueous solvent, confirming the effectiveness of model improvements. We report on the folding process at multiple levels; namely, the formation of secondary structural elements and tertiary interactions between secondary elements or across larger domains. We also observe significant enhancements in running times that make the folding simulation tractable for large molecules.
\end{abstract}
\vspace{5pt} \hrule \vspace{20pt}

\section{Introduction}

Proteins are large biomolecules that are responsible for a vast array of biological functions inside the cell, and appear in the form of enzymes, antibodies, motor proteins, transport proteins, etc. \cite{Kuriyan2012}. The function of a protein strongly depends on its 3D structure (i.e., `conformation') which in turn can be directly determined from the linear sequence of amino acids (AAs) linked together to form the protein chain (i.e., `configuration') \cite{Anfinsen1972}.\footnote{In the robotics literature, the term configuration is typically used to describe the complete set of kinematic variables. However, the term conformation is typically used for that purpose in molecular biology.}
Therefore, the computer-aided prediction of the folded structure of a protein from the knowledge of its sequence (referred to as `protein folding') is the key to understanding many biological processes in the cell. This knowledge is crucial toward the ultimate goal of modeling proper function or malfunction at molecular and cellular level (e.g., deadly diseases such as cancer, Alzheimer's, Parkinson's, etc.) and is central to a variety of bioengineering applications including `protein engineering' \cite{Van1988a,Chirikjian2005}.

\subsection{Related Work} \label{sec_lit}

There are several different computational approaches for protein folding prediction, ranging from knowledge-based techniques to methods starting from physical principles \cite{Echenique2007}.

\paragraph{Knowledge-Based Methods.}
The knowledge-based approaches predict the structure of a given protein using the information extracted from previously determined structures and known types of folds. They are generally more reliable than physics-based methods, but have limited applicability in predicting new types of folds. Examples of knowledge-based techniques are homology or comparative modeling \cite{Chothia1986,Marti2000,Krieger2003} and fold recognition or threading \cite{Bowie1991,Ginalski2005,Moult2005}. We refer the reader to \cite{Echenique2007} for a comprehensive review of such methods.

\paragraph{Physics-Based Methods.}
On the other hand, methods that utilize models formulated empirically or obtained from physical principles are less reliable and more time-consuming, but apply to a wider range of folding simulations \cite{Echenique2007}. These methods range from {\it de novo} \cite{Bradley2005,Schueler2005} to {\it ab initio} \cite{Osguthorpe2000,Bonneau2001} prediction techniques. Here we briefly review some of the common {\it ab initio} approaches, namely sampling methods and MD simulations \cite{Echenique2007}.

Sampling methods generate numerous samples in the conformation space, followed by an evaluation of their free energies. Different search algorithms are used to find the unchallenged global minimum of the free energy, assumed to be associated with the native structure according to the `thermodynamic hypothesis' \cite{Anfinsen1972}. These search methods include simulated annealing \cite{Hansmann1994,Simons1997}, basin hopping \cite{Li1987,David1997,Nayeem2004,Prentiss2008}, evolutionary algorithms \cite{Schug2004,Schug2006,Verma2007} and MC simulation with biased moves \cite{Abagyan1994,Abagyan1999,Carr2005}. A review of conformation sampling methods for protein folding can be found in \cite{Klenin2011}. Sampling methods have two major limitations, namely 1) they do not provide any information about the biological pathway; and 2) finding the global minimum is not guaranteed because of the finite number of samples.

On the other hand, MD approaches simulate the biological pathway using a model built upon physical principles. Standard MD techniques include Newtonian dynamics \cite{Gear1971,Beeman1976,Swope1982,Hockney1988}, Langevin dynamics \cite{Van1988,Ricci2003,Guarnieri2004,Ciccotti2004} and Brownian dynamics \cite{Rojnuckarin1998,Gabdoulline2001,Ando2005,Frembgen2009}. A review of MD simulation methods for protein folding is provided in \cite{Scheraga2007}. In order to keep the numerical algorithms stable, very small time steps (in the order of femtoseconds) along the simulation trajectory are required, which does not support folding simulation of typical proteins that span milliseconds except for small molecules \cite{Scheraga2007}.

\paragraph{Kinetostatic Compliance Method.}
The KCM was introduced in \cite{Kazerounian2004,Kazerounian2005} to overcome some of the key challenges in the aforementioned approaches. In this method, implemented in the software package {\sf Protofold} \cite{Kazerounian2004a,Kazerounian2004b,Kazerounian2005a}, the protein chain is modeled as a kinematic linkage which complies under the kinetostatic effect of the force-field obtained form intramolecular interactions between the atoms. The key contributions of KCM were
\begin{enumerate}
    \item modeling the constrained kinematics of the protein chain with significantly fewer degrees of freedom (DOF) than, for example, those of the `beads and springs' model used in many MD methods; and
    \item converging faster to the minimum energy state by using kinetostatic (i.e., 1st-order) variations rather than dynamic (i.e., 2nd-order) response.
\end{enumerate}
In KCM, each rotatable joint, used to model the constrained motion of the chemical bonds, changes by an amount proportional to the effective torque on that joint. It was shown that KCM is a faster and more stable alternative to the traditional dynamic simulation techniques \cite{Kazerounian2005a}.
The {\sf Protofold} platform has since provided a kinematic testbed for subsequent research activities. Examples are predicting hydrogen bond connectivity sub-graphs \cite{Shahbazi2010}, its application to the design of stable peptide nano-particles \cite{Shahbazi2010a}, the analysis of protein mobility (using the `pebble game' algorithm) \cite{Shahbazi2015}, the development of mechanical models for secondary structural elements \cite{Shahbazi2015a}, and nano-machanism synthesis from a `link soup' of pre-specified structural elements \cite{Tavousi2015,Tavousi2016}.

In the earlier stages of the development, the energetics were limited to intramolecular interactions in the gas-phase of the protein---e.g., Coulombic and van der Waals forces exchanged between atoms of the protein itself, ignoring the interactions with solvent molecules. However, an important class of biologically significant proteins are water-soluble, whose folding process is predominantly driven by the interactions with the solvent, particularly the so-called `hydrophobic effect'\footnote{The hydrophobic effect is explained as the strong tendency of nonpolar sidechains to pack together to form a hydrophobic core protected from the solvent by a hydrophilic surface \cite{Kuriyan2012}. This effect is formulated in terms of entropic changes in the solvent molecules surrounding the protein surface.} which was missing from {\sf Protofold I} \cite{Kazerounian2005a}.

\paragraph{Computing Solvation Effects.}
From a computational perspective, the solvation effects can be modeled in a number of different ways, broadly classified into `explicit' and `implicit' techniques.

The explicit methods use all-atom force field models such as SPC, SPC/E, TIP3P, TIP5P \cite{Mashayak2011,Jorgensen2005}, or coarse-grained (CG) models \cite{Wang2009,Izvekov2005} which are less structured representations of the solvent obtained by mapping two or more particles onto a single interaction site \cite{Mashayak2011}. A prohibitive computational cost is associated with the large number of solvent molecules required to model a bulk solution.

Alternatively, approximate schemes that include the solvent effects implicitly can provide useful quantitative estimates, yet remain computationally inexpensive \cite{Roux1999}.
The implicit models estimate the contribution of each solvent-exposed atom to the solvation free energy using empirical formulae, most commonly expressed as a linear function of the solvent accessible surface area (SASA) \cite{Eisenberg1986}. An exact computation of SASA requires obtaining the surface area of the envelope of overlapping spheres \cite{Richmond1984}, which is computationally expensive. Alternatively, approximate formulations have been developed to efficiently predict the {\it expected} (i.e., probabilistic average) SASA based on simplifying assumptions on the distribution of the coordinates of atoms (or groups of atoms) in the 3D space. For instance, CHARMM \cite{Brooks2004} uses the probabilistic approach from \cite{Wodak1980}, which estimates the SASA as a function of the distances between pairs of atoms or residues. A similar model with similar parametrization \cite{Fraternali1996} was used in GROMOS \cite{Van2002}, a recent improvement of which was given in \cite{Allison2011}. AMBER \cite{Weiner1984,Cornell1995} uses a fast linear combination of pairwise overlaps (LCPO) algorithm \cite{Weiser1999}, which improves the method in \cite{Wodak1980} by adding more terms to the approximation. Although being widely popular in well-known MD packages, these methods rely on simplifying assumptions that compromise accuracy. For example, the method in \cite{Wodak1980} assumes a uniform random spatial distribution of atoms or residues, which introduces bias into the simulation results.

The inclusion of the solvation free energy computed using an adequately accurate evaluation of the SASA \cite{Tavousi2013,Behandish2013} results in a more realistic energy- and force-model for simulating the natural behavior of water-soluble proteins.

\subsection{Outline \& Contributions}

In Section \ref{sec_form}, we introduce an improved free energy model making use of the linear implicit model given in \cite{Eisenberg1986} to compute the solvation free energy- and force-field from a knowledge of the SASA and its gradient for individual atoms at a given 3D conformation. We develop a simple offset surface enumeration technique that can approximate the SASA and its gradient up to any desired accuracy. Our method is significantly more accurate than the probabilistic methods such as those given in \cite{Wodak1980,Weiser1999} yet noteably faster than the exact method given in \cite{Richmond1984}, while the trade-off between speed and accuracy can be adjusted by a proper choice of the enumeration (i.e., surface sampling) density.

A second major contribution of this work is to develop significantly more efficient algorithms and data structures in Section \ref{sec_alg} to speed up the computations, and to implement them into {\sf Protofold II}:
\begin{enumerate}
    \item We use a 3D hash table data structure based on a uniform spatial grid that supports fast queries to identify pairs of proximal atoms. This helps speeding up the computations by eliminating negligibly small interactions associated with pairs of atoms that are farther than a so-called `cut-off distance'.
    \item We use a tree-based data structure to span the protein chain efficiently for the purpose of characterizing interaction types between pairs of atoms based on their distances along the topological structure of the chain.
    \item We develop sequential and parallel surface enumeration algorithms to compute the SASA and its gradient for individual atoms needed for the solvation energy and force computations, respectively, up to desired accuracy.
    \item We employ prefix sums \cite{Cole1989} to compute the joint torques on the kinematic linkage of the protein chain, given the resultant forces on the individual atoms.
\end{enumerate}
As a result, the numerical complexity for each KCM cycle, including the computation of resultant forces on the atoms and the links (except those resulting from solvation effects) and their conversion to joint torques, is reduced from $O(n^2)$ in {\sf Protofold I} \cite{Kazerounian2005a} to $O(n)$ in {\sf Protofold II}, where $n$ is the number of atoms in the protein molecule.\footnote{This is only the case under certain assumptions given in the Section \ref{sec_prox}, which are relatively reasonable for practical cases.}

The SASA evaluations for solvation force computations in our model turns out to be the bottleneck to the entire simulation---up to several orders of magnitude slower than the electrostatic and van der Waals force computations. Fortunately, the surface enumeration algorithm lends itself well to high-throughput data parallelism.
In Section \ref{sec_imp} we first present the CPU-parallel implementation using OpenMP, leading to moderate speed-up factors (up to one order of magnitude).
To leverage the massive processing power offered by the single-instruction multiple-thread (SIMT) architecture of the modern graphics hardware---onto which our data-parallel SASA enumeration algorithm maps perfectly---we present a GPU-parallel implementation and its optimization. The implementation takes advantage of the device memory hierarchy and hiding memory access latencies, in turn leading to larger speed-ups (up to two orders of magnitude).

\begin{figure*}
    \centering
    \includegraphics[width=0.75\textwidth]{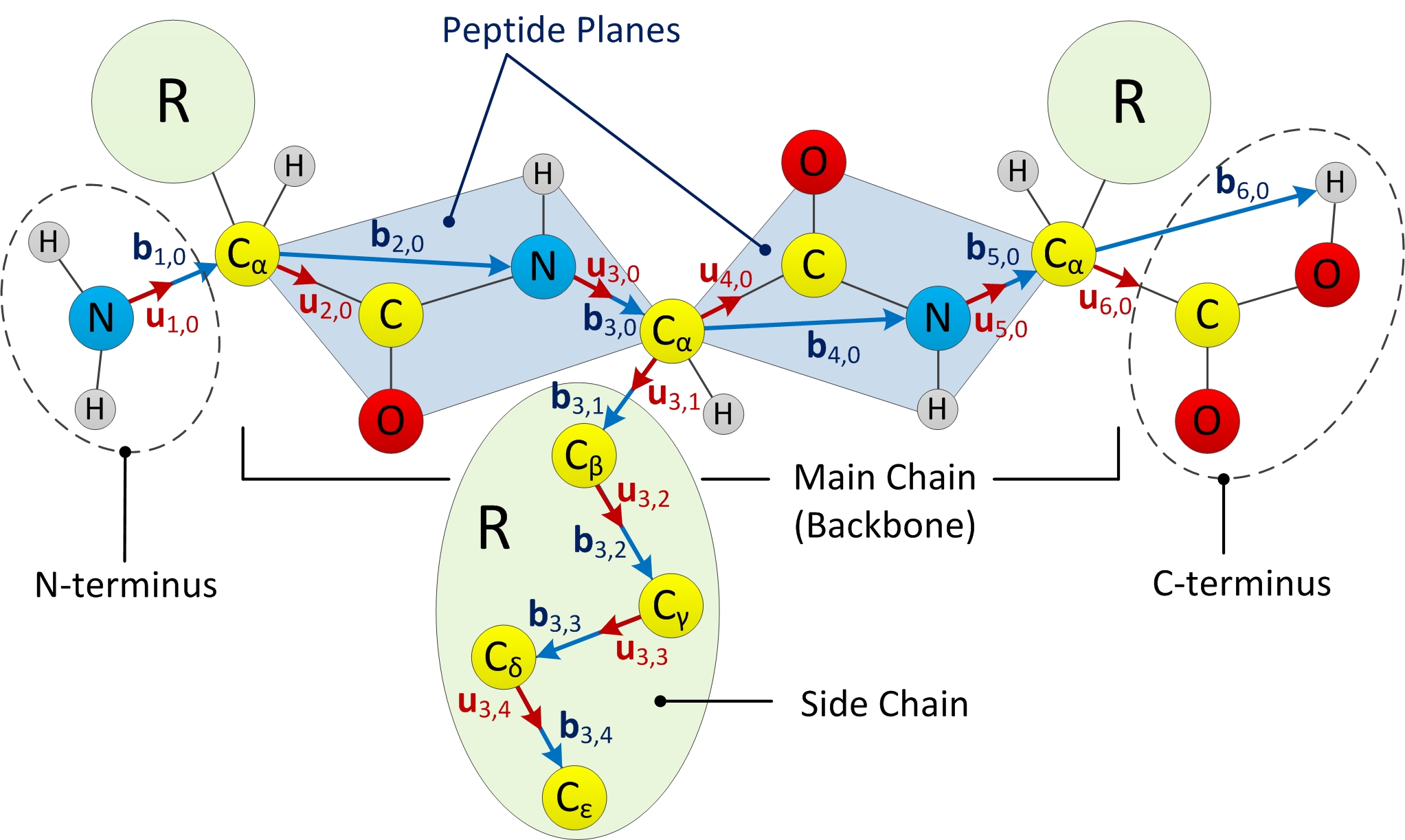}
    \caption{The polypeptide chain is modeled as a kinematic linkage, in which the peptide planes are assumed to be rigid.}
    \label{figure1}
\end{figure*}

\section{Formulation} \label{sec_form}

Section \ref{sec_kin} starts with an overview of the underlying kinematic principles of the KCM simulation first introduced in \cite{Kazerounian2004,Kazerounian2004a,Kazerounian2004b,Kazerounian2005,Kazerounian2005a}. The protein chain is modeled as an open kinematic linkage with reduced DOF in terms of dihedral and rotamer angles, which complies under the effect of interatomic and solvation forces. Next, the energy- and force-field formulation used in {\sf Protofold II} is described in Section \ref{sec_field}, with special emphasis on the newly introduced solvation effects. Lastly, the KCM optimization process is presented in Section \ref{sec_KCM}.

\subsection{Kinematic Model} \label{sec_kin}

Proteins are long polymeric chains made of AAs, which exist in only $20$ different types (except for few rare exceptions), joined together as a linear polypeptide chain \cite{Kuriyan2012}, structural details of which are summarized in Appendix \ref{app_peptide}. Here we restrict ourselves to the kinematic representation of the chain's conformation within the scope of KCM.

\paragraph{Linkage Parameterization.}
Figure \ref{figure1} schematically illustrates the repetitive sequence of $-$N$-$C$_\alpha-$C$-$ atoms,\footnote{Hereon, the notations C$_\alpha$ and C correspond to the alpha-carbon and carboxyl-carbon, respectively.}
called the `backbone' or the `main chain', with `side chains' resembling branches that extend out of it. As explained in Appendix \ref{app_peptide}, the backbone conformation can be specified to an adequate accuracy by two sets of dihedral angles; namely,
\begin{itemize}
    \item $-180^\circ \leq \phi_i < +180^\circ$ (around N$-$C$_\alpha$ in $AA_i$); and
    \item $-180^\circ \leq \psi_i < +180^\circ$ (around C$_\alpha-$C in $AA_i$);
\end{itemize}
for $1 \leq i \leq m$, where $m$ is the number of AA residues along the chain. The conformation of each side chain, on the other hand, can be specified by up to 4 extra angles $-180^\circ \leq \chi_{i,k} < +180^\circ$ for $1 \leq k \leq l_i$ where $0 \leq l_i \leq 4$ is the number of side chain links of $AA_i$, and the subscript $k$ corresponds to the bonds numbered in the obvious order along the side chain C and N atoms.

To set a reference for the angle measurements, the zero-reference position description (ZRPD) method \cite{Gupta1986} is used. The zero-position (ZP) for the protein chain is defined as the conformation of the serial linkage in which all peptide planes are coplanar (i.e., $\phi^0_i = \psi^0_i = -180^\circ$) and side chain dihedrals are set to default low energy values identified as `rotamers' \cite{Kazerounian2005}.

To unify the notations, all angular variables are denoted by $\theta_{j,k}~(1 \leq j \leq 2m, 0 \leq k \leq 4)$ where
\begin{align}
    &\theta_{2i-1,0} = \phi_i + 180^\circ, \quad~ 1 \leq i \leq m, \label{theta_1} \\
    &\theta_{2i,0} \quad= \psi_i + 180^\circ, \quad~ 1 \leq i \leq m,  \label{theta_2} \\
    &\theta_{2i-1,k} = \chi_{i,k} - \chi^0_{i,k}, \quad 1 \leq i \leq m, ~1 \leq k \leq l_i \leq 4, \label{theta_3}
\end{align}
where $0^\circ \leq \theta_{j,k} < 360^\circ$. The shifts in (\ref{theta_1}) and (\ref{theta_2}) by the intercept values of $\phi^0_i = \psi^0_i = -180^\circ$ and in (\ref{theta_3}) by the favorable rotamer angles $\chi^0_{i,k}$ ensure $\theta_{j,k} = 0^\circ$ at the ZP conformation.

A similar indexing scheme is used to identify the unit vectors along the rotation axes of revolute joints associated with these angles denoted by ${\bf u}_{j,k}~(1 \leq j \leq 2m, 0 \leq k \leq 4)$, i.e.,
\begin{itemize}
    \item ${\bf u}_{2i-1,0}~(1 \leq i \leq m)$ is the unit vector along the bond between N of $AA_i$ and C$_\alpha$ of $AA_i$;
    \item ${\bf u}_{2i,0}~(1 \leq i \leq m)$ is the unit vector along the bond between C$_\alpha$ of $AA_i$ to the C of $AA_i$; and
    \item ${\bf u}_{2i-1,k}~(1 \leq i \leq m, 1 \leq k \leq 4)$ are the unit vectors along the successive side chain C and N atoms.
\end{itemize}
Thus the kinematics of the linkage---which abstracts the protein conformation---can be completely specified in terms of the rigid body rotation transformations obtained from these rotation angles and rotation axes.

The spatial orientation of the rigid peptide planes can be described conveniently with a pair of base vectors whose linear combination spans the peptide plane. The so-called `body vectors' are denoted by ${\bf b}_{j,k}~(1 \leq j \leq 2m, 0 \leq k \leq 4)$, i.e.,
\begin{itemize}
    \item ${\bf b}_{2i-1,0}~(1 \leq i \leq m)$ is the base vector that connects the N of $AA_i$ to the C$_\alpha$ of $AA_i$;
    \item ${\bf b}_{2i,0}~(1 \leq i \leq m)$ is the base vector that connects the C$_\alpha$ of $AA_i$ to the N of $AA_{i+1}$; and
    \item ${\bf b}_{2i-1,k}~(1 \leq i \leq m, 1 \leq k \leq 4)$ are the base vectors along the successive side chain C and N atoms.
\end{itemize}
The first two sets of vectors listed above are called the `main chain body vectors'.
Every vector in the peptide plane that describes the relative positions of any two atoms can be obtained as a linear combination of these base vectors as $C_1 {\bf b}_{2i,0} + C_2 {\bf b}_{2i+1,0}$. The coefficients $C_1$ and $C_2$, referred to as `peptide plane constants', are invariant with respect to the rotations in the chain, thus can be precomputed prior to the KCM simulation. Different pairs of coefficients are used for vectors describing the relative positions of different pairs of atoms. Based on experimental evidence, it is a reasonable assumption that these coefficients are the same across all AAs \cite{Subramanian2007}, and the average values are given in Table \ref{table_coef}.\footnote{Nevertheless, in {\sf Protofold II} the user has the option to choose whether to use the values provided in Table \ref{table_coef} for all AAs, or to maintain the refined values when available---e.g., when the protein is imported from the protein data-bank (PDB).}

In addition to the main chain body vectors, the `side chain body vectors' (the third group listed above) are defined for the relative positions of the C and N atoms along the side chains.
We refer the reader to \cite{Subramanian2007} for more details about the molecular model of the peptide unit.

\begin{table}
    \caption{Peptide plane constants for bond vectors \cite{Subramanian2007}.} 
    \vspace{-0.5cm}
    \begin{center} \label{table_coef}
        \begin{tabular}{c c c | c c c}
            \hline
            BV & $C_1$ & $C_2$ & BV & $C_1$ & $C_2$ \\
            \hline
            $\overrightarrow{{\rm C}_{\alpha} {\rm C}}$  & $-0.2761$ & $+1.4488$ & $\overrightarrow{\rm CO}$  & $-1.3324$ & $+2.3401$ \\
            $\overrightarrow{\rm CN}$  & $+1.2761$ & $-1.4488$ & $\overrightarrow{\rm NH}$  & $+1.4103$ & $-2.5111$ \\
            \hline
        \end{tabular}
    \end{center}
\end{table}

For a protein chain with $m$ AA residues, the number of links can be obtained as
\begin{equation}
    l = \left( 2m + \sum_{i=1}^m l_i \right) \leq 6m = O(m),
\end{equation}
noting that $l_i \leq 4$. The term $2m$ counts two rigid links per each AA's backbone---one for $-$CO$-$NH$-$ and one for $-$C$_\alpha-$ in the peptide unit---in order to have each rigid link connected to the next with a single revolute joint along either N$-$C$_\alpha$ or C$_\alpha-$C, as depicted in Fig. \ref{figure1}. The second term accounts for the number of additional side chain links. As a result, the total DOF of the kinematic linkage is equal to the number of links. Table \ref{table_ub_angle} gives a complete description of dihedral angles, unit vectors, and body vectors for the entire protein chain.

\begin{table}
    \caption{Kinematic variables of the polypeptide linkage.} 
    \vspace{-0.5cm}
    \begin{center} \label{table_ub_angle}
        \begin{tabular}{p{1cm}  p{7cm}} 
            \hline
            Symbol & Description  \\ [0.5ex] 
            \hline
            $\theta_{2i-1,0}$  & {\small Torsion angle $\phi_i$ around main chain N$-$C$_\alpha$ in $AA_i$}  \\
            ${\bf u}_{2i-1,0}$  & {\small Unit vector along main chain N$-$C$_\alpha$ in $AA_i$}  \\
            ${\bf b}_{2i-1,0}$  & {\small Body vector from N to C$_\alpha$ in $AA_i$}  \\
            $\theta_{2i,0}$  & {\small Torsion angle $\psi_i$ around main chain C$_\alpha-$C in $AA_i$}  \\
            ${\bf u}_{2i,0}$  & {\small Unit vector along main chain C$_\alpha-$C in $AA_i$}  \\
            ${\bf b}_{2i,0}$  & {\small Body vector from C$_\alpha$ in $AA_i$ to N in $AA_{i+1}$$^\dag$} \\
            $\theta_{2i-1,k}$  & {\small Torsion angle $\chi_{i,k}$ of side chain C/Ns in $AA_i$} \\
            ${\bf u}_{2i-1,k}$  & {\small Unit vector along side chain C/Ns in $AA_i$}  \\
            ${\bf b}_{2i-1,k}$  & {\small Body vector connecting side chain Cs in $AA_i$} \\
            \hline
        \end{tabular}
    \end{center}
    \vspace{-0.2cm}
    {\footnotesize $^\dag$ The exception is ${\bf b}_{2m, 0}$ which connects C$_\alpha$ to the carboxyl H in $AA_{2m}$.}
\end{table}

\paragraph{Kinematic Equations.}
The instantaneous conformation of the protein chain is related to the ZP conformation with a set of rigid body transformations. Given a particular conformation in terms of $\theta_{j,k}~(1 \leq j \leq 2m, 0 \leq k \leq 4)$, the unit vectors and body vectors are transformed as follows:
\begin{align}
    [{\bf u}_{j,k}] &= [M_{j,k}] [{\bf u}^0_{j,k}], \quad 1 \leq j \leq 2m, ~ 0 \leq k \leq 4, \label{eq_kin_02} \\
    [{\bf b}_{j,k}] &= [M_{j,k}] [{\bf b}^0_{j,k}], \quad 1 \leq j \leq 2m, ~ 0 \leq k \leq 4, \label{eq_kin_03}
\end{align}
where the superscript $0$ indicates the reference ZP conformation. $[M_{j,k}]$ is the $3 \times 3$ matrix representation of the rigid body transformation $M_{j,k} \in \mathrm{SO}(3)$ that maps the ZP unit and body vectors ${\bf u}^0_{j,k}$ and ${\bf b}^0_{j,k}$ to their transformed orientations ${\bf u}_{j,k}$ and ${\bf b}_{j,k}$, respectively. These vectors are expressed using $3 \times 1$ column matrices. The transformation matrix for the main chain vectors $(k = 0)$ can be calculated as a product of successive rotations around individual joints in the main chain:
\begin{equation}\label{eq_kin_04}
    [M_{j,0}] = \prod_{r=1}^j [R_{r,0}], \quad 1 \leq j \leq 2m,
\end{equation}
while the transformation matrix for the side chain vectors $(k \geq 1)$ is defined as a product of rotations around joints in the main chain, and those around the side chain:
\begin{equation}\label{eq_kin_05}
    [M_{2i-1,k}]= \prod_{r=1}^{2i-1} [R_{r,0}] \prod_{s=1}^k[R_{2i-1,s}], \quad 1 \leq i \leq m, ~ 1 \leq k \leq 4,
\end{equation}
where $[R_{r,s}]$ is the $3 \times 3$ matrix representation of the joint rotation transformation $R_{r,s} \in \mathrm{SO}(3)$ around the ZP unit vector ${\bf u}^0_{r,s}$ with an angle $\theta_{r,s}~(1 \leq r \leq i, 0 \leq s \leq k)$ \cite{Kazerounian2005}, using the right-hand rule to choose the direction.

Once the body vectors are obtained using (\ref{eq_kin_03}) for a given conformation, the moved atom center positions can be computed for the individual atoms. Assuming that the N atom at the amino-terminus is fixed at the origin, the coordinates of the main chain N and C$_\alpha$ atoms are obtained as
\begin{equation}
    [{\bf r}_{j,0}] = \sum_{r = 1}^{j} [{\bf b}_{r, 0}], \quad 1 \leq j \leq 2m-1, \label{eq_r1}
\end{equation}
where the index $j = 2i - 1$ corresponds to the C$_\alpha$ atom of residue $AA_i$ while the index $j = 2i$ corresponds to the N atom of the residue $AA_{i+1}$ for $1 \leq i \leq m$. The coordinates for the other atoms in the peptide group, namely H, C and O, are obtained from those for C$_\alpha$ and N, and a linear combination $C_1 {\bf b}_{2i,0} + C_2 {\bf b}_{2i+1,0}$ of main chain body vectors using the coefficients $C_1$ and $C_2$ given in Table \ref{table_coef}. For the side chain C and N atoms, the coordinates are similarly obtained as
\begin{equation}
    [{\bf r}_{2i-1, k}] = \sum_{r = 1}^{2i-1} [{\bf b}_{r, 0}] + \sum_{s = 1}^k [{\bf b}_{2i-1, s}], \quad 1 \leq i \leq m, ~ 1 \leq k \leq 4, \label{eq_r2}
\end{equation}
where $k = 1, 2, 3$, and $4$ corresponds to the successive side chain C and N atoms in the residue $AA_i$. The coordinates for all other side chain atoms are obtained similarly from vectors along the side chain bonds subjected to the same set of motions \cite{Kazerounian2004}.

The atom position vectors obtained from (\ref{eq_r1}) and (\ref{eq_r2}) at each iteration are used in Section \ref{sec_field} to compute the energies, forces, and torques that will determine the motion for the subsequent iteration.

\subsection{Force Model} \label{sec_field}
The interatomic effects can be classified into covalent and noncovalent interactions. The covalent interactions need not be considered explicitly in the force-field, since they are implicitly introduced in terms of the kinematic constraints innate to the kinematic chain model adopted in Section \ref{sec_kin}. The noncovalent forces, which are responsible for conformational changes in the protein chain, can be derived from the free energy formulation that follows.

For a protein chain made of $n$ atoms, we assign each atom with a unique index $1 \leq i \leq n$, and its center coordinates $\mathbf{r}_i \in \mathds{R}^3$ obtained from dihedral angles using kinematic equations (\ref{eq_r1}) and (\ref{eq_r2}).\footnote{Note the slight change of notations from Section \ref{sec_kin}, where the subscript $j =2i-1$ or $2i$ referred to the AA index $1 \leq i \leq m$, while in Section \ref{sec_field} the single subscript $1 \leq i \leq n$ refers to the atom index.}
Each atom is identified by a unique tuple $a_i = (i, \mathbf{r}_i, R_i, q_i, \epsilon_i, \gamma_i, \cdots)~(1 \leq i \leq n)$, containing its index, position, radius, charge, well depth parameter, solvation parameter, and other atomic constants, to be introduced shortly. The set of all the atoms in the molecule is denoted by $\mathbb{A} = \{a_1, a_2, \cdots, a_n \}$. The aggregated free energy of all the atoms in $\mathbb{A}$ can be decomposed into the following terms:
\begin{equation}\label{eq_01}
    G^{\rm tot} (\mathbb{A}) = G^{\rm elec}(\mathbb{A}) + G^{\rm vdw}(\mathbb{A}) + G^{\rm cav}(\mathbb{A}),
\end{equation}
where $G^{\rm elec}(\mathbb{A})$ is the electrostatic energy, including intramolecular charge interactions, hydrogen bonding effects, and the induced polarization in the solvent when the molecule is dissolved. $G^{\rm vdw}(\mathbb{A})$ is the sum of intramolecular van der Waals energies, also called `steric effects', resulted from induced dipoles in the molecule. The sum of the first two terms has been accounted for in {\sf Protofold I} \cite{Kazerounian2004a,Kazerounian2004b,Kazerounian2005a} using the AMBER force-field model \cite{Weiner1984}.
$G^{\rm cav}(\mathbb{A})$ is the nonpolar solvation free energy, the free energy change resulting from transfering a molecule from vacuum to solvent, i.e., the entropic change due to the formation of the cavity occupied by the instantaneous 3D shape of the protein \cite{Bajaj2010}. Experimental results have shown that many water-soluble protein folding reactions are predominantly driven by a favorable reduction in $\Delta G^{\rm cav}(\mathbb{A})$ \cite{Kuriyan2012}, hence we incorporated this term into the improved energy formulation for {\sf Protofold II}.

\paragraph{Electrostatic Interactions.}
The charge interactions are formulated using the modified form of Coulomb's law \cite{Weiner1984}:
\begin{equation}\label{eq_02}
    G^{\rm elec} (\mathbb{A}) = \sum_{a_i \in \mathbb{A}} ~ \sum_{a_j \in \mathbb{A} - \{a_i\} } \frac{1}{4 \pi \varepsilon_{i,j}} \frac{q_i q_j}{d_{i,j}},
\end{equation}
where $d_{i,j} = \|{\bf r}_i - {\bf r}_j\|_2$ is the interatomic center distance, $q_i$ and $q_j$ are the electrostatic charges, and ${\bf r}_i$ and ${\bf r}_j$ are the position vectors of the pair of atoms $a_i, a_j \in \mathbb{A}$, respectively. $\varepsilon_{i,j} = \kappa_{i,j} \varepsilon_0$ is the `dielectric constant' and is generally larger than vacuum permittivity $\varepsilon_0 \approx 8.854 \times 10^{-12}$ Farads (i.e., $\kappa_{i,j} > 1$). Thus (\ref{eq_02}) can be used to calculate the interactions between charges in the solvent, if a continuum dielectric model is used \cite{Kuriyan2012}. The dielectric constant reflects the ability of the environment to attenuate electrostatic interactions, e.g., $\kappa_{i,j} \sim 80$ for aqueous solvent and $\kappa_{i,j} \sim 2$--$4$ for the interior of the protein \cite{Kuriyan2012}, where the larger value for the former is due to the induced polarization of water molecules. It is worthwhile noting that because of the nonuniformity of the dielectric medium, the most accurate computation of the electrostatic energy requires solving Poisson-Boltzman (PB) differential equations \cite{Fogolari2002}. However, solving PB for every cycle of the KCM simulation is computationally expensive, and approximate alternatives such as generalized Born (GB) model can be used \cite{Still1990,Onufriev2002}. A simple distance-dependent dielectric constant is used here (following the convention in \cite{Kazerounian2005a}) to mimic the polarization effect, with closer interactions weighted more heavily \cite{Weiner1984}.
The resultant Coulombic force ${\bf F}^{\rm elec}_{i} = -\nabla_{{\bf r}_i} G^{\rm elec}$ applied on the atom $a_i$ by other atoms is then obtained as
\begin{align}\label{eq_02_1}
    {\bf F}^{\rm elec}_i (\mathbb{A}) &= \sum_{a_j \in \mathbb{A} - \{a_i\} } \frac{1}{4 \pi \varepsilon_{i,j}} \frac{q_i q_j}{d^2_{i,j}} {\bf e}_{i,j},
\end{align}
where ${\bf e}_{i,j} = ({\bf r}_i-{\bf r}_j)/d_{i,j}$ is the unit vector along the line of centers of the pair of atoms $a_i, a_j \in \mathbb{A}$. Since ${\bf F}^{\rm elec}_i \propto 1/d^2_{i,j}$, electrostatic interactions between atoms that are farther than a so-called cut-off distance $d^{\rm elec}_{\rm cut} := 9.0~\angstrom$ are usually deemed negligible in the literature \cite{Kuriyan2012}.\footnote{Our experiments with larger molecules show that $9.0~\angstrom$ is not always a proper cut-off distance and larger values need to be used, as demonstrated in Section \ref{sec_real}.}
Therefore (\ref{eq_02_1}) is approximated as follows to reduce the number of pairwise computations between all the atoms:
\begin{equation}\label{eq_02_2}
    {\bf F}^{\rm elec}_i (\mathbb{A}^{\rm elec}_i) \approx \sum_{a_j \in \mathbb{A}^{\rm elec}_i} \frac{1}{4 \pi \varepsilon_{i,j}} \frac{q_i q_j}{d^2_{i,j}} {\bf e}_{i,j},
\end{equation}
where $\mathbb{A}^{\rm elec}_i = \{ a_j \in \mathbb{A} - \{a_i\} ~|~ d_{i,j} \leq d^{\rm elec}_{\rm cut} \}$ is referred to as the neighborhood of atom $a_i$ associated with the electrostatic force-field, and consists of all the atoms whose distance to $a_i$ are bounded by the cut-off distance $d^{\rm elec}_{\rm cut}$.

\paragraph{Van der Waals Interactions.}
The van der Waals interactions are formulated using the empirical Lennard-Jones 6-12 potential function formula \cite{Weiner1984}:
\begin{equation}\label{eq_03}
    G^{\rm vdw} (\mathbb{A}) = \sum_{a_i \in \mathbb{A}} ~ \sum_{a_j \in \mathbb{A} - \{a_i\} } \epsilon_{i,j} \left[ \left( \frac{D_{i,j}}{d_{i,j}} \right)^{12} - 2 \left( \frac{D_{i,j}}{d_{i,j}} \right)^6 \right],
\end{equation}
where $d_{i,j} = \|{\bf r}_i - {\bf r}_j\|_2$ is the interatomic center distance, $D_{i,j} = R_i + R_j$ is the `van der Waals distance' in which $R_i, R_j$ are the van der Waals radii of the atoms $a_i, a_j \in \mathbb{A}$, respectively. $\epsilon_{i,j} = \sqrt{\epsilon_i \epsilon_j}$ is the `depth of potential well' for the particular pair of atoms. It is worthwhile noting that the van der Waals effects have the same origin as the electrostatic forces, and reflect the induced dipoles due to transient fluctuations in electron clouds of the interacting atoms \cite{Kuriyan2012}.
The resultant van der Waals force ${\bf F}^{\rm vdw}_{i} =  -\nabla_{{\bf r}_i} G^{\rm vdW}$ on the atom $a_i$ by other atoms is then obtained as
\begin{align}\label{eq_03_1}
     {\bf F}^{\rm vdw}_i (\mathbb{A}) &= \sum_{a_j \in \mathbb{A} - \{a_i\} } 12 \epsilon_{i,j} \left( \frac{D_{i,j}^{12}}{d_{i,j}^{13}} - \frac{D_{i,j}^6}{d_{i,j}^7} \right) {\bf e}_{i,j},
\end{align}
where ${\bf e}_{i,j} = ({\bf r}_i-{\bf r}_j)/d_{i,j}$ is the unit vector along the line of centers of the pair of atoms $a_i, a_j \in \mathbb{A}$. The van der Waals forces have a much smaller effect radius and are significant only when the atoms approach each other very closely. The repulsive component becomes very large when the two atoms penetrate into each other, an effect widely known as the `steric clash'. The attractive component known as the `London dispersion' force, on the other hand, is dominant when the atoms are farther than the van der Waals distance $D_{i,j}$ \cite{Kuriyan2012}. These interactions decay much faster than Coulombic forces, hence a smaller cut-off distance of $d^{\rm vdw}_{\rm cut} := 5.0~\angstrom$ is sufficient \cite{Kuriyan2012} resulting in the following approximation:
\begin{equation}\label{eq_03_2}
    {\bf F}^{\rm vdw}_i (\mathbb{A}^{\rm vdw}_i) \approx \sum_{a_j \in \mathbb{A}^{\rm vdw}_i} 12 \epsilon_{i,j} \left( \frac{D_{i,j}^{12}}{d_{i,j}^{13}} - \frac{D_{i,j}^6}{d_{i,j}^7} \right) {\bf e}_{i,j},
\end{equation}
where $\mathbb{A}^{\rm vdw}_i = \{ a_j \in \mathbb{A} - \{a_i\} ~|~ d_{i,j} \leq d^{\rm vdw}_{\rm cut} \}$ is referred to as the neighborhood of the atom $a_i$ associated with the van der Waals force-field, and consists of all the atoms whose distance to $a_i$ are bounded by the cut-off distance $d^{\rm vdw}_{\rm cut}$.

\paragraph{Interaction Classification.}
The interactions discussed so far are between the pairs of atoms that are {\it not} covalently bonded, thus (\ref{eq_02_2}) and (\ref{eq_03_2}) have to be modified to eliminate the terms corresponding to the pairs of bonded atoms (i.e., `1-2 interactions'). Furthermore, it is a common convention in molecular mechanics to modify the electrostatic and van der Waals interactions between the pairs of atoms bonded to a common atom, i.e., atoms that are 2 bonds apart along the chain (i.e., `1-3 interactions'), as well as the atoms that are 3 bonds apart along the chain (i.e., `1-4 interactions') \cite{Rappe1997}. Consequently, the empirical forms of (\ref{eq_02_2}) and (\ref{eq_03_2}) are modified as
\begin{align}\label{eq_03_8}
   {\bf F}^{\rm elec}_i (\mathbb{A}^{\rm elec}_i) &\approx \sum_{a_j \in \mathbb{A}^{\rm elec}_i} \frac{w^{\rm elec}_{i,j}}{4 \pi \varepsilon_{i,j}} \frac{q_i q_j}{d^2_{i,j}} {\bf e}_{i,j}, \\
   {\bf F}^{\rm vdw}_i (\mathbb{A}^{\rm vdw}_i) &\approx \sum_{a_j \in \mathbb{A}^{\rm vdW}_i} 12 w^{\rm vdw}_{i,j} \epsilon_{i,j} \left( \frac{D_{i,j}^{12}}{d_{i,j}^{13}} -  \frac{D_{i,j}^6}{d_{i,j}^7}  \right) {\bf e}_{i,j}\label{eq_03_9},
\end{align}
where $w^{\rm elec}_{i,j}$ and $w^{\rm vdw}_{i,j}$ are the weight factors for the electrostatic and van der Waals forces, respectively, for the pair of atoms $a_i, a_j \in \mathbb{A}$ depending on their interaction type. The weights are set to $w_{i,j} = 0$ for 1-2 interactions, and $0 \leq w_{i,j} \leq 1$ for 1-3 and 1-4 interactions, whose values vary across different force models \cite{Weiner1984,Cornell1995,Van2002,Brooks2004}. $w_{i,j} = 1$ for all other situations. In other words, the atoms that have at least 4 bonds in between them along the graph of covalent bonds are far enough to be considered unaffected by the covalent electron clouds, as originally formulated in (\ref{eq_02_2}) and (\ref{eq_03_2}).

\begin{figure}
    \centering
    \includegraphics[width=0.48\textwidth]{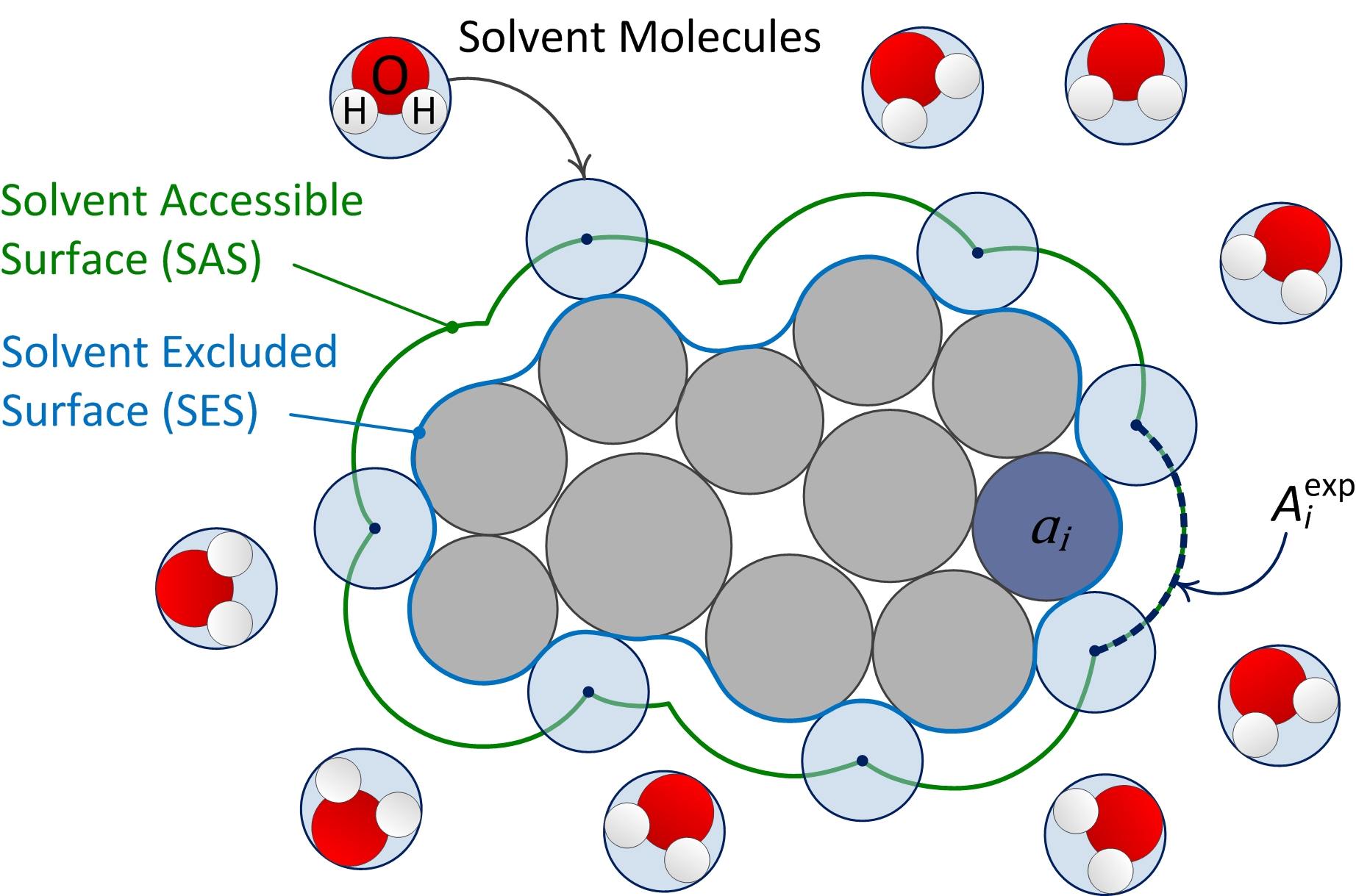}
    \caption{The solvent-accessible and -excluded surfaces.}
    \label{figure2}
\end{figure}

\begin{table}
    \caption{Atomic solvation parameters adopted from \cite{Wesson1992} for different adjustments by \cite{Kyte1982,Sharp1991}. Units are in $kcal~mol^{-1}~\angstrom^{-2}$.} 
    \vspace{-0.5cm}
    \begin{center}
        \begin{tabular}{p{2.2cm} l l}
            \hline
            Atom type & \multicolumn{1}{c}{Adjustment in \cite{Kyte1982}} & \multicolumn{1}{c}{Adjustment in \cite{Sharp1991}} \\
            \hline
            C  & $+0.004 \pm 0.003$ & $+0.012 \pm 0.003$  \\
            O/N  & $-0.113 \pm 0.014$ & $-0.116 \pm 0.013$  \\
            S  & $-0.017 \pm 0.022 $ & $-0.018 \pm 0.021$  \\
            O$^{-}$  & $-0.166 \pm 0.038$ & $-0.175 \pm 0.036$  \\
            N$^{+}$  & $-0.169 \pm 0.031$ & $-0.186 \pm 0.022$  \\
            \hline
        \end{tabular}
    \end{center} \label{tab_SASA}
\end{table}

\paragraph{Nonpolar Solvation Effects.}
The hydrophobic free energy of solvation, which reflects the entropy changes in the solvent molecules due to cavity creation, is formulated using the linear empirical formulation in \cite{Wesson1992}:
\begin{equation}\label{eq_04}
     G^{\rm cav} (\mathbb{A}) = \sum_{a_j \in \mathbb{A}} \gamma_j ~ A^{\rm exp}_j,
\end{equation}
where $\gamma_j$ is the atomic solvation parameter and $A^{\rm exp}_j$ is the Lee and Richards SASA for the atom $a_j \in \mathbb{A}$ \cite{Lee1971}. To obtain the atomic SASA at a given snapshot, a probe radius of $R_{\rm H_2 O} = 1.2-1.4~\angstrom$ is used to offset the van der Waals surfaces of the individual atoms as illustrated in Fig. \ref{figure2}. These offset spheres are overlapped to obtain the contributions of different atoms to the total SASA. The atomic solvation parameter $\gamma_j$ is an estimate of the free energy per unit area required to transfer the atom from vacuum to water, and depends on the atom type \cite{Wesson1992}. Table \ref{tab_SASA} shows the values of this parameter for different atom types (namely, C, uncharged O or N, S, O$^{-}$, and N$^{+}$) obtained in \cite{Wesson1992} based on the experimental results in \cite{Wolfenden1981} adjusted by \cite{Kyte1982,Sharp1991}.
The hydrophobic interaction forces ${\bf F}^{\rm cav}_{i} =  -\nabla_{{\bf r}_i} G^{\rm cav}$ on the atom $a_i$ by other atoms is then obtained as
\begin{equation}\label{eq_04_1}
     {\bf F}^{\rm cav}_i (\mathbb{A}) = -\sum_{a_j \in \mathbb{A}} \gamma_j \nabla_{{\bf r}_i} A^{\rm exp}_j,
\end{equation}
where $\nabla_{{\bf r}_i} A^{\rm exp}_j$ is the gradient of the exposed area of the atom $a_j$ due to an infinitesimal displacement of $a_i$. It is important to note that, unlike the force formulae presented earlier for the electrostatic and van der Waals effects in (\ref{eq_02_1}) and (\ref{eq_03_1}), the summation in (\ref{eq_04_1}) for the solvation free energy gradient iterates over all $a_j \in \mathbb{A}$ {\it including} $a_i$ itself.

Considering the case when $i = j$, one realizes that displacing the atom $a_i$ in any direction has the same effect on the geometry of the protein surface as displacing all the other atoms in the opposite direction. Therefore
\begin{equation}
    \nabla_{{\bf r}_i} A^{\rm exp}_i = - \sum_{a_j \in \mathbb{A} - \{a_i\}} \nabla_{{\bf r}_j} A^{\rm exp}_i.
\end{equation}
Substituting for this term in (\ref{eq_04_1}) leads to the following symmetric form, whose range of summation excludes $a_i$ itself, similar to (\ref{eq_02_1}) and (\ref{eq_03_1}):
\begin{align}
     {\bf F}^{\rm cav}_i (\mathbb{A}) &= - \gamma_i \nabla_{{\bf r}_i} A^{\rm exp}_i - \sum_{a_j \in \mathbb{A} - \{a_i\}} \gamma_j \nabla_{{\bf r}_i} A^{\rm exp}_j  \nonumber \\
     &= \sum_{a_j \in \mathbb{A} - \{a_i\}} \left( \gamma_i \nabla_{{\bf r}_j} A^{\rm exp}_i - \gamma_j \nabla_{{\bf r}_i} A^{\rm exp}_j \right). \label{eq_04_2}
\end{align}
We show in Section \ref{sec_imp} that (\ref{eq_04_2}) is computationally preferable over (\ref{eq_04_1}). To further simplify (\ref{eq_04_2}), note that for a pair of atoms $a_i$ and $a_j$ the infinitesimal displacement of one does not affect the overlapped solvent exposed area of the other if their offset spheres (i.e., the spheres with radii $R^{\rm off}_i = R_i + R_{\rm H_2 O}$ and $R^{\rm off}_j = R_j + R_{\rm H_2 O}$, respectively) do not intersect, i.e., $\nabla_{{\bf r}_i} A^{\rm exp}_j = \nabla_{{\bf r}_j} A^{\rm exp}_i = 0$ if $d_{i,j} > R_i + R_j + 2 R_{\rm H_2 O}$. Therefore, the number of terms that contribute a nonzero value to the summation of (\ref{eq_04_2}) is significantly reduced:
\begin{equation}\label{eq_04_3}
     {\bf F}^{\rm cav}_i (\mathbb{A}^{\rm cav}_i) = \sum_{a_j \in \mathbb{A}^{\rm cav}_i} \left( \gamma_i \nabla_{{\bf r}_i} A^{\rm exp}_j - \gamma_j \nabla_{{\bf r}_j} A^{\rm exp}_i \right),
\end{equation}
where $\mathbb{A}^{\rm cav}_i = \{ a_j \in \mathbb{A} - \{a_i\} ~|~ d_{i,j} \leq R_i + R_j + 2 R_{\rm H_2 O} \}$ is referred to as the neighborhood of the atom $a_i$ associated with the nonpolar solvent effects. For practical reasons that will be explained in Section \ref{sec_prox}, we use a larger neighborhood redefined as $\mathbb{A}^{\rm cav}_i = \{ a_j \in \mathbb{A} - \{a_i\} ~|~ d_{i,j} \leq d^{\rm cav}_{\rm cut} \}$ using the more conservative (but constant across all pairs of atoms) cut-off distance of $d^{\rm cav}_{\rm cut} := 2 (R_{\rm max} + R_{\rm H_2 O})$, where $R_{\rm max} = \max_{a_i \in \mathbb{A}} \{ R_i \}$. A value of $d^{\rm cav}_{\rm cut} := 8.0~\angstrom$ is typically safe. Note that unlike the case with (\ref{eq_02_1}) and (\ref{eq_03_1}), eliminating pairwise interactions with $d_{i,j} > 8.0~\angstrom$ from (\ref{eq_04_3}) does not impart an approximation error.

\subsection{Kinetostatic Simulation} \label{sec_KCM}
We use the KCM (presented in \cite{Kazerounian2004,Kazerounian2004a,Kazerounian2004b,Kazerounian2005,Kazerounian2005a} for protein folding) to explicitly integrate the conformational changes of the linkage model under the kinetostatic effect of the force-field computed in Section \ref{sec_field}.

\paragraph{Link Forces and Torques.}
For a protein chain with a total of $l = O(m)$ links, where $m$ is the number of AA residues, the resultant force and torque applied to the $j^\text{th}$ link $(1 \leq j \leq l)$ are computed as
\begin{align}
    {\bf F}^{\rm link}_j &= \sum_{a_i \in \mathbb{L}_j} \left( {\bf F}^{\rm elec}_i + {\bf F}^{\rm vdw}_i + {\bf F}^{\rm cav}_i \right), \label{eq_KCM_01} \\
    {\bf T}^{\rm link}_j &= \sum_{a_i \in \mathbb{L}_j} {\bf r}_i \times \left( {\bf F}^{\rm elec}_i + {\bf F}^{\rm vdw}_i + {\bf F}^{\rm cav}_i \right), \label{eq_KCM_02}
\end{align}
where ${\bf r}_i$ is the absolute center position vector of the atom $a_i \in \mathbb{A}$ obtained from (\ref{eq_r1}) and (\ref{eq_r2}) in Section \ref{sec_kin} (with different index notation), and $\mathbb{L}_j \subset \mathbb{A}$ is the subset of atoms that belong to the $j^\text{th}$ link along the chain. Note that the origin of the absolute coordinate system (arbitrarily picked the same as the N-terminus anchor of the chain) is selected as the torque center for all links.

\begin{figure}
    \centering
    \includegraphics[width=0.48\textwidth]{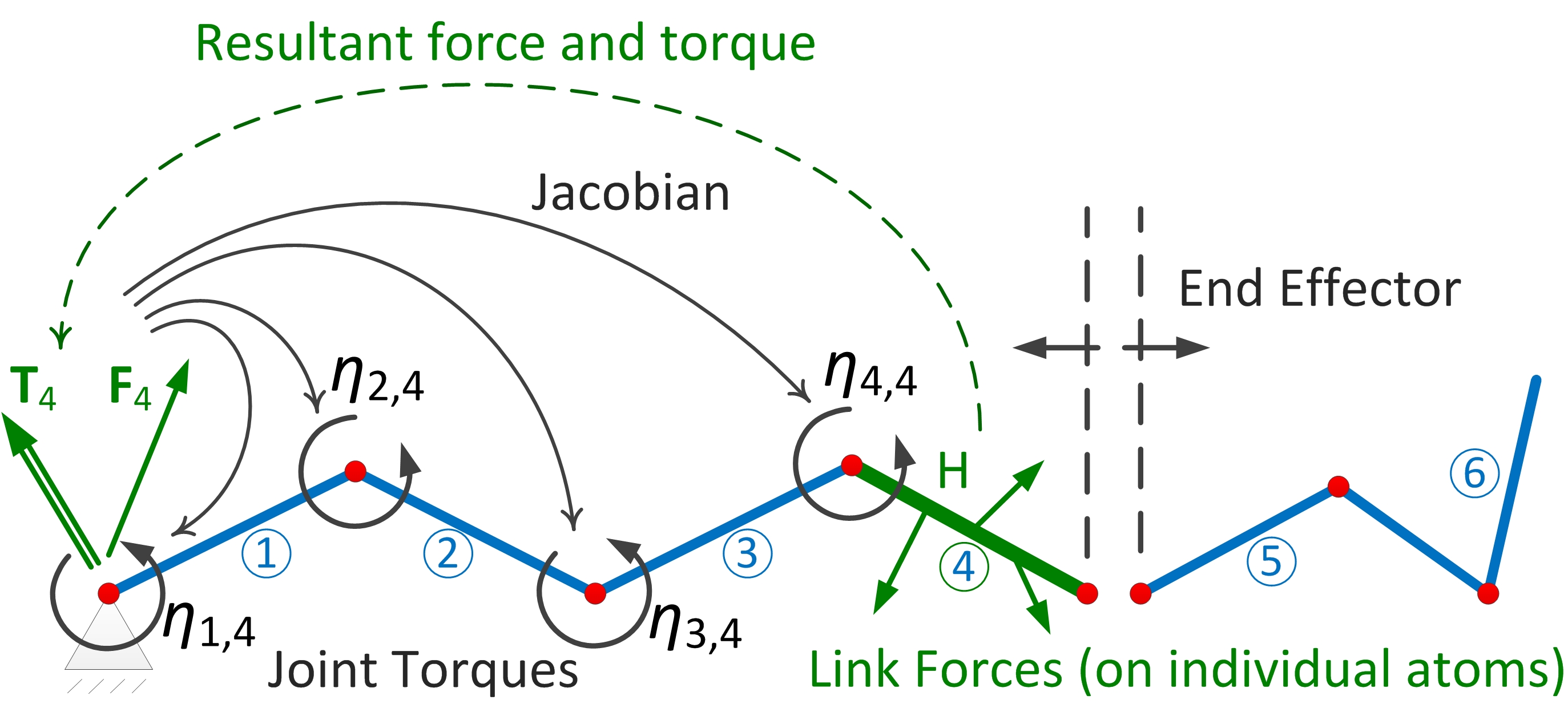}
    \caption{The forces on each link are converted into an equivalent set of joint torques on the preceding joints in the chain.}
    \label{figure3}
\end{figure}

\paragraph{Equivalent Joint Torques.}
Since the revolute joints are assumed to be frictionless, the action of the link forces and torques can be replaced by an equivalent set of torques acting on the joints \cite{Kazerounian2005a}, as shown in Fig. \ref{figure3}. For a given rigid link, one can trace a unique serial chain of $h$ successive links $(1 \leq h \leq l)$ starting from the N-terminus and ending at the link under consideration, which is equivalent to a path along the graph of the open linkage. The contribution of the force ${\bf F}^{\rm link}_h$ and the torque ${\bf T}^{\rm link}_h$ applied to the end-effector link (i.e., the $h^\text{th}$ link along the serial chain) to the joint torque at the $k^\text{th}$ joint along the chain preceding the end-effector, denoted as $\eta_{k,h}~(1 \leq k \leq h, 1 \leq h \leq l)$, can be computed using the conventional manipulator Jacobian matrix $[J]$ \cite{Kazerounian2005a}:
\begin{equation}\label{eq_KCM_04}
    [ \bm{\upeta}_h ] = [J]^{\rm T} \left[ \begin{array}{c} {\bf T}^{\rm link}_h \\ {\bf F}^{\rm link}_h \end{array} \right],
\end{equation}
where $[\bm{\upeta}_h] = [\eta_{1,h}, \eta_{2,h}, \cdots, \eta_{h,h}]^{\rm T}$ represents an $h \times 1$ array of joint torques that the end-effector force and torque will induce on the different joints preceding the end-effector along the serial chain.
$[J]^{\rm T}$ is the transpose of the $6 \times h$ Jacobian matrix $[J] = [ {\bf J}_1, {\bf J}_2, \cdots, {\bf J}_h ]$ for a given configuration of the chain \cite{Kazerounian2005a}. The $k^\text{th}$ column of the Jacobian associated with the $k^\text{th}$ revolute joint is given by
\begin{equation}\label{eq_KCM_05}
    [ {\bf J}_k ] = \left[ \begin{array}{c} {\bf u}_k \\ {\bf u}_k\times ({\bf p}_h - {\bf p}_k) \end{array} \right],
\end{equation}
where ${\bf u}_k$ is the unit vector along the $k^\text{th}$ joint and $({\bf p}_h - {\bf p}_k)$ is a vector connecting a point ${\bf p}_k$ along the $k^\text{th}$ joint's axis to the point ${\bf p}_h$ where the end-effector force ${\bf F}_h$ applies---i.e., the atom positions obtained from (\ref{eq_r1}) and (\ref{eq_r2}). This process is repeated for all links to calculate the contribution of each link on the joints preceding that link in the serial chain. The total torque for each joint is obtained from a summation of these contributions \cite{Kazerounian2005a}:
\begin{equation}\label{eq_KCM_06}
    \tau_k = [{\bf J}_k]^{\rm T} \sum_{h = k}^l \left[ \begin{array}{c} {\bf T}^{\rm link}_h \\ {\bf F}^{\rm link}_h \end{array} \right],
\end{equation}
where the indexing $h = k, k+1, \cdots, l$ of the links is ordered along the main chain from amino-terminus to carboxyl-terminus, and can branch along the side chains. The range of the summation in (\ref{eq_KCM_06}) implies that each joint torque $\tau_k ~(1 \leq k \leq l)$ depends on the forces ${\bf F}^{\rm link}_h$ and torques ${\bf T}^{\rm link}_h$ on the succeeding links $(k \leq h \leq l)$.

\paragraph{Kinetostatic Simulation.}
Making use of the assumption in \cite{Kazerounian2005a} that the inertia forces at the atomic scale have negligible effects on the dynamics of folding compared to interatomic forces, the successive kinetostatic fold compliance relates the joint torques to the changes in the dihedral angles as follows:
\begin{equation}\label{eq_KCM_07}
    \Delta \theta_j = \kappa \frac{\tau_j}{|\tau_{max}|},
\end{equation}
where $\Delta \theta_j$ and $\tau_j$ are the angular change and the joint torque of the $j^\text{th}$ revolute joint $(1 \leq j \leq l)$, respectively. $|\tau_{max}|$ is the maximum joint torque throughout the entire chain, used to normalize the torques to the interval $[0, 1]$, and the coefficient $\kappa$ is chosen small enough to avoid large changes in the angles, and to achieve numerical stability. One can notice that the conformational change computed in (\ref{eq_KCM_07}) is analogous to taking a step along the steepest-descent direction of the free energy gradient in the conformation landscape.

The computed changes in the joint angles are applied to the kinematic chain, and the entire process is repeated on the updated conformation until a convergence criteria is reached, as described in more detail is Section \ref{sec_imp}.

It is worthwhile noting that once the chain conformation (i.e., optimization variables) is modeled as in Section \ref{sec_kin} and the energy-field (i.e., objective function) is formulated as in Section \ref{sec_field}, the search for local or global minima of the free energy in (\ref{eq_01}) can be undertaken using a variety of classical (e.g., conjugate-gradients) and stochastic (e.g., genetic algorithm) optimization methods. Since the focus of this article is mainly on force-field modeling and computing, we skip a detailed treatment of the search phase.

\section{Algorithms} \label{sec_alg}

This section presents efficient algorithms and data structures to speed up the force field computation during kinetostatic iterations. To leverage the proximity information between the atoms, we use a 3D hash table data structure based on a uniform spatial grid in Section \ref{sec_prox}. To classify the interatomic interaction types based on chain topology to compute the electrostatic and van der Waals effects, we use a tree-based data structure in Section \ref{sec_tree} that replaces the adjacency matrix used in {\sf Protofold I} \cite{Kazerounian2004a,Kazerounian2004b,Kazerounian2005a}. To compute the solvation effects, we develop an approximate (yet adequately accurate) surface enumeration technique in \ref{sec_enum}, efficient CPU- and GPU-parallel implementations of which will be detailed in Section \ref{sec_imp}. Finally, we compute joint torques by aggregating contributions of different links (traversed along different paths in the linkage graph) on the joints along the chain, using the well-known `prefix computation' in Section \ref{sec_pref} which can also be implemented efficiently in parallel \cite{Cole1989}. We show that the computational complexity of all steps is decreased from $O(n^2)$ in {\sf Protofold I} \cite{Kazerounian2004a,Kazerounian2004b,Kazerounian2005a} to expected $O(n)$ in {\sf Protofold II} for a protein chain with a total of $n$ atoms.

\subsection{Rigid Transformations} \label{sec_rigid}
At every snapshot $t \geq 0$ of the KCM, the protein conformation is described by a set of dihedral angles $\theta^t_{j,k}$ defined in (\ref{theta_1}) through (\ref{theta_3}).
\begin{itemize}
    \item At the first iteration $(t = 0)$, all angles are initialized as $\theta^0_{j,k} = 0$ (ZP conformation).
    \item At subsequent iterations $(t \geq 1)$, for $1 \leq j \leq 2m$ (where $m$ is the number of AA residues) and $0 \leq k \leq l_i $ (where $0 \leq l_i \leq 4$ is the number of side chain links of the residue $AA_i$), the angles are obtained as $\theta^t_{j,k} = \theta^{t-1}_{j,k} + \Delta \theta^{t-1}_{j,k}$, where the increment $\Delta \theta^{t-1}_{j,k}$ is computed using (\ref{eq_KCM_07}) from the previous step's configuration and joint torques.
\end{itemize}
Once the dihedral angles are known, the transformation matrices $[M^t_{j,k}]$ are obtained from (\ref{eq_kin_04}) and (\ref{eq_kin_05}) using sequential matrix multiplication traversing the linkage tree from the anchored amino-terminus to the open carboxyl-terminus. Next, the unit vectors ${\bf u}^t_{j,k}$ and the body vectors ${\bf b}^t_{j,k}$ are computed from (\ref{eq_kin_02}) and (\ref{eq_kin_03}). Since the number of links is clearly less than the number of atoms, these computations take $O(n)$ time.
The Cartesian coordinates of the individual atom center positions ${\bf r}_i \in \mathds{R}^3 ~(1 \leq i \leq n)$ are obtained from the body vectors using (\ref{eq_r1}) and (\ref{eq_r2}), which also takes $O(n)$. Hereon, we assume that both dihedral angles and atom center positions are known for the purpose of computing the next snapshot's energies, forces, and torques.

\subsection{Proximity Queries} \label{sec_prox}
The brute-force approach for obtaining the proximity information at each snapshot is to check center distances against the cut-off distance for all possible pairs of atoms, which takes $O(n^2)$ time. Using this method, the approximate truncated formulae given in (\ref{eq_02}), (\ref{eq_03}), and (\ref{eq_04}) would take the same asymptotic time as the exact all-pairs formulae given in (\ref{eq_02_1}), (\ref{eq_03_1}), and (\ref{eq_04_2}), respectively, which is $O(n^2)$. Geometric hashing provides a simple solution to speed up proximity queries.

\paragraph{3D Hash Table.}
We use a 3D hash table data structure based on a uniform Cartesian grid, which bounds the current 3D structure of the protein and arranges the atom indices into groups based on their center positions. Each 3D grid cell is associated with a so-called `bucket' that stores the indices of the atoms whose centers are located inside the cell in a linked-list. The grid dimensions are set dynamically to adapt to the shape of the protein's bounding box at the current snapshot.

The grid cells are chosen to be cubic, i.e., with equal edge length $s_c$ along all 3 Cartesian axes. Given the min/max corner coordinates of the bounding box of the atom centers ${\bf r}_{\rm min}, {\bf r}_{\rm max} \in \mathds{R}^3$---which can be obtained in $O(n)$ by scanning through the $n$ atom center coordinates---we choose $s_c$ in such a way that it results in $\lceil \alpha n \rceil$ grid cells/buckets, where $\alpha > 0$ is an arbitrary constant. More precisely, we choose $s_c = [v_{\rm BB}(\mathbb{A})/(\alpha n)]^\frac{1}{3}$ where $v_{\rm BB}(\mathbb{A})$ is the protein bounding box volume. The dimensions of the grid bounding box are then chosen as $\lceil ({\bf r}_{\rm max} - {\bf r}_{\rm min}) / s_c \rceil s_c$ (slightly larger than the dimensions of the protein bounding box ${\bf r}_{\rm max} - {\bf r}_{\rm min}$), where the operator $\lceil \cdot \rceil$ is applied componentwise along the 3 Cartesian axes.
Before we proceed with presenting the complexity analysis, we make the following assumptions:

\begin{assumption} \label{assum_1}
    {\rm
    Due to the extremely strong repulsive van der Waals forces, the atoms that are not covalently bonded cannot penetrate into each other, and those covalently bonded intersect over a small volume. Given any arbitrary subset of atoms $\mathbb{A}' \subseteq \mathbb{A}$ with $R_{\rm min} = \min_{a_i \in \mathbb{A}'} R_i$ and $R_{\rm max} = \max_{a_i \in \mathbb{A}'} R_i$, let the maximum penetration volume between any pair of covalently bonded atoms $a_i, a_j \in \mathbb{A}'$ be upper-bounded by $\epsilon \min \{v_i, v_j \}$ where $v_i = \frac{4\pi}{3} R^3_i$ is the volume of the atom $a_i$, and $0 \leq \epsilon < \frac{1}{4}$ is a small number. Since each atom makes at most 4 covalent bonds, the unpenetrated volume for the atom $a_i$ is lower-bounded by $(1-4\epsilon)v_i$, hence it is safe to assume that $(1-4\epsilon) > 0$. Then the volume $v(\mathbb{A}')$ occupied by the union of all the atoms in $\mathbb{A}'$ is bounded as $\frac{4\pi}{3} (1-4\epsilon) |\mathbb{A}'| R_{\rm min}^3 \leq v(\mathbb{A}') \leq \frac{4\pi}{3} |\mathbb{A}'| R_{\rm max}^3$. Consequently, there exists an `average' radius $\bar{R}(\mathbb{A}')$ bounded as $(1-4\epsilon)^\frac{1}{3} R_{\rm min} \leq \bar{R}(\mathbb{A}') \leq R_{\rm max}$, such that $v(\mathbb{A}') = \frac{4\pi}{3} |\mathbb{A}'| \bar{R}^3(\mathbb{A}')$, where typically $\bar{R} \sim 1~\angstrom$.
    }
\end{assumption}

\begin{assumption} \label{assum_2}
    {\rm
    It is also reasonable to assume that if the protein is either in an extended conformation aligned with one of the Cartesian axes (which is the case near the ZP conformation) or in a globular conformation (which is the case for most water-soluble proteins at their folded conformation), the empty space inside the bounding box is not extremely larger than the space occupied by the protein atoms, i.e., $v_{\rm BB}(\mathbb{A}) / v(\mathbb{A}) = O(1)$. Supported by experimentation, we assume this to be the case in the intermediate conformations as well, to simplify the analysis. However, there are possible conformations (e.g., extended along a diagonal direction in the axis-aligned bounding box) that would violate this assumption and result in slightly larger running times than predicted here, in spite of the low probability.
    }
\end{assumption}

Letting $\mathbb{A}' := \mathbb{A}$ (hence $|\mathbb{A}'| = n$) in Assumption \ref{assum_1}, and noting from the definition that $v_{\rm BB}(\mathbb{A}) = (\alpha n) s_c^3$, from Assumption \ref{assum_2} it follows that $\frac{3\alpha}{4\pi} (s_c/\bar{R})^3 = O(1)$. Therefore, if we choose the grid cell size $s_c \sim 1~\angstrom$ then $\alpha = O(1)$ and the number of buckets will be $\lceil \alpha n \rceil = O(n)$.

\begin{figure}
    \centering
    \includegraphics[width=0.48\textwidth]{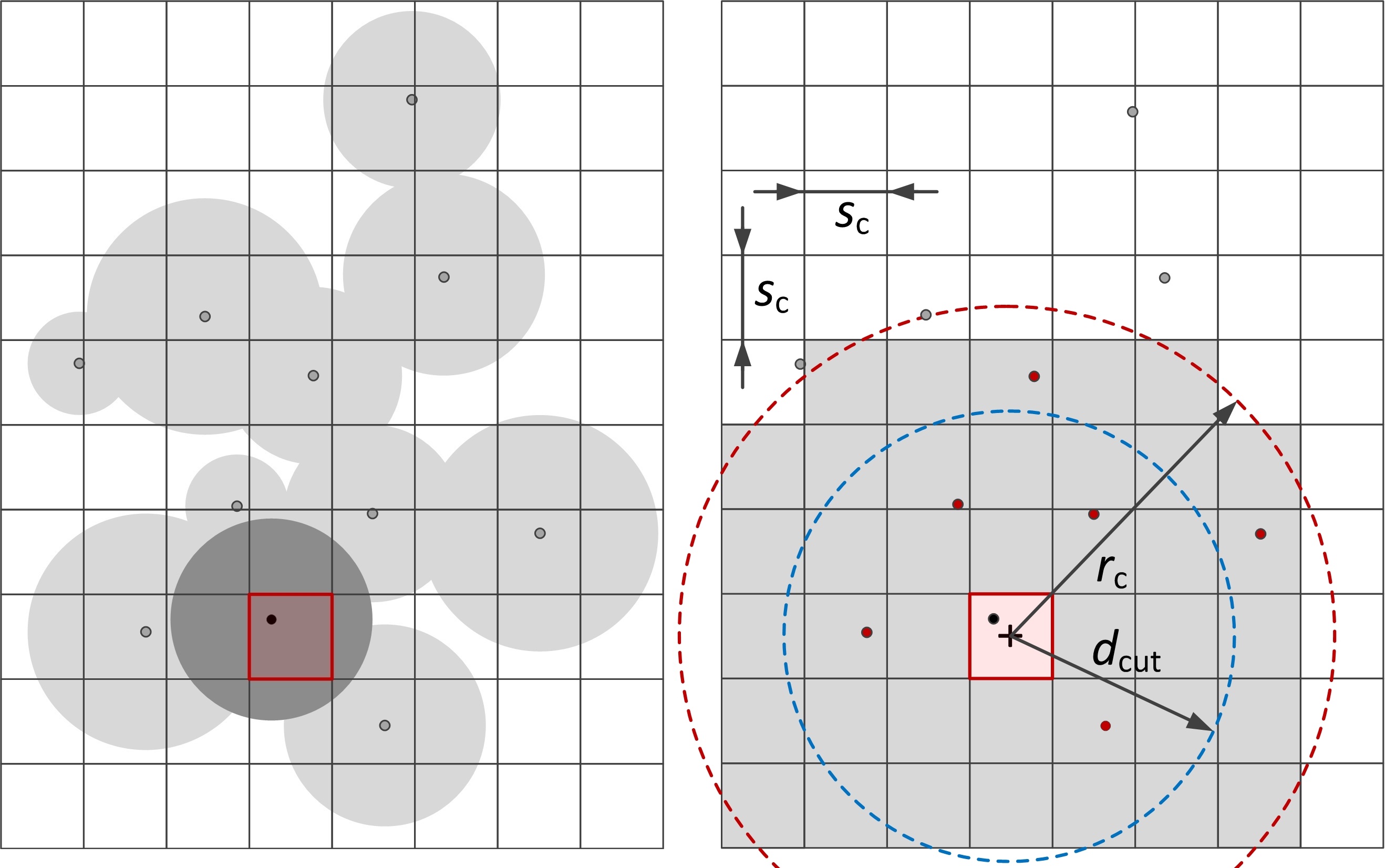}
    \caption{Atom centers are hashed into a 3D grid and the neighbors are selected within a cut-off distance.}
    \label{Figure4}
\end{figure}

\paragraph{Table Construction.}
At each snapshot of the simulation, the algorithm scans through all the atoms with updated positions, and the deterministic hash function simply maps the atom center positions ${\bf r}_i \in \mathds{R}^3$ that lie inside a grid cell into the corresponding 3D array of buckets. The 3D index ${\bf k} \in \mathds{Z}^3 \cap [0, +\infty]^3$ of the bucket to which a given atom $a_i \in \mathbb{A}$ belongs is determined in $O(1)$ time as ${\bf k} = \lfloor ({\bf r}_{i} - {\bf r}_{\rm min}) / s_c \rfloor$, where the operator $\lfloor \cdot \rfloor$ is applied componentwise along the 3 Cartesian axes. Therefore, scanning through the atoms and constructing the 3D grid data structure is expected to $O(n)$ time and to requires $O(n)$ space.

\paragraph{Neighbor Queries.}
Once the atoms are arranged into the buckets, the algorithm iterates through the grid cells and scans through the linked-lists within the buckets. For each atom $a_i$ in a given bucket associated with the grid cell index ${\bf k}$, the set of `neighbor atoms' defined as
\begin{equation} \label{eq_neigh}
    \mathbb{A}_i = \{ a_j \in \mathbb{A} ~|~ \| {\bf r}_i - {\bf r}_j \|_2 \leq d_{\rm cut} \}, \quad d_{\rm cut} \in \{d_{\rm cut}^{\rm elec}, d_{\rm cut}^{\rm vdw}, d_{\rm cut}^{\rm cav}\}
\end{equation}
can be identified rapidly for a given cut-off distance $d_{\rm cut}$ associated with any of the energetic interactions explained in Section \ref{sec_field}.
As illustrated in Fig. \ref{Figure4}, a spherical region of radius $r_c = d_{\rm cut} + \sqrt{3} s_c$ is considered around the (center point of) each grid cell to look for the (center point of) `neighbor cells', defined as the collection of cells which {\it completely} lie inside this spherical region.
The cut-off distance $d_{\rm cut}$ is offset by the diagonal size of the cells $\sqrt{3} s_c$ which takes into account the worst-case difference of the distance between cell centers and the distance between atom centers.
This guarantees that the set of all the atoms inside this collection of covered cells (denoted as $\mathbb{A}_i'$) contains the set of all neighbor atoms, i.e., $\mathbb{A}_i' \supseteq \mathbb{A}_i$, where $\mathbb{A}_i$ is one of the neighbor sets $\mathbb{A}^{\rm elec}_i$, $\mathbb{A}^{\rm vdw}_i$, or $\mathbb{A}^{\rm cav}_i$.
Letting $\mathbb{A}' := \mathbb{A}_i'$ in Assumption \ref{assum_1}, the volume occupied by this set of atoms is $v(\mathbb{A}_i') = \frac{4\pi}{3} |\mathbb{A}_i'| \bar{R}^3(\mathbb{A}_i')$. Noting that $\mathbb{A}_i'$ is contained inside the spherical region of radius $r_c$, $v(\mathbb{A}_i') \leq \frac{4\pi}{3} r_c^3$ hence $|\mathbb{A}_i'| \leq (r_c/\bar{R})^3 = [(d_{\rm cut} + \sqrt{3} s_c)/\bar{R}]^3 = O(1)$, since $\bar{R}, s_c \sim 1~\angstrom$ and $d_{\rm cut} \sim 10~\angstrom$. As a result, it is expected to take $O(1)$ time to scan through the atoms inside the collection of neighbor cells, and $O(n)$ total time to traverse all pairs of atoms using the 3D hash table.

For parallel implementation purposes, we construct (and dynamically maintain) a `neighborhood matrix' composed of an array of $n$ $O(1)-$sized linked-lists (one list per atom), where the $i^\text{th}$ list $(1 \leq i \leq n)$ contains the indices of the neighbor atoms $\mathbb{A}_i'$. Constructing this data structure is expected to take $\sum_{i = 1}^n |\mathbb{A}_i'| = \sum_{i = 1}^n O(1) = O(n)$ time and space, and accessing each atom's neighbors is expected to take $|\mathbb{A}_i'| = O(1)$ time.

\paragraph{Energy and Force Computations.}
Once the pairs of neighbors are identified, computing their electrostatic and van der Waals forces can be done in $O(1)$ time per pair, using the analytical truncated equations given in (\ref{eq_03_8}) and (\ref{eq_03_9}), respectively.
It is important to note that such interactions are strictly {\it pairwise}, i.e., they depend on the relative positions of pairs of atoms $a_i\in \mathbb{A}$ and $a_j \in \mathbb{A}^{\rm elec}_i$ or $\mathbb{A}^{\rm vdw}_i$, and the presence of any third atom $a_k \in \mathbb{A}~(k \neq i,j)$ does not affect the force exchanged between $a_i$ and $a_j$. Therefore, the computation algorithm is straightforward: it iterates over all the atoms for $1 \leq i \leq n$ (sequentially or in parallel) and for each atom, it computes ${\bf F}^{\rm elec}_i (\mathbb{A}_i')$ and ${\bf F}^{\rm vdw}_i (\mathbb{A}_i')$, by sequentially aggregating the contributions of the hashed neighbor atoms $a_j \in \mathbb{A}_i'$, using (\ref{eq_03_8}) and (\ref{eq_03_9}), respectively.

Unfortunately, this is {\it not} the case for solvation force computation using (\ref{eq_04_2}) or (\ref{eq_04_3}), which requires computing the gradients of the atomic SASA with respect to the coordinates of the set of neighbor atoms. The SASA variations in one atom $a_i\in \mathbb{A}$ with respect to an infinitesimal change in the position of another atom $a_j \in \mathbb{A}^{\rm cav}_i$, can be affected by the presence of a third atom $a_k \in \mathbb{A}^{\rm cav}_i~(k \neq i,j)$, thus cannot be obtained in a pairwise fashion. This is because the overlaps of the pairs of offset spheres are not mutually disjoint. A segment of the offset surface can be overlapped with more than one neighbor sphere simultaneously, thus displacing one of the overlapping spheres may or may not affect the SASA.
We return to this subject in Section \ref{sec_enum} where a surface sampling algorithm is proposed as a simple solution.

\subsection{Bonds Tree/Graph} \label{sec_tree}

To decide the weight factors $w^{\rm elec}_{i,j}$ in (\ref{eq_03_8}) and $w^{\rm vdw}_{i,j}$ in (\ref{eq_03_9}), one needs to quickly identify the types of interactions based on the number of bonds between pairs of atoms as described in Section \ref{sec_field}. An $n \times n$ look-up table was used in {\sf Protofold I} \cite{Kazerounian2004a,Kazerounian2004b,Kazerounian2005a} for all pairs of $n$ atoms which required a preprocessing step with $O(n^2)$ time and space. This can be improved by constructing a tree/graph data structure that stores the combinatorial structure of the chain, in which the vertices are atoms and edges are the covalent bonds between them.
By excluding a single edge from the loops associated with the rare aromatic groups in certain side chains (e.g., imidazole in His and indole in Trp), this graph can be converted to a tree whose root is arbitrarily chosen as the N atom of the amino-terminus. The interaction types are then identified by the shortest path lengths between atoms.

The algorithm starts from the root and visits all vertices using a standard tree traversal routine. For each vertex, it stores a pointer to its parent and the index of the corresponding atom's AA residue. During the force computation in each KCM iteration, the residue indices for a pair of atoms of interest are checked. If the atoms are farther than a residue apart (i.e., if AA indices are neither identical nor consecutive), the weights in (\ref{eq_03_8}) and (\ref{eq_03_9}) are simply set to 1 (i.e., 1-4 interactions or beyond). Otherwise, the algorithm checks 1) if one atom is the parent of the other (i.e., 1-2 interaction); 2) if one atom is the grand parent or sibling of the other (i.e., 1-3 interaction); or 3) if one atom is the great-grand parent or sibling of parent of the other (1-4 interactions). This requires $O(n)$ time and space for preprocessing and $O(1)$ query time during the KCM iterations.

\subsection{Surface Enumeration} \label{sec_enum}

As mentioned in Section \ref{sec_lit}, several attempts have been made to approximate the SASA and its derivative by a pairwise treatment of the overlaps, including the probabilistic methods \cite{Wodak1980,Fraternali1996,Weiser1999,Allison2011}, popular in many molecular simulation software such as CHARMM \cite{Brooks2004}, GROMOS \cite{Van2002}, and AMBER \cite{Weiner1984,Cornell1995}. In an early attempt to add solvation effects to {\sf Protofold II}, we used these pairwise approximate formulae, which made it possible to compute the solvation forces with running times comparable to those of the electrostatic and van der Waals force computations. However, a comparison with the exact method \cite{Richmond1984} showed that when the distribution of the atoms deviates from that assumed in the the probabilistic methods, prohibitively large errors can be introduced into the resultant effects. The exact method \cite{Richmond1984} takes $O(|\mathbb{A}_i'|)$ operations for computing the SASA and its gradient for $a_i \in \mathbb{A}$ by using the coordinates of all neighbor atoms $\mathbb{A}^{\rm cav}_i \subseteq \mathbb{A}_i'$. This is asymptotically $O(1)$ time per atom (since we reasoned earlier that $|\mathbb{A}_i'| = O(1)$ under Assumption \ref{assum_1}), but in practice it is not nearly as fast as using the pairwise formulae. Alternatively, we use an approximate method that relies on an enumeration of the surface area, in which the deviations from the exact results can be controlled to a desired precision in a trade-off with computation time.

\paragraph{Offset Sphere Sampling.}
For a given atom $a_i \in \mathbb{A}$ of van der Waals radius $R_i$, an offset sphere of radius $R^{\rm off}_i = R_i + R_{\rm H_2 O}$ concentric with the atom sphere is considered. The atom's SASA is obtained by measuring the area $A^{\rm exp}_i$ of the portion of the offset surface that is not overlapped by the offset sphere of any neighbor atom $a_j \in \mathbb{A}^{\rm cav}_i$ (hence exposed to the solvent). To approximate the SASA, one can generate a large but finite `quasi-uniform' set of sample points denoted as $Q_i~(1 \leq i \leq n)$ on the surface of the offset sphere of the atom $a_i \in \mathbb{A}$, by which we mean a sampling that allows approximating the exposed fraction of the surface by the ratio of the number of exposed sample points to the total number of sample points.
In other words, if we let
\begin{equation}
    Q^{\rm exp}_i = \{ {\bf q} \in Q_i ~|~ \nexists a_j \in \mathbb{A}_i' ~:~ \| {\bf q} - {\bf r}_j \|_2 \leq R^{\rm off}_j \}
\end{equation}
be the subset of the solvent-exposed sample points, i.e., the points that are outside the offset spheres of all neighbors, then $\lim_{|Q_i| \rightarrow \infty} |Q^{\rm exp}_i| / |Q_i| = A^{\rm exp}_i/ A^{\rm off}_i$. If we define the `exposure ratio' as $f^{\rm exp}_i = |Q^{\rm exp}_i| / |Q_i|$, the SASA can be approximated as $A^{\rm exp}_i \approx f^{\rm exp}_i A^{\rm off}_i$ where $A^{\rm off}_i = 4\pi (R^{\rm off}_i)^2$, using a large enough sample size $|Q_i| \gg 1$.\footnote{An alternative approach is uniform random sampling, e.g., using the simple method in \cite{Muller1959}. Random sampling is easier to implement in parallel since every sample point in $Q_i$ would be independent from others, and results in $A^{\rm exp}_i = [f^{\rm exp}_i] A^{\rm off}_i$ where $[f^{\rm exp}_i]$ is the {\it expected} ratio of the exposed sample points in probabilistic terms. However, it requires much larger sample sizes to approach the expectation and to achieve adequate accuracy in practice.}

There are different ways to obtain a quasi-uniform deterministic sampling on a sphere with consistent incremental quality \cite{Yershova2004}. For example, one could use a triangular spherical meshing algorithm, which starts from an icosahedron approximation of the sphere and recursively creates successive triangular subdivisions projected back on the sphere. Alternatively, one could use a polar geodesic sampling algorithm, which starts from a set of orbits with uniform angular distribution and samples a number of points uniformly on each orbit proportional to the orbit's circumference.
We take the latter approach whose details are presented in \cite{Behandish2013}.

To improve the efficiency, one could always precompute the coordinates for a single sampling $Q$ on a unit sphere centered at the origin, and map it into individual atoms with different offset sphere center positions ${\bf r}_i$ and radii $R^{\rm off}_i$ using the mapping $Q_i = T_i(Q)$ where $T_i({\bf q}) = {\bf r}_i + R^{\rm off}_i{\bf q}$ for $1 \leq i \leq n$ and ${\bf q} \in Q$. This implies an equal number of sample points $N = |Q|$ for all the atoms, selected as $N = 4\pi (R^{\rm off}_{\rm max})^2 / \delta A$ where $R^{\rm off}_{\rm max} = R_{\rm max} + R_{\rm H_2 O}$ is the maximum offset sphere radius, and $\delta A$ is the desired characteristic area element carried by each sample point. Hence the sampling takes $O(nN)$ operations for all the atoms regardless of the sampling technique. For implementation purposes, we assign an arbitrary ordering to the sample points, letting $Q = \{ {\bf q}_1, {\bf q}_2, \cdots, {\bf q}_N\}$ and denoting transformed sample points as ${\bf q}_{i,k} = T_i({\bf q}_k)$.

\paragraph{Energy and Force Approximations.}
Substituting for $A^{\rm exp}_i \approx f^{\rm exp}_i A^{\rm off}_i = 4\pi f^{\rm exp}_i (R^{\rm off}_i)^2$ into (\ref{eq_04}), the total solvation free energy can be computed directly from
\begin{equation} \label{eq_G_cav_patch}
    G^{\rm cav}(\mathbb{A}) \approx \sum_{a_j \in \mathbb{A}} 4\pi \gamma_j f^{\rm exp}_j (R^{\rm off}_j)^2.
\end{equation}
To obtain the solvation force on any atom $a_i \in \mathbb{A}$, the energy must be differentiated with respect to the atom's center coordinates ${\bf r}_i \in \mathds{R}^3$, giving rise to (\ref{eq_04_2}). It is very important to note that an infinitesimal displacement of the atom $a_i$ can change the SASA of $a_i$ itself, as well as that of the neighbor atoms $a_j \in \mathbb{A}^{\rm cav}_i \subseteq \mathbb{A}_i'$. However, we showed in Section \ref{sec_field} that the two effects are geometrically dependent, yielding the symmetric form in (\ref{eq_04_3}). From a computational perspective, (\ref{eq_04_3}) is preferred over (\ref{eq_04_3}), because
\begin{enumerate}
    \item it eliminates the need for computing the gradient $\nabla_{{\bf r}_j} A^{\rm exp}_i$ for the cases when $i = j$, hence decreasing the number of such computations from $n^2$ to $n(n-1)$; and
    \item its symmetric form lends itself to a data-parallel implementation that is balanced between computation and data sharing tasks, as we show in Section \ref{sec_imp}.
\end{enumerate}
This follows from the fact that for a given pair of indices $i \neq j$, the first term in the formula for ${\bf F}^{\rm cav}_i$ is identical to the second term in the formula for ${\bf F}^{\rm cav}_j$ (both given by (\ref{eq_04_3})), cutting the number of required SASA gradient computations to $n(n-1)/2$. Substituting for $A^{\rm exp}_i$ and $A^{\rm exp}_j$ in (\ref{eq_04_3}), the solvation forces can be approximated as
\begin{align}
    {\bf F}^{\rm cav}_i (\mathbb{A}_i') &\approx 4 \pi \gamma_i (R^{\rm off}_i)^2 \sum_{a_j \in \mathbb{A}_i'} \nabla_{{\bf r}_j} f^{\rm exp}_i \nonumber \\
    &- 4 \pi \sum_{a_j \in \mathbb{A}_i'} \gamma_j(R^{\rm off}_j)^2 \nabla_{{\bf r}_i} f^{\rm exp}_j, \label{eq_F_cav_patch}
\end{align}
where $\nabla_{{\bf r}_j} f^{\rm exp}_i$ and $\nabla_{{\bf r}_i} f^{\rm exp}_j$ can be approximated using the forward-difference method from finite variations of $f^{\rm exp}_i$ and $f^{\rm exp}_j$ with respect to the positions of the atoms $a_i$ and $a_j$, respectively:
\begin{equation} \label{eq_finitediff}
    \nabla_{{\bf r}_j} f^{\rm exp}_i \approx \sum_{s = 1,2,3} \frac{f^{\rm exp}_{i,j,s} - f^{\rm exp}_i}{\delta r} {\bf e}_s,
\end{equation}
where $\delta r > 0$ is the finite difference, ${\bf e}_s ~(s = 1, 2, 3)$ are the unit vectors along the 3 Cartesian axes, and $f^{\rm exp}_{i,j,s}$ are the exposure ratio of the atom $a_i \in \mathbb{A}$ after changing the position of the neighbor atom $a_j \in \mathbb{A}_i'$ from the current value ${\bf r}_j$ to a hypothetical variant ${\bf r}_j + \delta r {\bf e}_s$.

\paragraph{Enumeration Algorithm.}
In order to compute the exposure ratio and its finite-difference gradient, we use a binary enumeration function $B: \mathbb{A} \times Q \rightarrow \{0, 1\} ~(1 \leq i \leq n)$ such that $B(a_i, \mathbf{q}_k) = 1$ if the sample point ${\bf q}_{i,k} \in T_i(Q)$ on the offset sphere of the atom $a_i \in \mathbb{A}$ is overlapped by at least one neighbor offset sphere (i.e., if $\exists a_j \in \mathbb{A}_i'$ such that $\| T_i({\bf q}_k) - {\bf r}_j \|_2 \leq R^{\rm off}_j$) and $B(a_i, \mathbf{q}_k) = 0$ if the sample point is exposed to the solvent. The algorithm iterates over all the atoms for $1 \leq i \leq n$ and all the sample points for $1 \leq k \leq N$ (sequentially or in parallel). For each sample point, the indicator $B_{i,k} := B(a_i, \mathbf{q}_k)$ is initialized to $0$, and each point is tested against the set of neighbors $\mathbb{A}_i'$, scanned sequentially. As soon as one overlapping neighbor is found, $b_{i,k}$ is set to $1$ and there is no need to test the rest of the neighbors. The exposure ratio is then computed as
\begin{equation}
    f^{\rm exp}_i = 1 - \sum_{\mathbf{q}_k \in Q} \frac{B(a_i, \mathbf{q}_k)}{|Q|} = 1 - \sum_{k = 1}^{N} \frac{B_{i,k}}{N}.
\end{equation}
In the worst case, this takes $|\mathbb{A}_i'| N$ tests and $N$ binary sums per atom where $N = |Q|$ is the sample size, which adds to $O(N)$ basic operations per atom (since we reasoned earlier that $|\mathbb{A}_i'| = O(1)$ under Assumption \ref{assum_1}), and a total of $O(nN)$ time for all the atoms.

\begin{algorithm}
    \SetKwInput{KwInput}{Input}
    \SetKwInput{KwOutput}{Output}
    \KwInput{${\bf r}_i, R^{\rm off}_i, \gamma_i, Q_i, \text{ and } \mathbb{A}_i'$ for all $a_i \in \mathbb{A}~(1 \leq i \leq n)$;}
    \KwOutput{$G^{\rm cav}_i \text{ and } {\bf F}^{\rm cav}_i$ for all $a_i \in \mathbb{A}~(1 \leq i \leq n)$;}
    \vspace{-0.1in}
    \hrulefill\\
    \For{$1 \leq i \leq n$ {\rm (seq. or in $\parallel$)}}
    {
        {\bf Step 1: Energy Computation:}\\
        initialize $f^{\rm exp}_i \leftarrow 1$\;
        initialize $G^{\rm cav}_i \leftarrow G^{\rm cav}_{i,0} \leftarrow 4\pi \gamma_i (R^{\rm off}_i)^2$\;
        \For{$1 \leq k \leq |Q_i|$ {\rm (seq. or in $\parallel$)}}
        {
            initialize $C_{i,k} \leftarrow 0$; $j^{\rm over}_{i,k} \leftarrow -1$\;
            \For{$j =$ indices of atoms in $\mathbb{A}_i'$ {\rm (seq.)}}
            {
                \If{$\| {\bf q}_{i,k} - {\bf r}_j \|_2 \leq R^{\rm off}_j$}
                {
                    increment $C_{i,k} \leftarrow C_{i,k} + 1$\;
                    \If{$C_{i,k} = 1$}
                    {
                        //Save critical neighbor index: \\
                        write $j^{\rm over}_{i,k} \leftarrow j$\;
                        atomic read+modify+write \\
                        $\qquad f^{\rm exp}_i \leftarrow f^{\rm exp}_i - 1/|Q_i|$\;
                        $\qquad G^{\rm cav}_i \leftarrow G^{\rm cav}_i - G^{\rm cav}_{i,0}/|Q_i|$\;
                    }
                    {\bf else}\\
                    \If{$C_{i,k} \geq 2$}
                    {
                        write $j^{\rm over}_{i,k} \leftarrow -1$\;
                        {\bf break}\;
                    }
                }
            }
        }
        {\bf Step 2: Force Computation:}\\
        initialize $f^{\rm exp}_{i,1} \leftarrow f^{\rm exp}_{i,2} \leftarrow f^{\rm exp}_{i,3} \leftarrow f^{\rm exp}_i$\;
        initialize $F^{\rm cav}_{i,1} \leftarrow F^{\rm cav}_{i,2} \leftarrow F^{\rm cav}_{i,3} \leftarrow 0$\;
        Synchronize for all $1 \leq i \leq n$\;
        \For{$1 \leq k \leq |Q_i|$ {\rm (seq. or in $\parallel$)}}
        {
            \For{$1 \leq s \leq 3$ {\rm (seq. or in $\parallel$)}}
            {
                \If{$C_{i,k} = 0$}
                {
                   \For{$j =$ indices of atoms in $\mathbb{A}_i'$ {\rm (seq.)}}
                    {
                        \If{$\| {\bf q}_{i,k} - ({\bf r}_j + \delta r {\bf e}_s) \|_2 \leq R^{\rm off}_j$}
                        {
                            atomic read+modify+write \\
                            {
                                $\qquad f^{\rm exp}_{i,j,s} \leftarrow f^{\rm exp}_{i,j,s} - 1/|Q_i|$\;
                                $\qquad F^{\rm cav}_{i,s} \leftarrow F^{\rm cav}_{i,s} - G^{\rm cav}_{i,0} / (|Q_i| \delta r)$\;
                                $\qquad F^{\rm cav}_{j,s} \leftarrow F^{\rm cav}_{i,s} + G^{\rm cav}_{i,0} / (|Q_i| \delta r)$;$^\dag$\\
                            }
                            {\bf break}\;
                        }
                    }
                }
                {\bf else}\\
                \If{$C_{i,k} = 1$}
                {
                    write $j \leftarrow j^{\rm over}_{i,k}$; //note: $j^{\rm over}_{i,k} \neq -1$\\
                    \If{$\| {\bf q}_{i,k} - ({\bf r}_j + \delta r {\bf e}_s) \|_2 > R^{\rm off}_j$}
                    {
                        atomic read+modify+write \\
                        {
                            $\qquad f^{\rm exp}_{i,j,s} \leftarrow f^{\rm exp}_{i,j,s} + 1/|Q_i|$\;
                            $\qquad F^{\rm cav}_{i,s} \leftarrow F^{\rm cav}_{i,s} + G^{\rm cav}_{i,0} / (|Q_i| \delta r)$\;
                            $\qquad F^{\rm cav}_{j,s} \leftarrow F^{\rm cav}_{i,s} - G^{\rm cav}_{i,0} / (|Q_i| \delta r)$;$^\dag$\\
                        }
                        {\bf break}\;
                    }
                }
            }
        }
        write ${\bf F}^{\rm cav}_i \leftarrow (F^{\rm cav}_{i,1}, F^{\rm cav}_{i,2}, F^{\rm cav}_{i,3})$\;
    }
    {\footnotesize //$^\dag$ The instructions that require architecture-specific mutex.}
    \caption{SASA enumeration algorithm for solvation free energy and force computation.} \label{alg_enum}
\end{algorithm}

For every sample point, the sequential inner loop of the algorithm can be repeated $3|\mathbb{A}_i'|$ times for computing the variations $f^{\rm exp}_{i,j,1}$, $f^{\rm exp}_{i,j,2}$, and $f^{\rm exp}_{i,j,3}$ used in (\ref{eq_finitediff}), after introducing the finite difference to the 3 Cartesian coordinates (one at a time) of each neighbor atom $a_j \in \mathbb{A}_i'$. This takes $3|\mathbb{A}_i'|^2 N$ more tests per atom, still asymptotically $O(N)$ but not fast enough in practice.
There is a notably more efficient way to do the latter computation by ruling out the subset of sample points that cannot possibly contribute to $f^{\rm exp}_{i,j,s} - f^{\rm exp}_i ~ (s = 1, 2, 3)$ in (\ref{eq_finitediff}) during the first iteration when computing $B_{i,k}$ indicators. In particular, if a sample point is overlapped by more than one neighbor, displacing any neighbor does {\it not} affect its exposure state (from overlapped to exposed or vice versa), hence it does not contribute to $f^{\rm exp}_{i,j,s} - f^{\rm exp}_i$. To leverage this property, we expand the binary definition of the state function to $C: \mathbb{A} \times Q \rightarrow \mathds{Z} \cap [0, \infty)$ such that $C(a_i, \mathbf{q}_k)$ counts the actual number of neighbors $a_j \in \mathbb{A}_i'$ that overlap the sample point ${\bf q}_{i,k} \in T_i(Q)$.
Three different states for a sample point are observed in terms of the changes in $C_{i,k} := C(a_i, \mathbf{q}_k)$:

\begin{enumerate}
    \item `Not overlapped' or `exposed' ($C_{i,k} = 0$). In this case, displacing any neighbor either keeps the state at $C_{i,k} = 0$ or changes it to $C_{i,k} = 1$, where the latter case affects the contribution to SASA. Hence the inner loop needs to be repeated for all neighbors (i.e., for $3|\mathbb{A}_i'|$ times).
    \item `Critically overlapped' ($C_{i,k} = 1$). In this case, the only neighbor whose displacement may change the sample point's state to $C_{i,k} = 0$ is the one that originally overlapped it, and displacing any other neighbor either keeps the state at $C_{i,k} = 1$ or changes it to $C_{i,k} = 2$, both of which correspond to overlapped states that does {\it not} affect the contribution to SASA. Hence the inner loop is repeated only 3 times after displacing that critical neighbor along the 3 Cartesian axes.
    \item `Multiply overlapped' ($C_{i,k} \geq 2$). In this case, displacing any neighbor either keeps the state at $C_{i,k} \geq 2$ or changes it to $C_{i,k} = 1$, both of which correspond to overlapped states that does {\it not} affect the contribution to SASA. Hence the inner loop need not be repeated at all.
\end{enumerate}

Therefore, the only changes that contribute a nonzero value to $f^{\rm exp}_{i,j,s} - f^{\rm exp}_i ~ (s = 1, 2, 3)$ are those from exposed $(C_{i,k} = 0)$ to critically overlapped $(C_{i,k} = 1)$ and vice versa, thus a significant amount of computation time can be saved by early detection of the rest. An atom $a_j \in \mathbb{A}_i'$ is called a `critical neighbor' of the atom $a_i \in \mathbb{A}$ with respect to a sample point ${\bf q}_{i,k} \in Q_i$ along a particular direction ${\bf e}_s~(s = 1, 2, 3)$, if a finite displacement $\delta r {\bf e}_s$ results in such a change.
As a direct consequence of geometry, if $a_j$ is a critical neighbor of $a_i$ along $+{\bf e}_s$, then $a_i$ is also a critical neighbor of $a_j$ along $-{\bf e}_s$, both with respect to the same sample point. Therefore, a pair of neighbor atoms $a_i, a_j \in \mathbb{A}$ exchange a solvation force $\pm \delta {\bf F}^{\rm cav} = \pm 4\pi \gamma_i (R^{\rm off}_i)^2 / (N\delta r) {\bf e}_s$ due to their overlap at the sample point ${\bf q}_{i,k}$ if and only if they are critical neighbors with respect to ${\bf e}_s$.
The improved algorithm (based on the integer-valued $C_{i,k}$) is different from the original (based on the binary-valued $B_{i,k}$) in that the first iteration of the sequential inner loop for computing $C_{i,k}$ terminates after the {\it second} (rather than the {\it first}) overlap is encountered, because all $C_{i,k} \geq 2$ have equivalent implications according to the above rules.\footnote{Hence one could redefine to $C: \mathbb{A} \times Q \rightarrow \{0, 1, \text{``2 or more''} \}$ to implement the same trick with only 3 distinct flags, as in Algorithm \ref{alg_enum}.}
During this step, the value of $f^{\rm exp}_i$ is initialized to $1$ for each atom, and every time a sample point with $C_{i,k} = 1$ or $2$ is discovered, $f^{\rm exp}_i$ is decremented by $1/N$.
The next 3 repetitions of the inner loop per neighbor atom displacement depend on the aforementioned rules based on the value of $C_{i,k}$. The 3 variants of the exposure ratio $f^{\rm exp}_{i,j,s}~(s = 1, 2, 3)$ are initialized to $f^{\rm exp}_i$ for each atom with respect to displacements in all of its neighbors. Every time a sample point with $C_{i,k} = 0$ or $1$ is encountered, the inner loop is repeated with displaced neighbor coordinates to discover the critical neighbors, each adding $\pm 1/N$ to $f^{\rm exp}_{i,j,s}$.

Significant speed-ups are achieved in terms of the average time, a rigorous analysis of which is not possible without assumptions on the spatial distribution of atoms. However, the worst case time complexity is still $O(nN)$ for the sequential algorithm.  One could easily parallelize the algorithm at the outer loops over the atoms and sample points, while the inner loops over the neighbor atoms is best implemented sequentially. On a simple concurrent-read concurrent-write (CRCW) parallel random-access machine (PRAM) with common conflict resolution (briefly introduced in Appendix \ref{app_PRAM}),
the parallel running time of $O(nN/P)$ can be achieved in theory using $P$ processors (i.e., linear speed-up), which leads to $O(n)$ if we have $P = O(N)$ processors at our disposal---not far from reality when using GPUs. However, there are more complications to the machine architecture in practice, as will be addressed in Section \ref{sec_imp}.

The complete process is described in pseudo-code in Algorithm \ref{alg_enum}. The instructions marked by a dagger ($^\dag$) modify variables that belong to different atoms iterated in parallel by the outer-most loop, hence require a mutex with nuances that depend on the architecture as described in Section \ref{sec_imp}.

\subsection{Prefix Sum Calls} \label{sec_pref}

There are multiple references in {\sf Protofold II} to the generic prefix sum routine---explained in Appendix \ref{app_pref}, which can be performed using optimal sequential and parallel algorithms in linear number of steps---that emerge naturally as a consequence of the linear topology of the polypeptide backbone:
\begin{enumerate}
    \item Computing the link transformations from the successive matrix multiplications in (\ref{eq_kin_04}) and (\ref{eq_kin_05}), in which the domain is $\Sigma := \mathrm{SO}(3)$ (represented by $3 \times 3$ rotation matrices) and the operator $\oplus$ is the matrix multiplication.
    \item Computing the atom center coordinates from the successive vector summations in (\ref{eq_r1}) and (\ref{eq_r2}), in which the domain is $\Sigma := \mathds{R}^3$ (represented by $3 \times 1$ column matrices) and the operator $\oplus$ is the vector summation.
    \item Computing the joint torques from the successive superposition of the contributions of each link on the preceding joints in the chain using (\ref{eq_KCM_04}) and (\ref{eq_KCM_06}), in which the domain is $\Sigma := \mathds{R}^6$ (represented by $6 \times 1$ column matrices) and the operator $\oplus$ is the inner product.
\end{enumerate}
The first item clearly takes $O(n)$ steps, while the latter two take $O(l) = O(m)$ steps (which is also $O(n)$). To explain the last item further, let $[J_k]$ be the $k^\text{th}$ column of the Jacobian matrix $[J]$ and $[{\bf P}_h] := [{\bf T}^{\rm link}_h ~ {\bf F}^{\rm link}_h]^{\rm T}$ be the so-called generalized force on the right-hand side of (\ref{eq_KCM_04}) on the $h^\text{th}$ link along the chain, both of which are $6 \times 1$ column matrices. The contribution of ${\bf P}_h$ on the $k^\text{th}$ joint is obtained as the inner product of the two matrices $\eta_{k, h} = [{\bf J}_k]^{\rm T} [{\bf P}_h]$ arranged into the following matrix:
\begin{equation}\label{eq_04_9}
    [\eta] =
    \left[
        \begin{array}{cccccc}
            {\bf J}^{\rm T}_1 {\bf P}_1 & {\bf J}^{\rm T}_1 {\bf P}_2 & \cdots & {\bf J}^{\rm T}_1 {\bf P}_l \\
            0               & {\bf J}^{\rm T}_2 {\bf P}_2 & \cdots & {\bf J}^{\rm T}_2 {\bf P}_l \\
            \vdots  & \vdots  & \ddots & \vdots  \\
            0       & 0       & \cdots & {\bf J}^{\rm T}_l {\bf P}_l
        \end{array}
    \right],
\end{equation}
where $[\eta]$ is an $l \times l$ upper-triangular matrix made of the torque contributions $\eta_{k, h}$, whose $h^\text{th}$ column's upper nonzero elements form the $h \times 1$ column matrix $[\bm{\upeta}_h]$ introduced in (\ref{eq_KCM_04}).
Note that each link only affects the preceding joints in the chain, hence $\eta_{k, h} = 0$ for all $h \leq k - 1$. The total torque joints $\tau_k (1 \leq k \leq l)$ can be obtained as a summation over the rows of the above matrix via (\ref{eq_KCM_06}). In {\sf Protofold I} \cite{Kazerounian2004a,Kazerounian2004b,Kazerounian2005a} this was accomplished by scanning through the terms along the columns in (\ref{eq_04_9}), which took $l (l+1) /2 = O(l^2)$ operations. In {\sf Protofold II} we perform row scanning of the matrix, starting from the bottom row and moving upwards. More specifically, by factoring out the Jacobian terms $[{\bf J}^{\rm T}_k]$ in each row of (\ref{eq_04_9}) and aggregating the generalized forces into $[{\bf P}^{\rm agg}_k] = \sum_{h=k}^l [{\bf P}_h]$, (\ref{eq_KCM_06}) yields $\tau_k = [{\bf J}_k]^{\rm T} [{\bf P}^{\rm agg}_k]$ as the sum of each row. Then $[{\bf P}^{\rm agg}_k]$ can be obtained in $O(1)$ time from $[{\bf P}^{\rm agg}_{k+1}]$ as $[{\bf P}^{\rm agg}_k] = [{\bf P}_k] + [{\bf P}^{\rm agg}_{k+1}]$, which leads to a total of $O(l)$ prefix computation steps.

\begin{figure}
    \centering
    \includegraphics[width=0.48\textwidth]{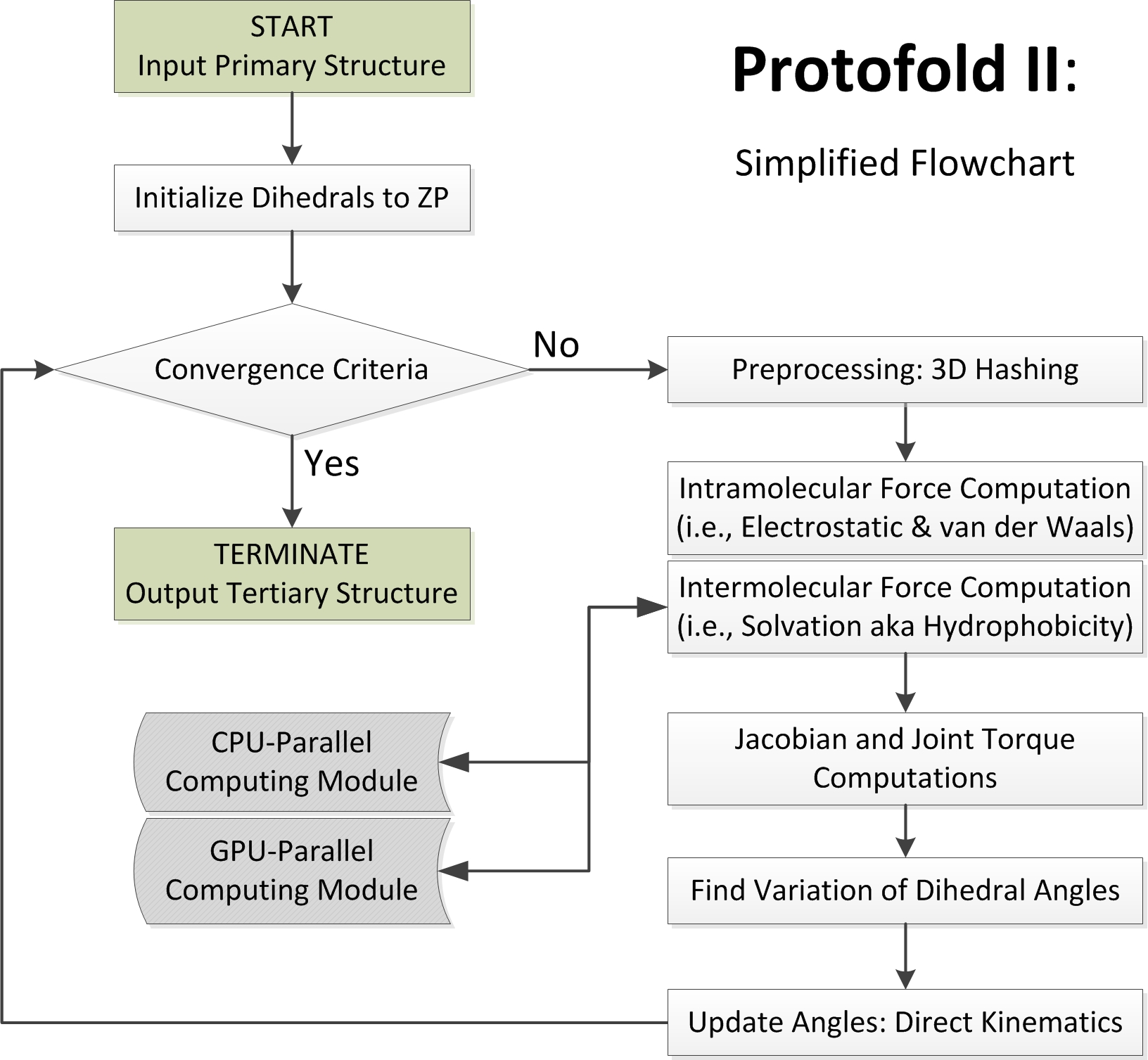}
    \caption{The main process flowchart of {\sf Protofold II}}
    \label{figure5}
\end{figure}

\section{Implementation} \label{sec_imp}

Figure \ref{figure5} is a schematic of the {\sf Protofold II} architecture elaborated in Section \ref{sec_arch}.
The remainder of this section is dedicated to the parallel implementation of Algorithm \ref{alg_enum} on the CPU in Section \ref{sec_CPU} and on the GPU in Section \ref{sec_GPU}, which are identified by the alternative shaded modules in Fig. \ref{figure5}.

\subsection{\textbf{\textsf{Protofold II}} Architecture} \label{sec_arch}

Unlike {\sf Protofold I} \cite{Kazerounian2004a,Kazerounian2004b,Kazerounian2005a} that was programmed in Matlab$^\circledR$, {\sf Protofold II} is reprogrammed with a new architecture in C++. The CPU- and GPU-parallel algorithms are implemented as external modules and linked to the main application thread as dynamic link libraries (DLL), which can be integrated into other folding packages.

As depicted in Fig. \ref{figure5}, a typical KCM simulation in {\sf Protofold II} can be summarized into the following steps:
\begin{enumerate}
    \item {\bf Input:} The user specifies a primary structure (i.e., AA sequence information) to the interface.
    \item {\bf Preprocessing:} The program constructs the AA chain using the structural assumptions given in Section \ref{sec_kin} to arrange the atoms into the consecutive peptide planes. The double-bond angles are all set to the fixed values of $\omega_i = 0^\circ$ (cis) or $-180^\circ$ (trans) and the body vectors are assigned with the values given in Table \ref{table_coef}.
    \item {\bf Initialization:} The main chain dihedral angles are initialized as $\phi^0_i = \psi^0_i = -180^\circ$ and the side chain dihedral angles are initialized to rotameric default values $\chi^0_{i,k}$ \cite{Kazerounian2005} for $1 \leq i \leq n, 1 \leq k \leq l_i \leq 4$---i.e., set all $\theta^0_{j,k} = 0^\circ$ in (\ref{theta_1}) through (\ref{theta_3}) referred to as ZP initial conditions.
    \item {\bf Forward Kinematics:} The conformation variables summarized in Table \ref{table_ub_angle} are converted to the Cartesian coordinates of the individual atoms by using the sequence of rigid body transformations described in Section \ref{sec_kin}.
    \item {\bf Coordinate Hashing:} Using the 3D grid data structure presented in Section \ref{sec_prox}, the atom coordinates are arranged into buckets for fast neighborhood queries based on the cut-off distances.
    \item {\bf Force Computations:} The free energy- and force-fields are computed from the atom coordinates using the equations given in Section \ref{sec_field}. This is where the CPU- or GPU-parallel modules are called for computing the solvation effects.
    \item {\bf Torque Computations:} The forces on the atoms are converted to joint torques using the Jacobian transformation described in Section \ref{sec_field}.
    \item {\bf KCM Stepping:} The kinetostatic effect of the joint torques are computed using the simple steepest-descent stepping explained in Section \ref{sec_KCM}.
    \item {\bf Termination:} If the convergence criteria is met, the program terminates; otherwise it repeats the steps 4 through 8 above.
    \item {\bf Output:} The intermediate (every several frames) and final conformations in PDB format, the variations of the dihedral angles and free energy terms, and the performance measures (e.g., running times of different steps) are exported by the program.
\end{enumerate}
These steps characterize the process of arriving from sequence configuration (i.e., primary structure) to stable 3D conformation (i.e., tertiary structure) without any additional assumptions. Although this is the ultimate goal of protein folding, it is rather ambitious to obtain results that are consistent with experimental observations except in the case of relatively short chains; e.g., folding simulation of $\alpha-$helix coiling described in section \ref{sec_fold}. This is due to a variety of reasons ranging from the sensitivity of the folding pathway to the physical parameters (e.g., adjusted coefficients in the empirical force-field equations) to the sensitivity of the spatial structure of long chains to simplifying geometric assumptions (e.g., the exact planarity of the peptide planes).

\paragraph{Additional Functionalities.}
In order to enable addressing certain computer-aided structural studies on real proteins effectively in spite of the aforementioned difficulties, we found it imperative to include the following additional functionalities in {\sf Protofold II}:
\begin{itemize}
    \item The user has the option to 1) specify only sequence data, from which the `canonical' peptide plane geometry (i.e., assuming exact planarity $\omega_i = 0^\circ$ (cis) or $-180^\circ$ (trans) and average lengths in Table \ref{table_ub_angle}); or 2) import the protein structure as a PDB file and retain the peptide group geometry as-read when constructing the rigid links.
    \item The user has the option to limit the mobility of the linkage by fixing as many dihedral angles as desired. This enables folding studies at multiple levels and different scales. For example, it is possible to group collections of AAs (e.g., secondary elements, motifs, domains, etc.) into presumed rigid bodies and limit the DOF to deformations at the loops connecting them.
    \item In addition to the ZP initial conditions, the user may choose to use other initial conditions, including but not limited to completely random initial conditions or the native conformation perturbed by arbitrary (deterministic or randomized) changes to certain dihedral angles.
    \item When importing PDB files, the program eliminates water molecules---since their effect is implicitly incorporated by the solvation energies---but retains other heteroatoms (e.g., metal ions, co-factors, substrates, etc.) and includes them among chain atoms when computing the force-field. This is crucial since the proper folding of many proteins is dependent on these agents.
\end{itemize}
In addition to the above features, the following need to be included in future versions:
\begin{itemize}
    \item The program currently supports monomeric protein folding in its simplest topology. It is desirable to enable multimeric protein folding by maintaining multiple chains bound together (i.e., quaternary structure) and more complex topologies induced by other effects (e.g., disulfide bonds, hydrogen bonds, lipidation, etc.)
    \item the simplistic steepest-descent search process presented in Section \ref{sec_KCM} has not evolved much since {\sf Protofold I} \cite{Kazerounian2004a,Kazerounian2004b,Kazerounian2005a}. Our numerical experiments suggest that better optimization algorithms such as a hybrid Monte Carlo sampling combined with steepest-descent or conjugate-gradients KCM\footnote{This module is already implemented into {\sf Protofold II} but not tested yet, as the focus of this article is on the improved model and implementation of the force-field.}
        could be more effective in avoiding local minima and enable faster convergence to the global minimum.
\end{itemize}

\paragraph{Parallel Implementations.}
As demonstrated in Section \ref{sec_enum}, the solvation energy and force computations using (\ref{eq_04_3}) and Algorithm \ref{alg_enum} are the most time-consuming steps of each KCM iteration, mainly due to the large number of sample points $|Q| = N \gg 1$ required to enumerate the offset sphere of each atom for an adequate approximation of SASA and its gradient.
To benefit from the single-instruction multiple-data (SIMD) characteristic of Algorithm \ref{alg_enum}, the variables pertaining to different atoms are assigned to different processors. The two terms on the right-hand side of (\ref{eq_04_3}) are computed concurrently by different processors assigned to $a_i, a_j \in \mathbb{A}$ and broadcasted to each other to minimize the computational work. An immediate consequence is an additional communication overhead and possible network contention due to concurrent write attempts. Such a trade-off between computation and communication intensities is a common characteristic of parallel algorithms \cite{Maggs1995}, and will be considered here for code optimization.

Here we focus on the implementation of the SIMD Algorithm \ref{alg_enum} using two parallel computing models; namely,
\begin{itemize}
    \item one that is designed for coarse-grained multiprocesser machines such as multi-core CPUs (Section \ref{sec_CPU}); and
    \item another that is designed for fine-grained multiprocesser machines such as many-core GPUs (Section \ref{sec_GPU}).
\end{itemize}

\begin{figure}
    \centering
    \includegraphics[width=0.48\textwidth]{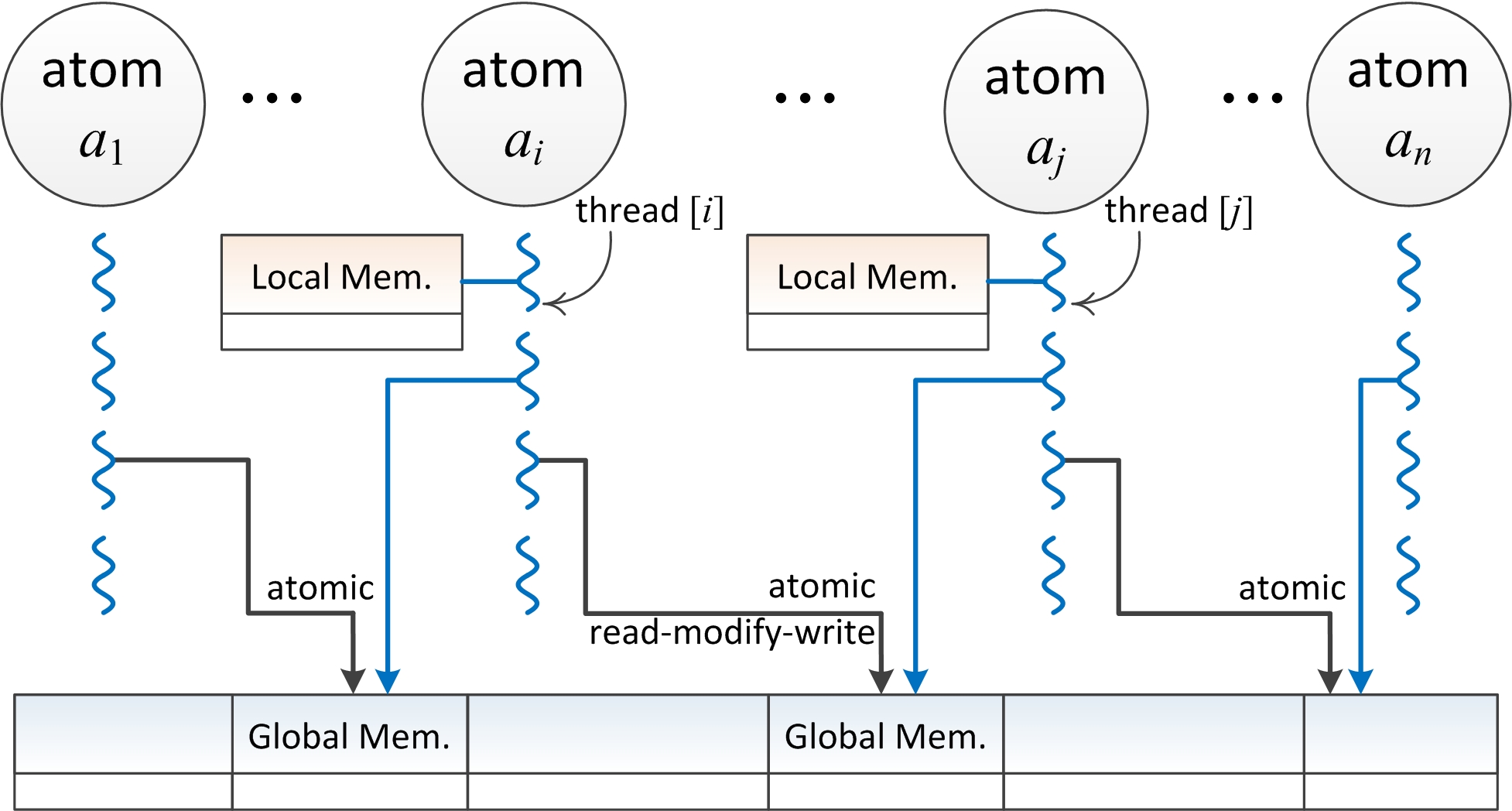}
    \caption{Thread execution model on the CPU.}
    \label{figure6}
\end{figure}

\subsection{CPU-Parallel Implementation} \label{sec_CPU}

The first parallel implementation targets a coarse-grained shared-memory multiprocessor machine, i.e., one with a multi-core multi-thread CPU. Given $n$ atoms $n$ threads are generated, assigning one thread per atom $a_i ~(1 \leq i \leq n)$. The neighborhood information (i.e., the list of indices of all $a_j \in \mathbb{A}_i$ defined in (\ref{eq_neigh}) for each atom $a_i \in \mathbb{A}$) is constructed and saved in the {\it global} memory shared among all processors, hence can be accessed concurrently from different threads. Each thread stores the sample point coordinates, the SASA and its gradient, and the resulting solvation energy and force components in the {\it local} memory of the processor. The thread iterates sequentially over the sample points on the offset sphere, and a counter variable that keeps track of the number of overlapped sample points is initialized within the scope of the thread. For each sample point, the coordinates are computed and tested sequentially against all neighbors to obtain $C_{i,k}~(1 \leq i \leq n, 1 \leq k \leq |Q_i|)$.

Once the exposure states are obtained for the original configuration of the neighbors, the thread loops over all neighbors one more time to examine the effects of their displacement along the 3 coordinate axes one at a time. If the criteria given in Section \ref{sec_enum} are met, the pair of force components $\pm \delta F^{\rm cav} = \pm 4\pi \gamma_i (R^{\rm off}_i)^2 / (N\delta r)$ need to be added to the total solvation forces of two neighbor atoms $a_i$ and $a_j$ along the proper coordinate axis, and in opposite directions. This results in two write operations per incidence, the first of which modifies ${\bf F}^{\rm cav}_i$ of $a_i$, which is safely assigned to the current thread and occurs in the local memory without any concern related to communication between the threads. The second write operation, on the other hand, modifies ${\bf F}^{\rm cav}_j$ of $a_j$, a variable assigned to a different thread. This requires communication between the two threads, and has to be implemented using atomic write operations into the global memory to guarantee mutual exclusion. Figure \ref{figure6} shows the multi-threading scheme for the CPU-parallel algorithm.
The algorithm is implemented using the OpenMP library. Although linear speed-up is expected in theory on an abstract CRCW PRAM, the actual speed-up is sublinear (as depicted in Section \ref{sec_timeres}) in practice due to bus traffic, network contention, cache invalidations, and serialized operations.

\paragraph{CPU Optimization.}
The number of CPU cores is generally much smaller that the number of atoms ($p \ll n$). Nevertheless, it is good practice to generate more threads than the number of cores to maximize the performance by keeping the processors saturated at all times with computational work. Accessing global memory incurs latency at the incidence of a cache miss and multithreading is a standard technique for hiding such latencies.
The computation instructions are interleaved with memory access instructions, hence every time one thread is accessing the global memory the processor can switch the context to a different thread. Other optimization attempts include using local memories instead of global memories whenever possible, and avoiding multiple computations of constant parameters or variables that are used repeatedly.

\begin{figure}
    \centering
    \includegraphics[width=0.48\textwidth]{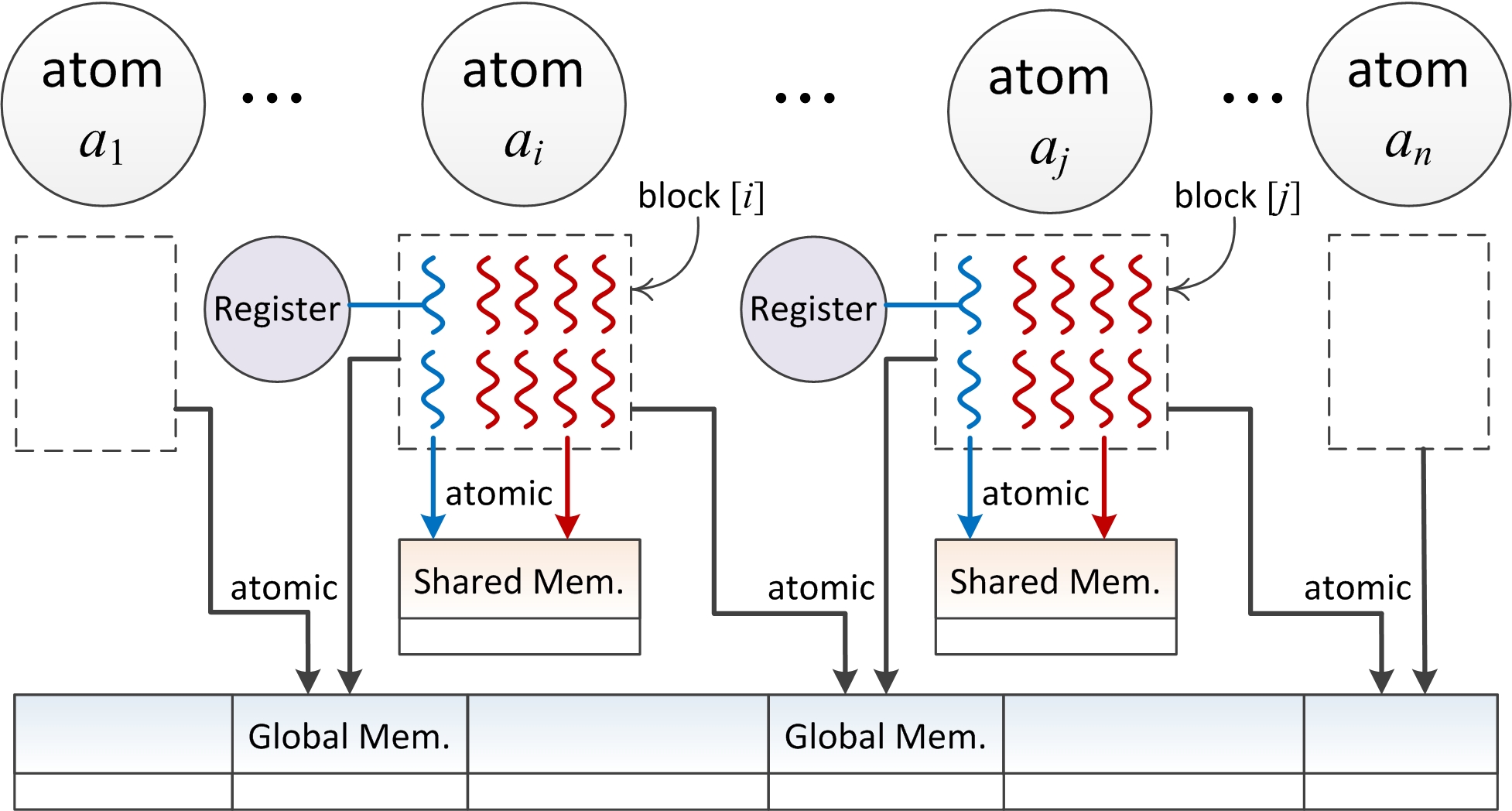}
    \caption{Thread execution model on the GPU.}
    \label{figure7}
\end{figure}

\subsection{GPU-Parallel Implementation} \label{sec_GPU}

The second parallel implementation targets a fine-grained machine with a hierarchical memory architecture, i.e., one with a many-core many-thread GPU. Given $n$ atoms, a linear grid of $n$ blocks is generated, assigning one block per atom $a_i ~(1 \leq i \leq n)$. Each block is further divided into $N = |Q|$ threads, assigning one thread per sample point $\mathbf{q}_k \in Q ~(1 \leq q \leq n)$ on the unit offset sphere.
Prior to GPU kernel execution, the neighborhood information (i.e., the list of indices of all $a_j \in \mathbb{A}_i$ defined in (\ref{eq_neigh}) for each atom $a_i \in \mathbb{A}$) is transferred from the CPU (i.e., {\it host}) memory to the GPU (i.e., {\it device}) memory.
For each thread, the iteration over different neighbors is performed sequentially, similar to the CPU-parallel code presented in \ref{sec_CPU}. The solvation energy and force components, a counter that keeps track of the number of overlapped sample points, and exposure states are initialized in the {\it shared} memory of the blocks, which require atomic operations for access safety by multiple threads, while sample point coordinates are stored in the {\it registers} that are local to each thread. Each sample point is tested sequentially against all neighbors to obtain $C_{i,k}~(1 \leq i \leq n, 1 \leq k \leq |Q_i|)$.

Once the exposure states are obtained for the original configuration of the neighbors, the thread loops over all neighbors one more time to examine the effect of their displacement along the 3 coordinate axes one at a time. Similar to the CPU-based implementation, whenever the pair of force components $\pm \delta F^{\rm cav} = \pm 4\pi \gamma_i (R^{\rm off}_i)^2 / (N\delta r)$ need to be added to the total solvation forces of two neighbor atoms $a_i$ and $a_j$ along the proper coordinate axis, the write operation that modifies ${\bf F}^{\rm cav}_i$ of $a_i$ happens atomically in the shared memory. This ensures mutual exclusion between threads of the same block. On the other hand, the write operation that modifies ${\bf F}^{\rm cav}_j$ of $a_j$ happens atomically in the global memory to ensure mutual exclusion between blocks of the same grid. Figure \ref{figure7} shows the multi-threading scheme for the GPU-parallel algorithm.
The algorithm is implemented using NVIDIA's compute-unified device architecture (CUDA). Kernel invocation is carried out synchronously within the default CUDA stream, hence synchronization between blocks is automatically guaranteed, while barrier synchronization is needed between threads of the same block.

\paragraph{GPU Optimization.}
The optimization attempts can be categorized as memory, execution, instruction, and flow-control optimization.

\begin{itemize}
    \item Memory optimization is the most effective of all, as demonstrated by the results in the Section \ref{sec_timeres}. In contrast to the CPU-parallel algorithm that makes most references through the cached global memory, the GPU-parallel algorithm transfers the coordinates, radii, solvation parameters, and neighbor index lists for each atom into the shared memory to minimize the number of global memory references. The variables that are exclusive to the threads, on the other hand, such as the exposure states or sample coordinates are allocated in the registers. However, the limited amount of shared memory and register resources on the streaming multiprocessor (SM) imposes a restriction on the number of resident blocks on the SM and can adversely affect thread occupancy at any time during the simulation. Therefore, one needs to avoid excessive variable definitions within the scope of the GPU kernels.
    \item For execution level optimization, the kernels should be executed with proper granularity to maximize SM thread occupancy. Specifying a larger number of threads per block generally contributes to latency hiding, but is limited by the architecture as well as the on-chip memory resources. The number of threads is the same as the sample size $N = |Q|$ a proper choice of which is a trade-off between accuracy and performance.
    \item For instruction level optimization, the transcendental math functions are converted to their intrinsic alternatives that are executed on the special function units (SFU) of the CUDA cores.
    \item Flow-control optimization is realized by avoiding multiple execution paths within the same block, which might lead to thread divergence and serialization within the same warp. In particular, when checking for overlaps between neighbor atoms and sample points, the conditional (e.g., if/else/then) statements are set in such a way that one of the two execution paths is always null.
\end{itemize}
The near-optimal conditions are reached by successive experimentation and modification of the code. For more information regarding the GPU architecture and terminology, see Appendix \ref{app_SIMT}.

\begin{figure*}
    \centering
    \includegraphics[width=\textwidth]{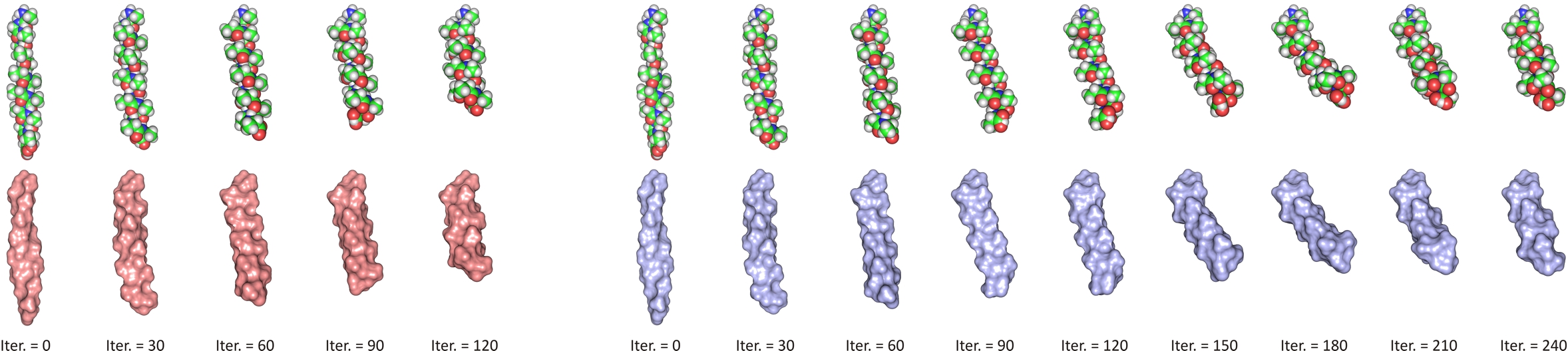}
    \caption{Left-handed $\alpha-$helix formation for a 15-residue polyalanine chain in vacuum (left) and in water (right) starting from $\phi^0_i = \psi^0_i = +10^\circ$ using {\sf Protofold II}. Initial conditions and solvation effects dramatically affect the folding pathway.}
    \label{figure8}
\end{figure*}
\begin{figure*}
    \centering
    \includegraphics[width=\textwidth]{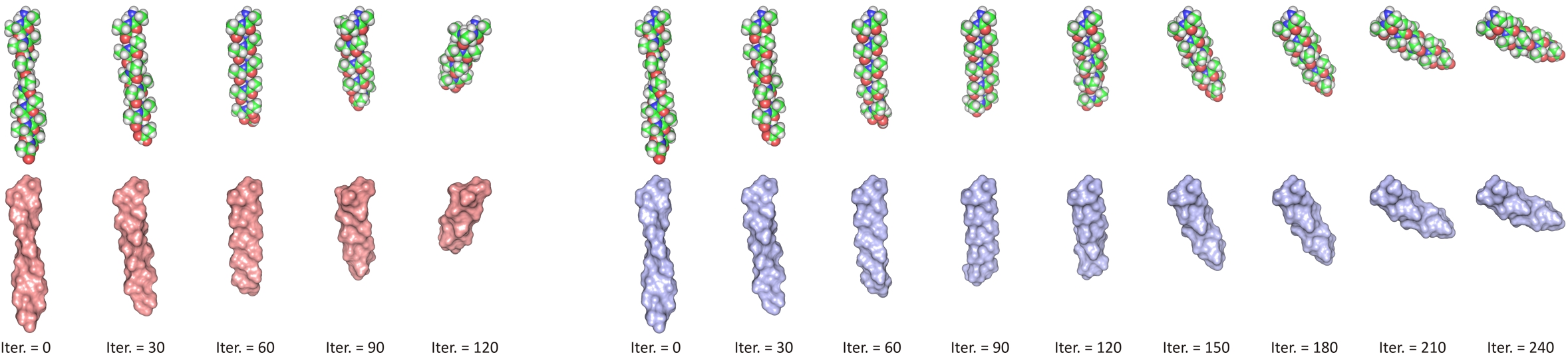}
    \caption{Right-handed $\alpha-$helix formation for a 15-residue polyalanine chain in vacuum  (left) and in water (right) starting from $\phi^0_i = \psi^0_i = -10^\circ$ using {\sf Protofold II}. Initial conditions and solvation effects dramatically affect the folding pathway.}
    \label{figure9}
\end{figure*}

\section{Results \& Discussion} \label{sec_res}

This section presents a preliminary assessment of the model and implementation enhancements from {\sf Protofold I} \cite{Kazerounian2004a,Kazerounian2004b,Kazerounian2005a} to {\sf Protofold II}. The folding process is simulated and assessed at multiple levels, ranging from the formation of secondary structural elements (e.g., $\alpha-$helix coiling or $\beta-$strands formation) from an open chain to tertiary interactions between secondary elements or across larger domains that can be assumed to be rigid in real protein examples.

In Section \ref{sec_fold} we discuss the impact of introducing solvation effects on the folding process of secondary structural elements starting from different initial conditions. We present some performance measures in Section \ref{sec_timeres} to validate the practical benefits of algorithmic improvements (e.g., coordinate hashing) as well as implementation improvements (i.e., CPU- and GPU-parallel computing). Finally, we look at a few real protein molecules in Section \ref{sec_real} and examine the energy variations in the neighborhood of the native structures.

\subsection{The Folding Process} \label{sec_fold}

The simplest structural elements that are ubiquitous across many protein domains are $\alpha-$helices and $\beta-$strands.
Here we start by considering simple test runs on relatively short (e.g., 10-20 residues long) peptide chains made of Ala residues (typically used as the benchmark AA type in its most common L-stereoisomeric form) to visualize their compliance into secondary structural elements.

\begin{figure}
    \centering
    \includegraphics[width=0.48\textwidth]{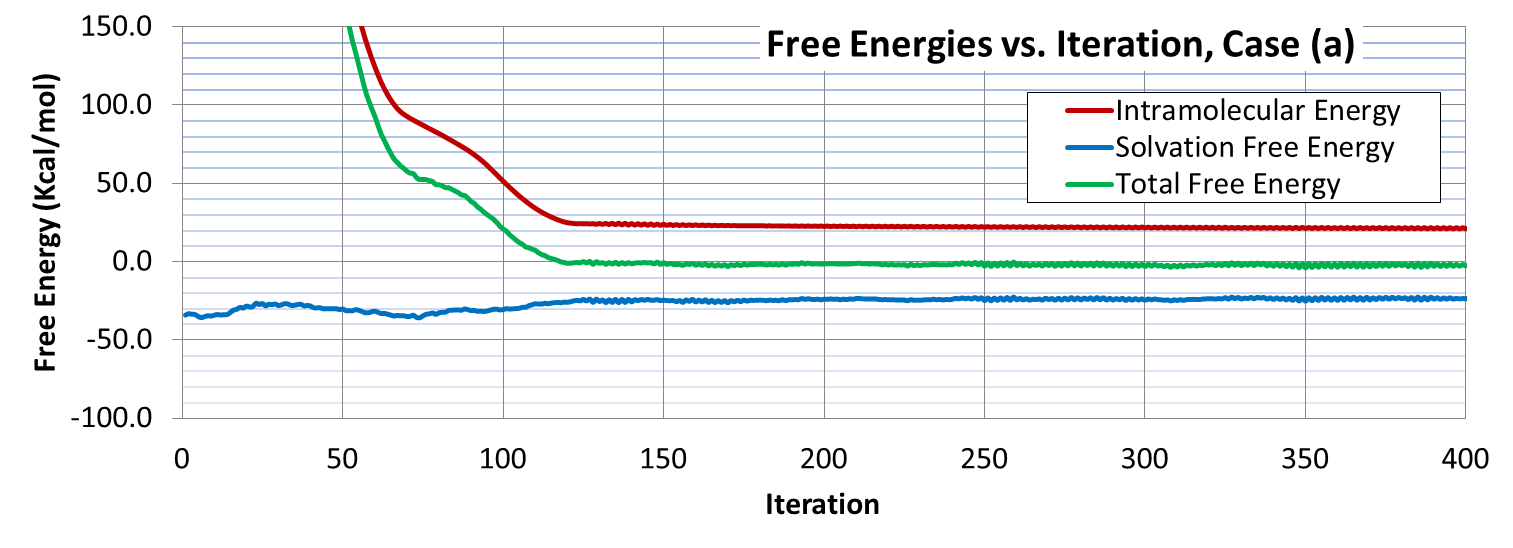}
    \includegraphics[width=0.48\textwidth]{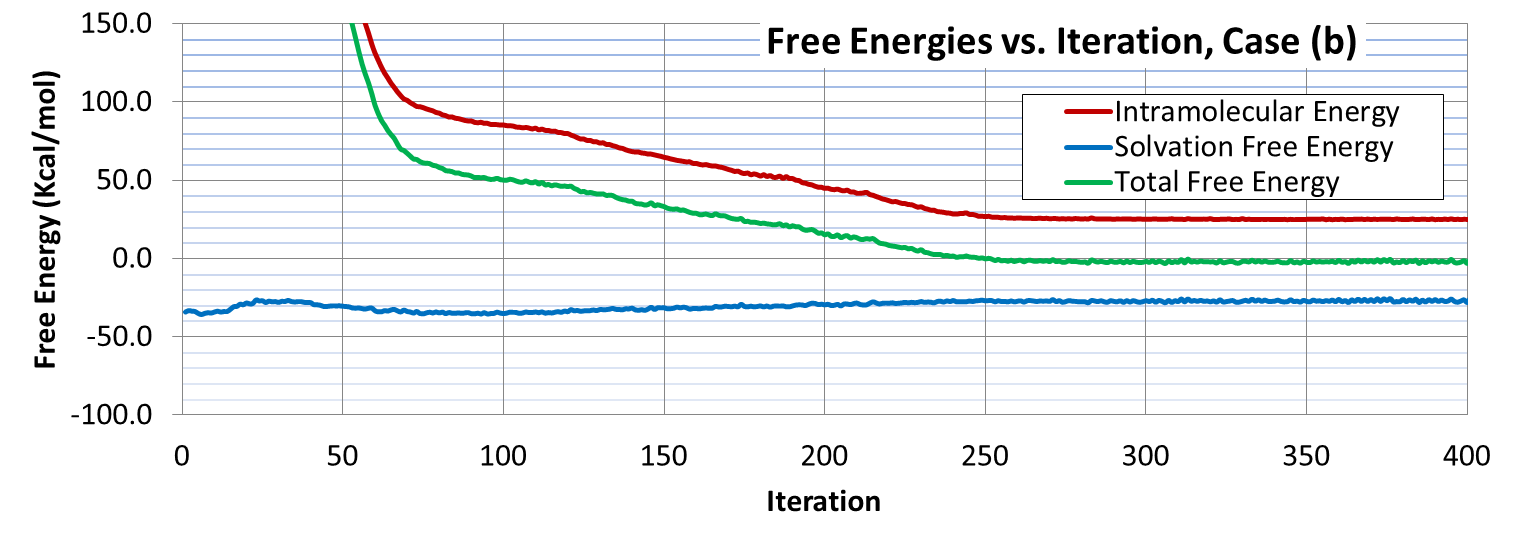}
    \caption{Free energy variations during folding of a 15-residue polyalanine chain into a left-handed $\alpha-$helix in vacuum (top) and in water (bottom) using {\sf Protofold II}.}
    \label{figure10}
\end{figure}

\begin{figure}
    \centering
    \includegraphics[width=0.48\textwidth]{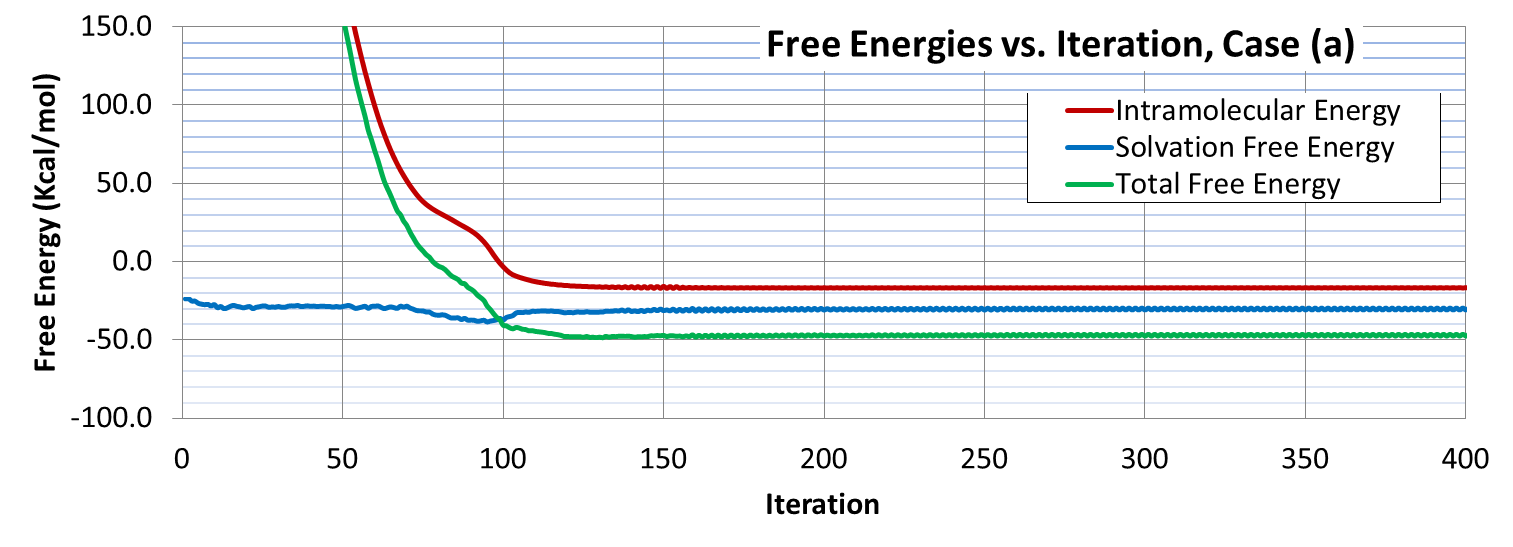}
    \includegraphics[width=0.48\textwidth]{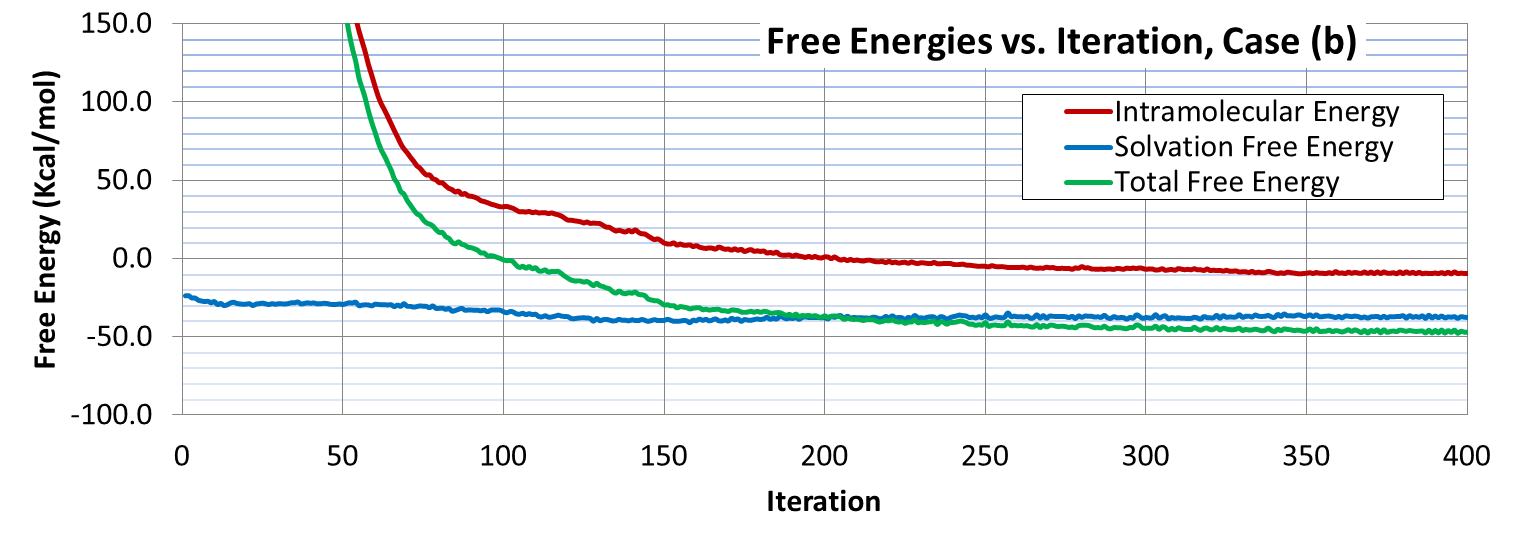}
    \caption{Free energy variations during folding of a 15-residue polyalanine chain into a right-handed $\alpha-$helix in vacuum (top) and in water (bottom) using {\sf Protofold II}.}
    \label{figure11}
\end{figure}

\paragraph{Alpha Helix Formation.}
First, we run four tests on a 15-residue chain starting from two different initial conditions, for both of which we simulate the folding process without and with solvation effects taken into account:
\begin{enumerate}
    \item Starting from the (slightly pre-coiled) initial conditions $\phi^0_i = \psi^0_i = +10^\circ$ for all $1 \leq i \leq 15$ the chain folds into a left-handed $\alpha-$helix as depicted in Fig. \ref{figure8}. The energy variation during KCM iterations is given in Fig. \ref{figure10}.
    \item Starting from the (slightly pre-coiled) initial conditions $\phi^0_i = \psi^0_i = -10^\circ$ for all $1 \leq i \leq 15$ the chain folds into a right-handed $\alpha-$helix as depicted in Fig. \ref{figure9}. The energy variation during KCM iterations is given in Fig. \ref{figure11}.
\end{enumerate}
The folding process in vacuum, i.e., without considering the solvation effects, emulates the behavior in the absence of the polar solvent, e.g., in membrane proteins extended along the nonpolar lipid bilayer or in secondary structural elements wrapped inside the hydrophobic core of globular proteins \cite{Kuriyan2012}. On the other hand, the folding process in water, i.e., with the presence of the solvation effects in addition to the intramolecular interactions, emulates the formation of elements that reside at the hydrophilic surface of globular proteins \cite{Kuriyan2012}.

In the case of $\alpha-$helix formations in Figs. \ref{figure8} and \ref{figure9}, the hand of the initial coil determines the hand of the final helix.\footnote{The surface visualizations can be deceiving where the right-handed helix appears to have a left-handed twist and vice versa. This is due to the transversal ridges and grooves formed in between the side chains \cite{Kuriyan2012}.}
This is due to the gradient descent nature of the KCM search algorithm that tends to converge to different local minima depending on the initial state. The effects of the solvation are hard to observe in these examples with the energetically unchallenged helical structures due to proper stacking of the atoms favored by all considered effects. In both cases the van der Waals and solvation effects work in the same direction until the steric clash prevents the helix to coil further.

Figures \ref{figure10} and \ref{figure11} are plots of the free energy variations versus KCM iteration number for the four runs described above. Note that in all four cases (top and bottom plots) the solvation energy is evaluated and plotted, but only in two of them (bottom plots) its effects are applied to deform the chain. For both left and right-handed helix formation, the inclusion of solvation effects clearly changes the folding pathway and increases the number of iterations before convergence from around 150 to 300. However, in either case the solvation free energy changes are not as significant as those of intramolecular (particularly van der Waals) effects.
Another important observation is that the right-handed $\alpha-$helix exhibits a notably more stable conformation than the left-handed $\alpha-$helix with about $\sim$ 40$-$50 kcal per mol lower total free energy state---to be accurate, 43.8 and 45.9 kcal per mol for the entire chain, i.e., 2.9 and 3.1 kcal per mol per AA residue, without and with solvation effects, respectively. Although this is qualitatively consistent with the expectation of right-handed coiling being favored by L-alanine chains, the energy differences are higher than the ones reported in earlier studies (e.g., MD results in \cite{Lins2006}). However, a meaningful comparison would require using identical simulation parameters, which is beyond the scope of this paper.

The final dihedral angles for all 15 Ala residues corresponding to the folded (i.e., stable) conformations, obtained after a large enough number of iterations, are given in Table \ref{tab_helix}.

\begin{figure*}
    \centering
    \includegraphics[width=0.75\textwidth]{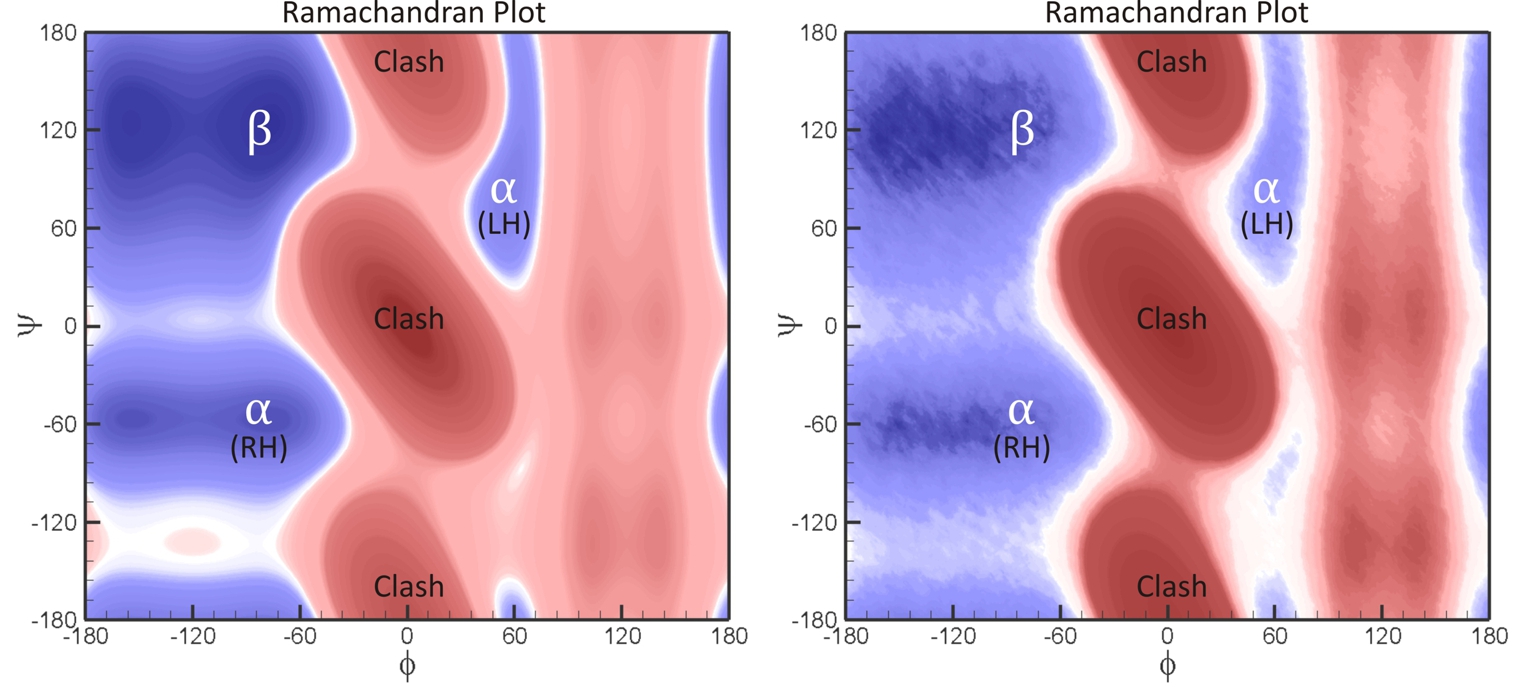}
    \caption{The Ramachandran plots for the energy variations of a pair of Ala residues in vacuum (i.e., without solvation effects) (left) and in water (i.e., with solvation effects) (right) using {\sf Protofold II}.}
    \label{figure12}
\end{figure*}

\paragraph{Ramachandran Plots.}
To examine the local effects of energetics, Ramachandran plots (for a pair of Ala residues in tandem) are generated by {\sf Protofold II} using the energy-field presented in Section \ref{sec_field}. The plots in Fig. \ref{figure12} show the energy variations across different pairs $(\phi, \psi)$ of dihedral angles, without and with considering the solvation effects. The vertical high-energy regions in the middle corresponds to prohibitive steric clashes between the atoms. The low-energy regions around it, on the other hand, correspond to the geometric relations between consecutive peptide planes that, when repeated for a segment of multiple residues along the chain, create secondary structural elements such as coiled $\alpha-$helices and flat $\beta-$strands. Although the solvation effects do not significantly alter the shape of the energy profile, there is a certain amount of noise added due to the discrete nature of the enumeration algorithm presented in Section \ref{sec_enum}.

\begin{figure*}
    \centering
    \includegraphics[width=0.75\textwidth]{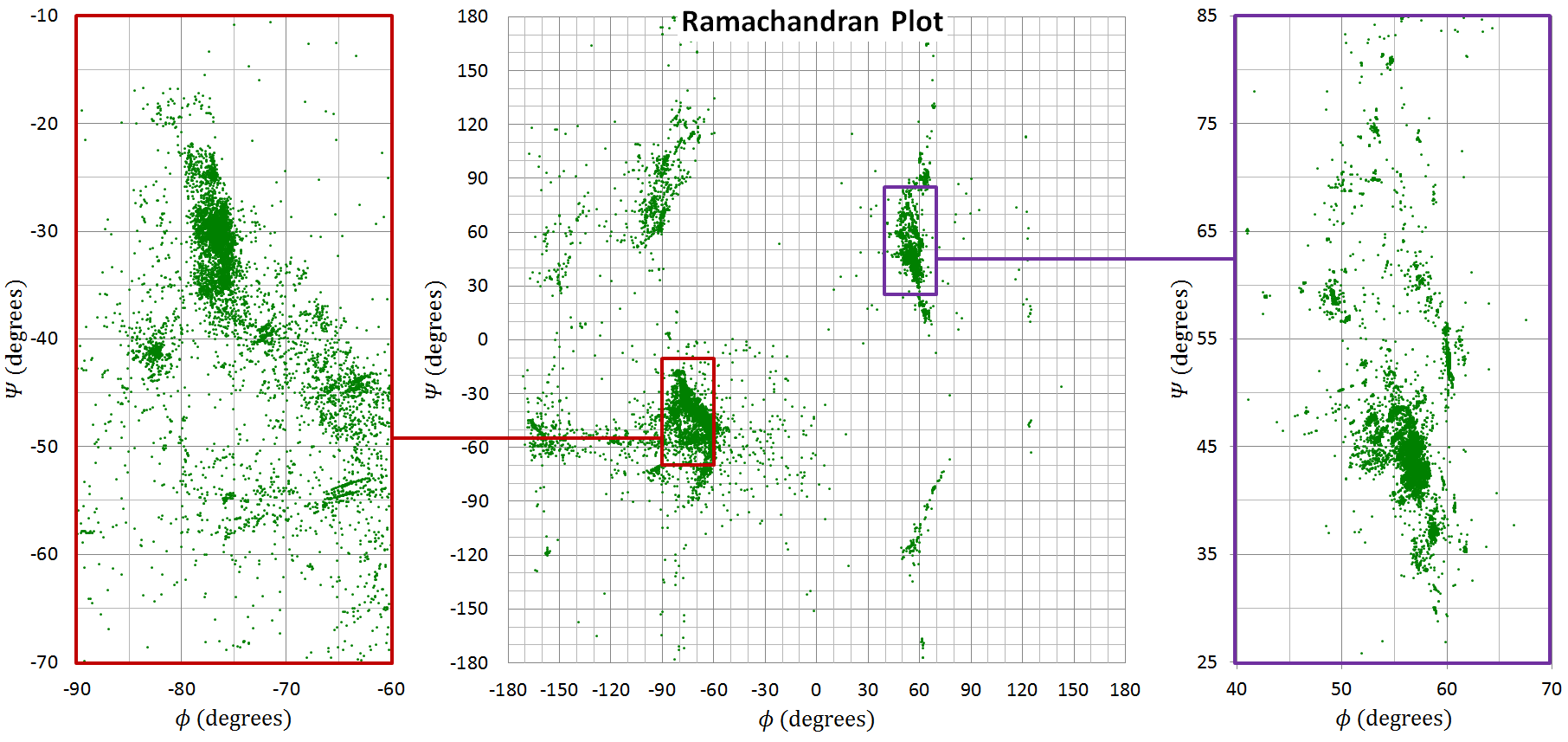}
    \includegraphics[width=0.75\textwidth]{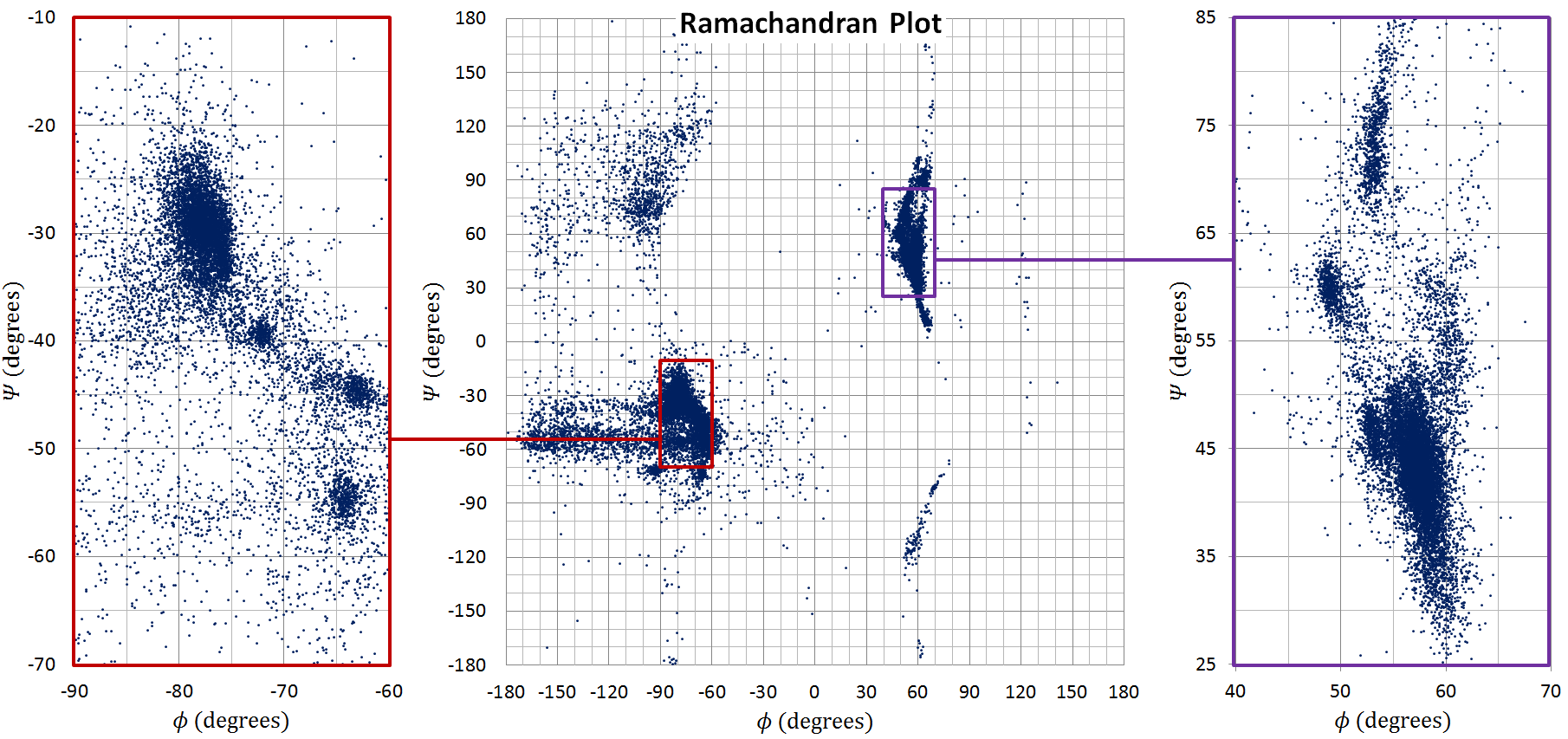}
    \caption{The Ramachandran plots for the folded backbone conformation of 2,000 10- to 20-residue polyalanine chains in vacuum (i.e., without solvation effects) (top) and in water (i.e., with solvation effects) (bottom) starting from random initial conditions $-90^\circ \leq \phi^0_i, \psi^0_i \leq +90^\circ$ using {\sf Protofold II}.}
    \label{figure13}
\end{figure*}

To observe the effects of solvation, one needs to carry out more extensive simulations on larger data sets with different chain lengths and various initial conditions. We carried out KCM runs on 2,000 independent polyalanine chains of random lengths in the range $10 \leq m \leq 20$ starting from random initial angles in the range $-90^\circ \leq \phi^0_i, \psi^0_i \leq +90^\circ$. The tests were run separately without and with considering the solvation effects using the same random seed. The resulting dihedral angles for all 27,717 Ala residues of all chains\footnote{The angles for the first AA residue of all chains are eliminated as outliers since they differ dramatically from those of the subsequent residues due to the anchoring at the N-terminus (see Table \ref{tab_helix}).}
are plotted in Fig. \ref{figure13}. One can observe multiple concentration areas that clearly correspond to left- and right-handed $\alpha-$helices and flat $\beta-$strands, the former two helical folds being more populated on the plots. Zooming further on the two $\alpha-$regions reveals multiple local minima where the points are concentrated more, which correspond to different subtypes of $\alpha-$helices. Comparing the two plots in Fig. \ref{figure13} reveals that the solvation effects do not significantly change the locations of the local minima. However, the energy profile is relaxed around the local minima where it has sharp cracks traced by the concentrated points along the valleys of the intramolecular energy landscape.

\begin{table}
    \caption{Final conformations of a 15-residue polyalanine chain folded in vacuum using {\sf Protofold II}.}
    \vspace{-0.5cm}
    \begin{center} \label{tab_helix}
        \begin{adjustbox}{max width=0.48\textwidth}
            \begin{tabular}{| r | r r | r r | r r | r r |} 
                \hline
                & \multicolumn{4}{|c|}{Left-handed coil $\phi^0_i = \psi^0_i = +10^\circ$ } & \multicolumn{4}{|c|}{Right-handed coil $\phi^0_i = \psi^0_i = -10^\circ$ } \\
                \hline
                & \multicolumn{2}{|c|}{In vacuum} & \multicolumn{2}{|c|}{In water} & \multicolumn{2}{|c|}{In vacuum} & \multicolumn{2}{|c|}{In water} \\
                \hline
                $i$ & $\phi_i (^\circ)$ & $\psi_i (^\circ)$ & $\phi_i (^\circ)$ & $\psi_i (^\circ)$ & $\phi_i (^\circ)$ & $\psi_i (^\circ)$ & $\phi_i (^\circ)$ & $\psi_i (^\circ)$  \\ [0.5ex] 
                \hline
                 1 & $+10.0$ & $+118 $ & $-7.20$ & $+21.5$ & $-10.0$ & $-59.1$ & $+15.1$ & $+54.4$ \\
                 2 & $+60.2$ & $+53.9$ & $+59.9$ & $+47.6$ & $-82.4$ & $-41.1$ & $-107 $ & $-47.1$ \\
                 3 & $+58.9$ & $+36.8$ & $+58.6$ & $+40.0$ & $-76.9$ & $-24.8$ & $-91.9$ & $+11.3$ \\
                 4 & $+56.8$ & $+45.4$ & $+57.9$ & $+45.1$ & $-75.7$ & $-34.4$ & $-81.1$ & $-34.4$ \\
                 5 & $+57.9$ & $+42.2$ & $+58.8$ & $+40.3$ & $-75.8$ & $-28.4$ & $-83.1$ & $-23.8$ \\
                 6 & $+56.7$ & $+43.7$ & $+57.8$ & $+43.6$ & $-75.3$ & $-31.3$ & $-76.7$ & $-32.1$ \\
                 7 & $+57.3$ & $+42.7$ & $+57.9$ & $+43.6$ & $-76.1$ & $-29.3$ & $-79.4$ & $-28.9$ \\
                 8 & $+56.8$ & $+44.0$ & $+58.1$ & $+41.6$ & $-75.6$ & $-30.5$ & $-77.1$ & $-28.9$ \\
                 9 & $+56.8$ & $+42.3$ & $+58.3$ & $+42.1$ & $-76.3$ & $-29.2$ & $-79.8$ & $-29.2$ \\
                10 & $+56.5$ & $+45.5$ & $+58.4$ & $+43.3$ & $-76.5$ & $-29.7$ & $-78.4$ & $-28.3$ \\
                11 & $+56.2$ & $+42.4$ & $+57.9$ & $+41.7$ & $-77.5$ & $-29.5$ & $-79.4$ & $-28.6$ \\
                12 & $+55.1$ & $+48.4$ & $+56.3$ & $+46.0$ & $-75.6$ & $-33.5$ & $-78.9$ & $-28.7$ \\
                13 & $+53.5$ & $+45.5$ & $+55.6$ & $+45.0$ & $-72.1$ & $-39.0$ & $-75.0$ & $-33.4$ \\
                14 & $+49.2$ & $+59.4$ & $+52.3$ & $+52.6$ & $-63.1$ & $-44.4$ & $-72.6$ & $-40.0$ \\
                15 & $+53.2$ & $+69.8$ & $+53.2$ & $+72.7$ & $-64.3$ & $-53.5$ & $-65.4$ & $-48.0$ \\
                \hline
                Ave. & $+53.0$ & $+52.0$ &  $+52.9$ & $+44.4$ & $-70.2$ & $-35.9$ & $-74.1$ & $-24.4$ \\
                \hline
            \end{tabular}
        \end{adjustbox}
    \end{center}
    \vspace{-0.2cm}
\end{table}

\subsection{Computation Times} \label{sec_timeres}

Next, we demonstrate the substantial performance improvements in {\sf Protofold II} as a result of introducing the algorithms and data structures presented in Section \ref{sec_alg}.
The running times are reported and compared on two computer systems, namely:
\begin{itemize}
    \item {\bf C-1}: Dell Precision T7500 workstation with an Intel$^\circledR$ Xeon$^\circledR$ E5645 CPU (12 cores, clock rate 2.40 GHz, and host memory 24 GB). The system is equipped with one NVIDIA Quadro$^\circledR$ 4000 GPU (256 CUDA cores with compute capability (CC) 2.0 and device memory 2 GB).
    \item {\bf C-2}: Dell Precision T7600 workstation with an Intel$^\circledR$ Xeon$^\circledR$ E5-2687W CPU (32 cores, clock rate 3.10 GHz, and host memory 64 GB). The system is equipped with two graphics cards: a NVIDIA Quadro$^\circledR$ K5000 GPU (1,536 CUDA cores with CC 3.0 and device memory 4 GB) and a NVIDIA Tesla$^\circledR$ K20C GPU (2,496 CUDA cores with CC 3.5 and device memory 5 GB).
\end{itemize}

\begin{figure}
    \centering
    \includegraphics[width=0.48\textwidth]{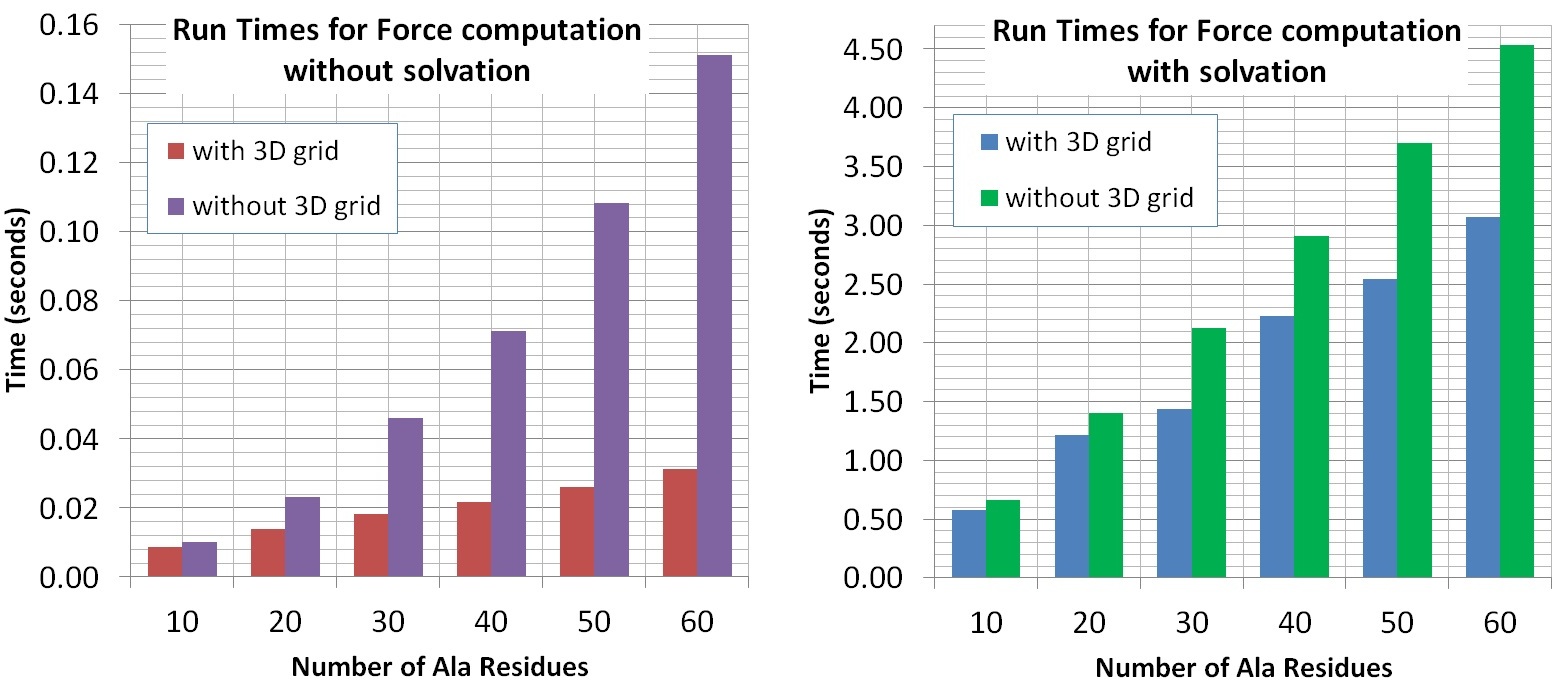}
    \caption{The effect of 3D hashing on the force computation times of a 60-residue polyalanine chain with and without solvation effects (on {\bf C-1}).}
    \label{figure14}
\end{figure}

\paragraph{Effect of Hashing.}
Figure \ref{figure14} shows the running times of the force computation step (on {\bf C-1}) in a single KCM iteration for folding polyalanine chains of different lengths with and without 3D hashing presented in Seciton \ref{sec_prox} both in vacuum (i.e., without considering solvation effects) and in water (i.e., with considering solvation effects). In both cases, the results show a significant reduction in the running times with hashing (e.g., up to $4.6\times$ in vacuum and $1.5\times$ in water for $m = 60$ Ala residues), and the difference scales with the size of the molecule. Nevertheless, the solvation force computations remain the bottleneck of the simulation (using sequential CPU implementation) and adversely affect the speed-up gained from hashing.

\begin{figure}
    \centering
    \includegraphics[width=0.48\textwidth]{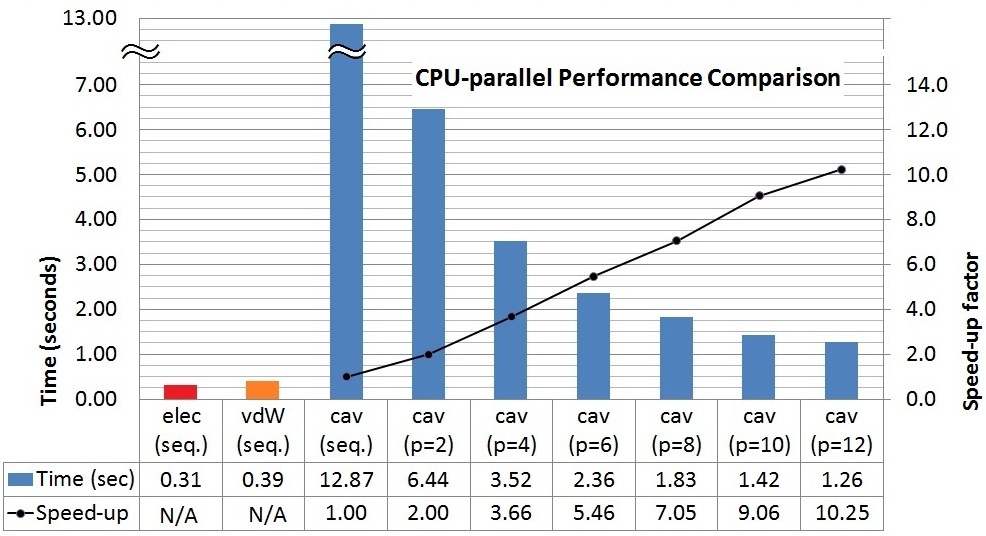}
    \caption{Sequential running times for electrostatic, van der Waals, and solvation forces, and CPU-parallel running times of the latter for a 1,200-residue polypeptide chain (on {\bf C-1}).}
    \label{figure15}
\end{figure}

\paragraph{Parallel Computing.}
The first 3 columns on the left in Fig. \ref{figure15} show the sequential CPU running times (on {\bf C-1}) for the electrostatic, van der Waals, and nonpolar solvation forces in a single KCM iteration. The results were obtained for a polypeptide chain composed of $1,200$ residues that are randomly selected from Ala, Cys, and Ser AAs---the latter two being equivalent to replacing the H in the nonpolar Ala side chain with SH and OH, respectively, resulting in polar side chains. It is clearly observed that with a single CPU core, the first two terms take a very small fraction of the total time (around 5$-$6\% of the total force computation time per iteration) and the solvent effects are clearly the bottleneck.

The same column chart in Fig. \ref{figure15} also shows the running times and corresponding speed-ups (on {\bf C-1}) for the CPU-parallel computation of the solvation force using up to 12 CPU cores. An almost linear speed-up is achieved by increasing the number of processors (e.g., $\sim 10\times$ with 12 cores). However, the solvation force calculation is still about an order of magnitude slower than that of the other two force types.


\begin{figure}
    \centering
    \includegraphics[width=0.48\textwidth]{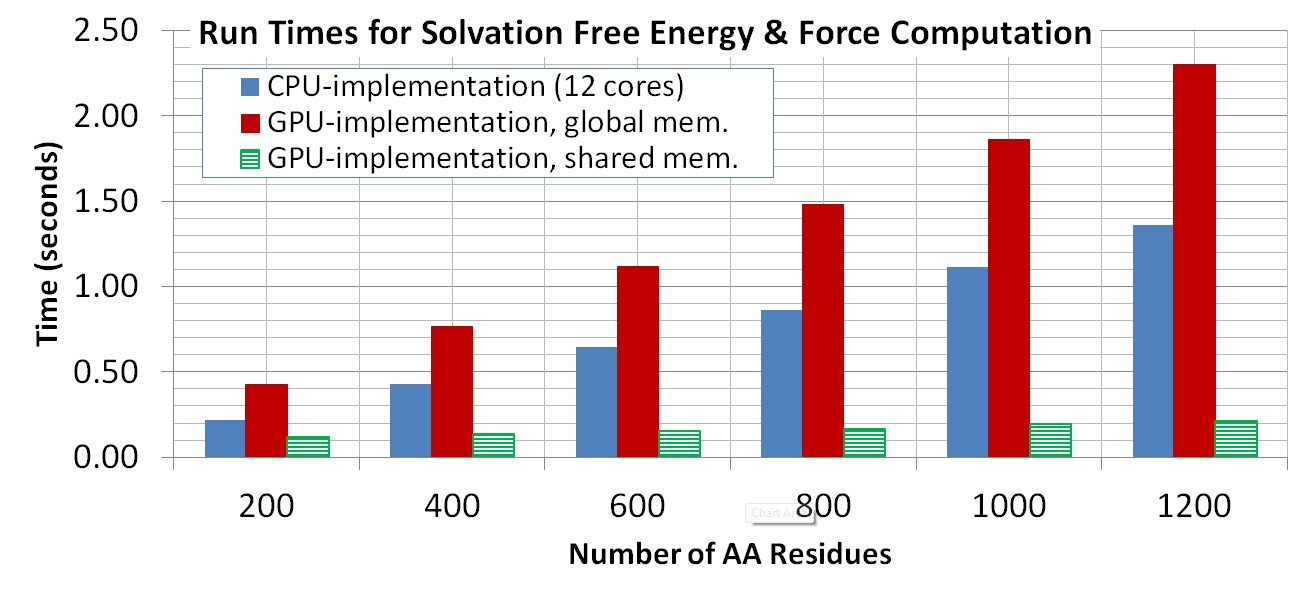}
    \includegraphics[width=0.48\textwidth]{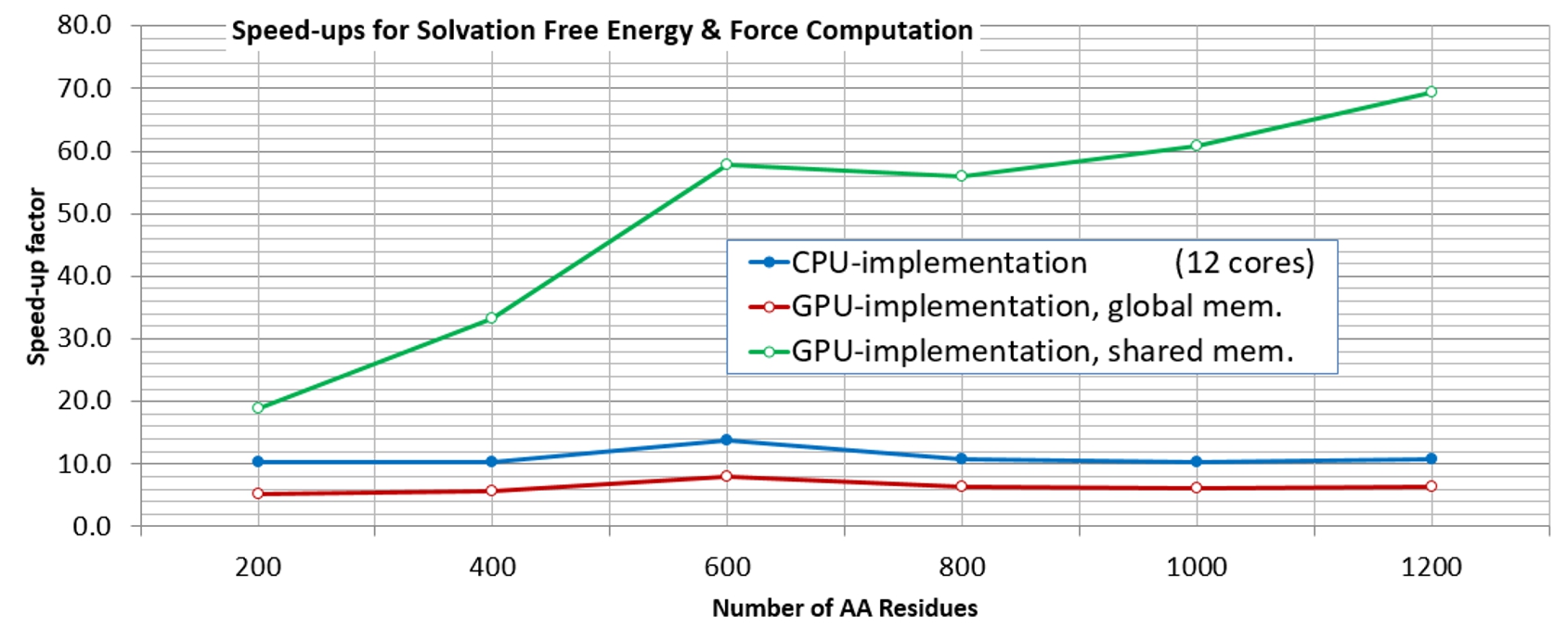}
    \caption{CPU- and GPU-parallel running times (top) and speed-ups (bottom) with and without memory optimization for polypeptide chains of various lengths (on {\bf C-1}).}
    \label{figure16}
\end{figure}

Figure \ref{figure16} compares the running times and corresponding speed-ups (on {\bf C-1}) of the CPU- and GPU-parallel implementations for polypeptides of different lengths, ranging from $m = 200$ (i.e., $\sim$ 2K atoms) to $1,200$ AAs (i.e., $\sim$ 13K atoms). To depict the importance of memory optimization presented in Section \ref{sec_GPU}, the running times are shown for two different cases; namely, one that uses global memory for communications between all threads of all blocks, and the optimized code making extensive use of shared memories for communications between threads within the same block. It is interesting to note that when the GPU shared memory is not utilized, the results do not show an improvement over the 12-core CPU implementation due to large global memory access latencies. However, proper shared memory usage results in a huge performance improvement that scales by protein size, e.g., from $20\times$ for $m = 200$ AAs to $70\times$ for $m = 1,200$ AAs with respect to the sequential run on a single CPU.

\begin{figure}
    \centering
    \includegraphics[width=0.48\textwidth]{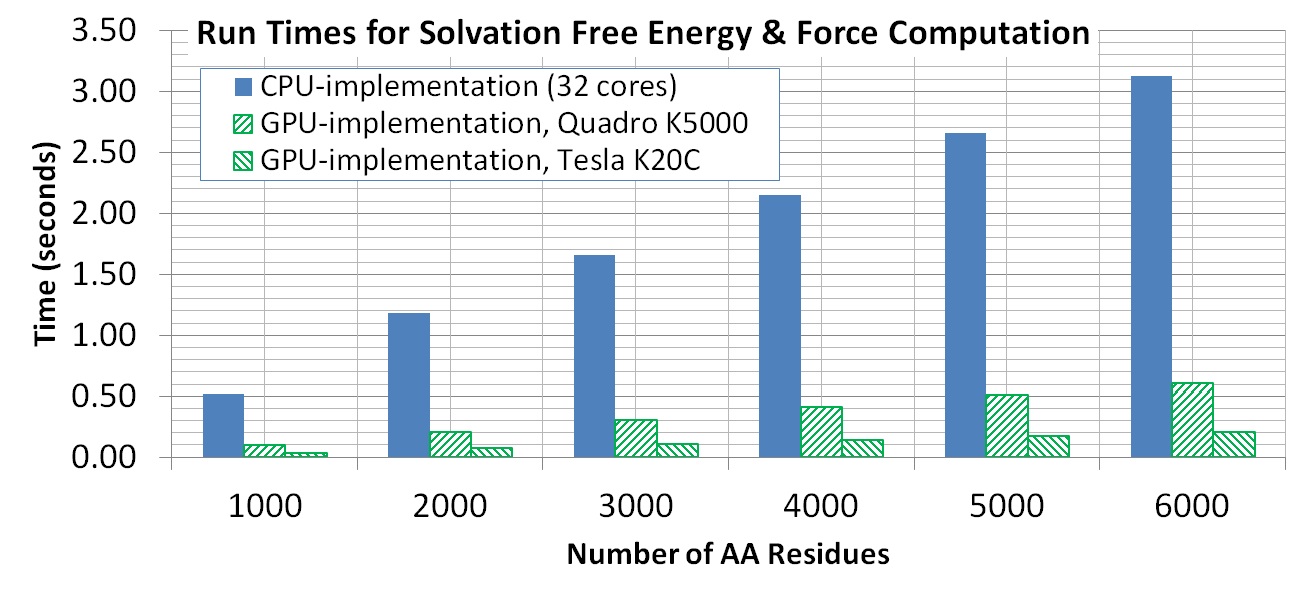}
    \includegraphics[width=0.48\textwidth]{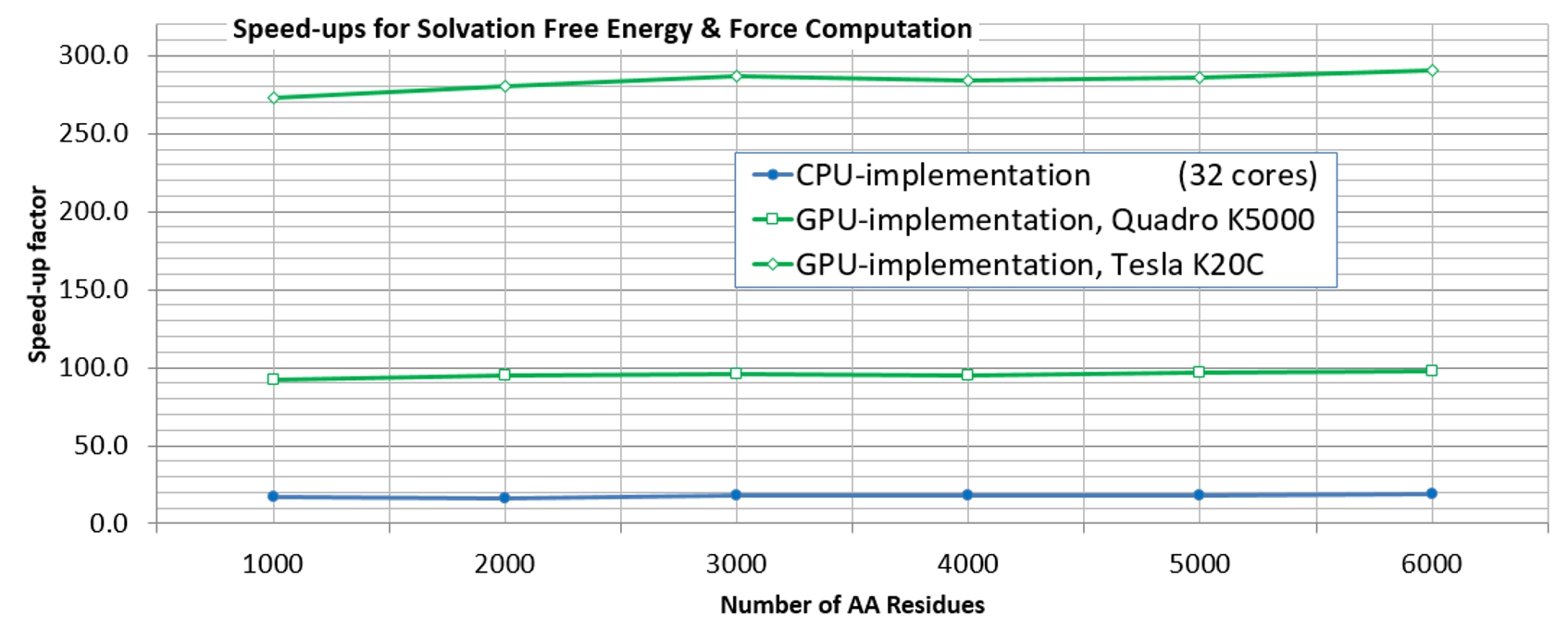}
    \caption{CPU- and GPU-parallel running times (top) and speed-ups (bottom) with memory optimization on two GPUs for polypeptide chains of various lengths (on {\bf C-2}).}
    \label{figure17}
\end{figure}

The computations are repeated for even larger molecules in Fig. \ref{figure17}, ranging from $m = 1,000$ AAs (i.e., $\sim$ 10K atoms) to $6,000$ AAs (i.e., $\sim$ 64K atoms), enabled by leveraging a more powerful machine (i.e., {\bf C-2}). For the case of $m = 1,000$ AAs, {\bf C-2} yields a two-fold speed-up on the CPU and a three-fold speed-up on the GPU compared to {\bf C-1}.
As depicted in Fig. \ref{figure17}, significantly higher and more consistent CPU speed-ups of $16$-$18\times$ and GPU speed-ups of $90$-$100\times$ (for Quadro$^\circledR$ K5000) and $270$-$290\times$ (for Tesla$^\circledR$ K20C) are observed. These observations imply proper scalability of the data-parallel implementation with molecular size.

Even for molecules with tens of thousands of atoms, each force-field computation takes only a fraction of a second per iteration. This enables fast KCM simulation of folding for large proteins over extended periods of time via {\sf Protofold II}, which wouldn't be tractable via {\sf Protofold I} \cite{Kazerounian2004a,Kazerounian2004b,Kazerounian2005a}.


\subsection{Real Examples} \label{sec_real}

Having considered the folding of secondary structural elements with complete flexibility (i.e., each peptide plane being treated as a separate rigid body), we proceed to study the tertiary interaction between larger rigid units in real proteins. We report on the following case studies:
\begin{itemize}
    \item The interactions between multiple rigid $\alpha-$helices that rotate around flexible loops within the containing motifs/domains are considered. Two examples from the PDB are used for this purpose: Myoglobin (PDB: 1TES) and Troponin-C (PDB: 2JNF).
    \item The interactions between multiple rigid domains that are connected via flexible loops within the containing monomeric unit are examined. The example of Gamma-B Crystallin (PDB: 1GCS) is used for this purpose.
\end{itemize}
These PDB structures that are obtained from X-ray crystallography (e.g., 1TES and 1GCS) do not contain the H atoms, hence are first preprocessed using Duke University's {\sf MolProbity} server \cite{Davis2007,Chen2009} which predicts and adds the H atoms to the coordinates information before importing the structure into {\sf Protofold II}. The PDB structures that are obtained from nuclear magnetic resonance (NMR) spectrometry (e.g., 2JNF), on the other hand, already contain the H atoms positions. The size of the molecules, i.e., the number of AA residues $m$ and the number of atoms $n'$ and $n$ with and without the H atoms included, respectively, are given in Table \ref{tab_molecules}.

\begin{table}
    \caption{The size of the analyzed protein examples.}
    \vspace{-0.5cm}
    \begin{center} \label{tab_molecules}
            \begin{tabular}{l l c c c} 
                \hline
                Protein Name & PDB Code & $m$ & $n'$ & $n$  \\ [0.5ex] 
                \hline
                Myoglobin & 1TES & 154 & 1,231 & 2,478 \\
                Troponin-C & 2JNF & 158 & 1,232 & 2,401 \\
                Gamma-B Crystallin & 1GCS & 174 & 1,474 & 2,844 \\
                \hline
            \end{tabular}
    \end{center}
    \vspace{-0.2cm}
\end{table}

\begin{figure*}
    \centering
    \includegraphics[width=0.75\textwidth]{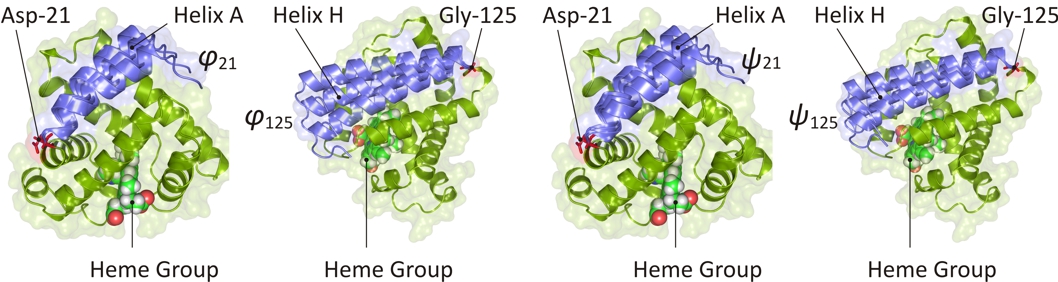}
    \caption{Myoglobin (PDB: 1TES) hinged at Asp-21 and Gly-125 for variations in $\phi_{21}, \phi_{125}$ (left) and $\psi_{21}, \psi_{125}$ (right).}
    \label{figure18}
\end{figure*}
\begin{figure*}
    \centering
    \includegraphics[width=0.75\textwidth]{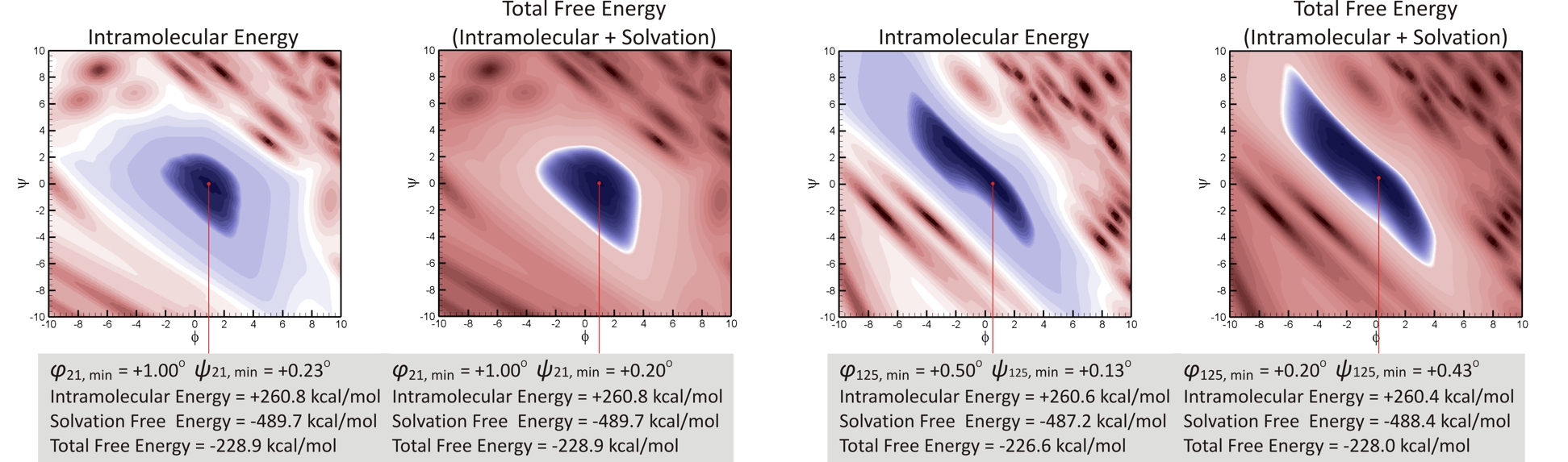}
    \caption{Intramolecular versus total (i.e., solvation included) free energy landscape for Myoglobin (PDB: 1TES) in the vicinity of the native conformation versus variations in $(\phi_{21}, \psi_{21})$ (left) and $(\phi_{125}, \psi_{125})$ (right).}
    \label{figure19}
\end{figure*}

\begin{figure*}
    \centering
    \includegraphics[width=0.75\textwidth]{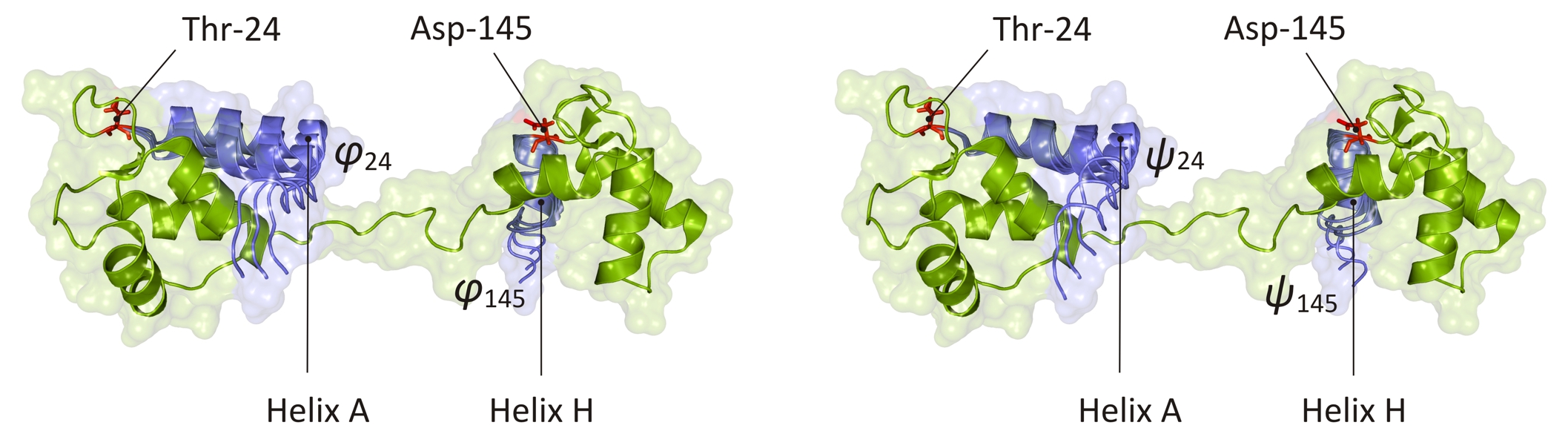}
    \caption{Troponin-C (PDB: 2JNF) hinged at Thr-24 and Asp-145 for variations in $\phi_{24}, \phi_{145}$ (left) and $\psi_{24}, \psi_{145}$ (right).}
    \label{figure20}
\end{figure*}
\begin{figure*}
    \centering
    \includegraphics[width=0.75\textwidth]{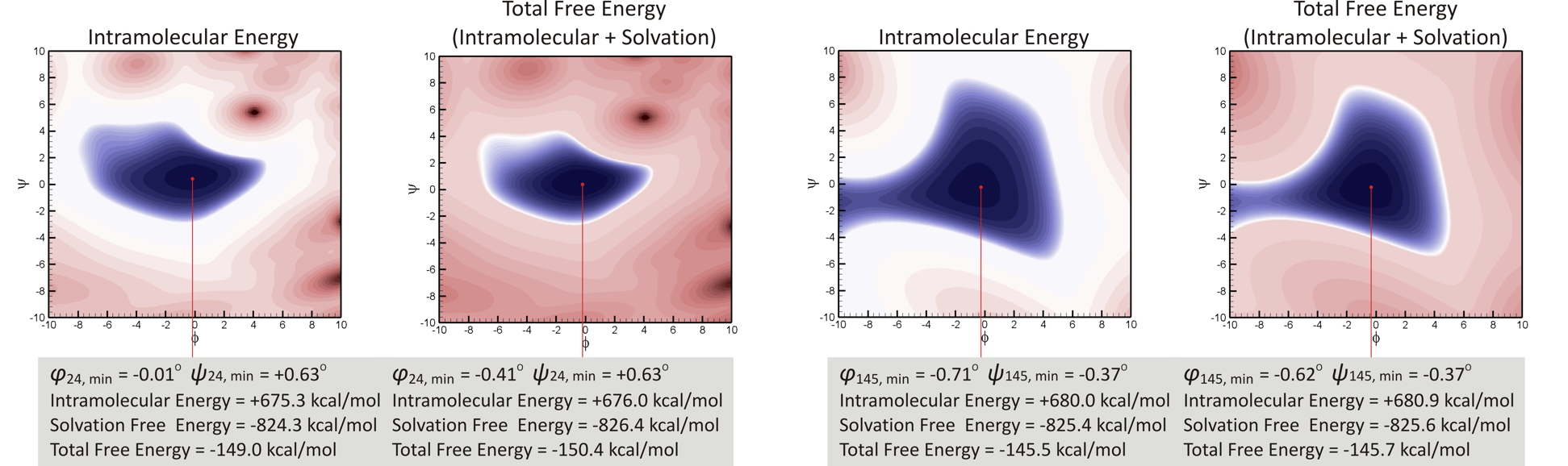}
    \caption{Intramolecular versus total (i.e., solvation included) free energy landscape for Troponin-C (PDB: 2JNF) in the vicinity of the native conformation versus variations in $(\phi_{24}, \psi_{24})$ (left) and $(\phi_{145}, \psi_{145})$ (right).}
    \label{figure21}
\end{figure*}

\paragraph{Secondary Structural Interactions.}
Let us first consider the energy variations when an $\alpha-$helix of an $\alpha-$domain is reoriented from its stable conformation with respect to the rest of the bundle, as illustrated in Figs. \ref{figure18} and \ref{figure20}.

Myoglobin (PDB: 1TES) is an oxygen binding muscle protein that is composed of a single `globin fold' domain, which is an $\alpha-$domain motif consisting of a bag of 8 $\alpha-$helices per domain (denoted A through H) arranged at $\sim$ $+90^\circ$ and $+50^\circ$ angles with respect to each other, as shown in Fig. \ref{figure18}. This arrangement creates a hydrophobic pocket in the interior that wraps the stabilizer co-factor known as `heme group' \cite{Petsko2004}.
Assuming that the helices are rigid, we examine the energy variations due to dihedral rotations at the loops that connect the two end $\alpha-$helices; namely, local changes in $(\phi_i, \psi_i)$ for $i = 21$ and $i = 125$ where the A and H helices are hinged, respectively.

Troponin-C (PDB: 2JNF) is a calcium binding muscle protein that is composed of two `EF-hand' domains, which are $\alpha-$domain motifs consisting of a bundle of 4 $\alpha-$helices per domain (denoted A through H) arranged in an up-and-down anti-parallel conformation \cite{Petsko2004}, as shown in Fig. \ref{figure20}. There is a long pseudo-helical segment that connects the two globular domains.
Assuming that the helices are rigid, we examine the energy variations due to dihedral rotations at the loops that connect the two end $\alpha-$helices; namely, local changes in $(\phi_i, \psi_i)$ for $i = 24$ and $i = 145$ where the A and H helices are hinged, respectively.

Figures \ref{figure19} and \ref{figure21} show the free energy variations in vacuum (i.e., without considering solvation effects) and in water (i.e., with considering solvation effects) for the two protein domains as the end $\alpha-$helices A and H of each are rotated within a range of $\pm 10^\circ$ of the native $(\phi_i, \psi_i)$ at the hinged loops.\footnote{The colormaps are generated using a nonlinear but consistent contouring scheme.}
In all four cases the energy model of {\sf Protofold II} exhibits a local minima within $\pm 1^\circ$ of the native conformation. We also observe that the solvation effects contribute a significant amount to the total free energy; however, the SASA variations are so small in the considered neighborhood that the location of the local minima is almost unchanged. The van der Waals effects appear to be dominant in this neighborhood, manifested as the shape complementarity between the ridges and grooves of the mobile $\alpha-$helix and those of the other helices in the bundle \cite{Eilers2002}. However, when variations across larger angular ranges are considered, the solvation effects are expected to play a more determining role.

\paragraph{Tertiary Structural Interactions.}
Lastly, we consider the energy variations when a rigid domains of a protein is reoriented with respect to another domain against which it is packed into a stable structure.

\begin{figure}
    \centering
    \includegraphics[width=0.48\textwidth]{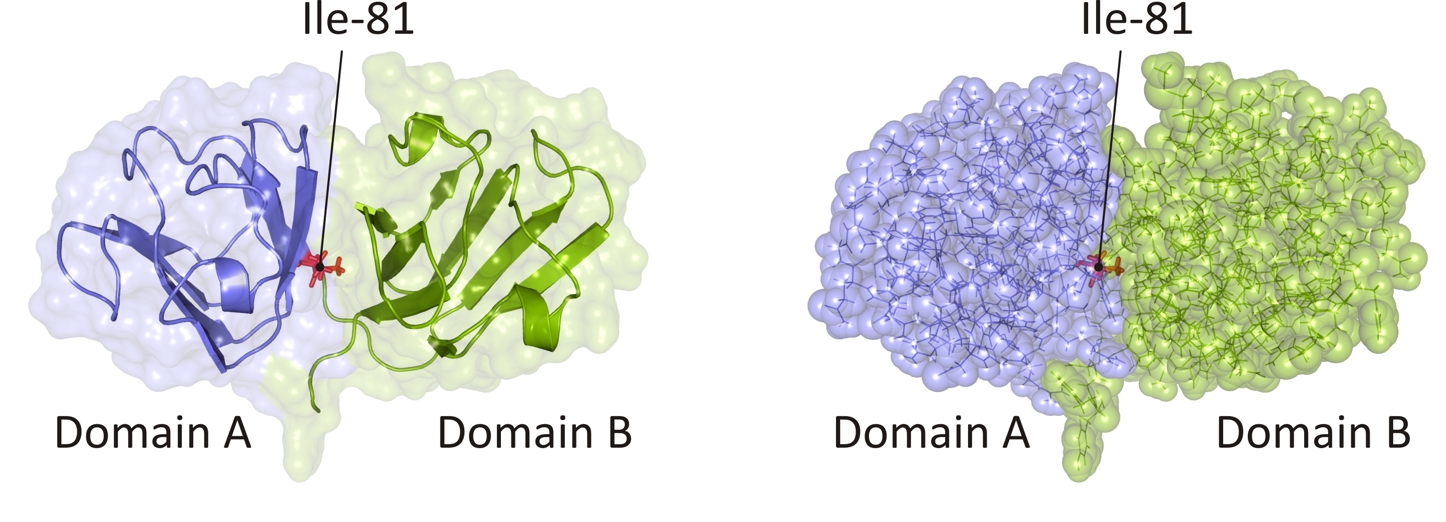}
    \caption{Gamma-B Crystallin (PDB: 1GCS) hinged at Ile-81 for variations in $\phi_{81}$ and $\psi_{81}$, one at a time.}
    \label{figure22}
\end{figure}

Gamma-B Crystallin (PDB: 1GCS) is an eye-lens protein that is made of two similar domains that are 40\% identical in sequence \cite{Petsko2004}. Each domain is composed of two anti-parallel $\beta-$sheets each made of 4 $\beta-$strands with the same arrangement topology \cite{Petsko2004}, as shown in Fig. \ref{figure22}.
Assuming that these domains are rigid, we analyze the energy variations due to rotating one domain with respect to the other at one of the residues that belong to the connecting loop (e.g., $i = 81$). To observe the global effects of solvation, we allow both dihedral angles $(\phi_{81}, \psi_{81})$ to vary over the entire range of $\pm 180^\circ$ from the native values. To facilitate visualization, this time we consider only one angle's variation at a time.

\begin{figure*}
    \centering
    \includegraphics[width=0.75\textwidth]{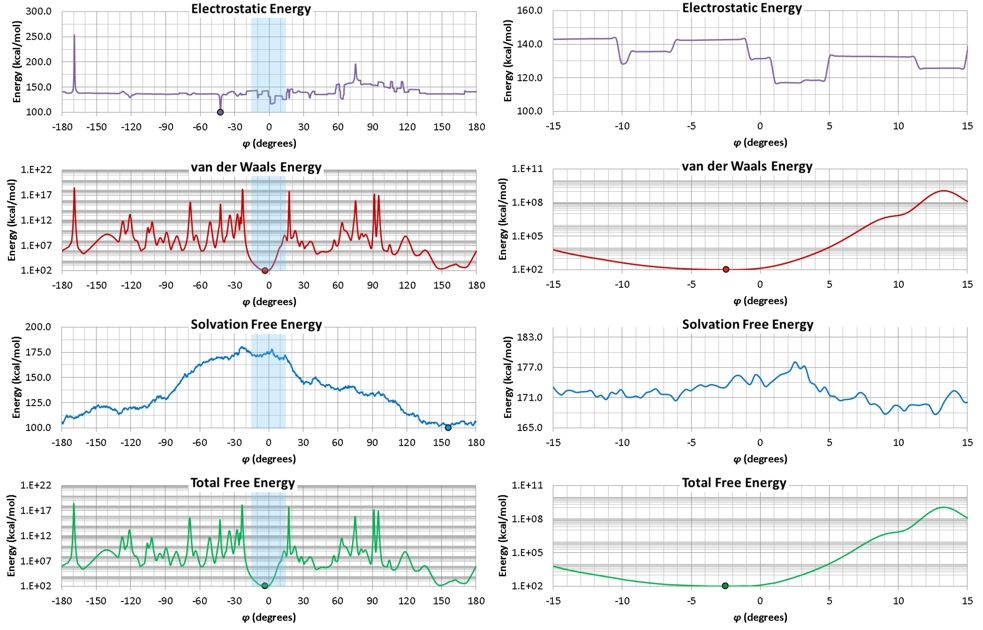}
    \caption{Energy variations for Gamma-B Crystallin (PDB: 1GCS) versus changes in $\phi_{81}$ for fixed native $\psi_{81}$.}
    \label{figure23}
\end{figure*}

\begin{figure*}
    \centering
    \includegraphics[width=0.75\textwidth]{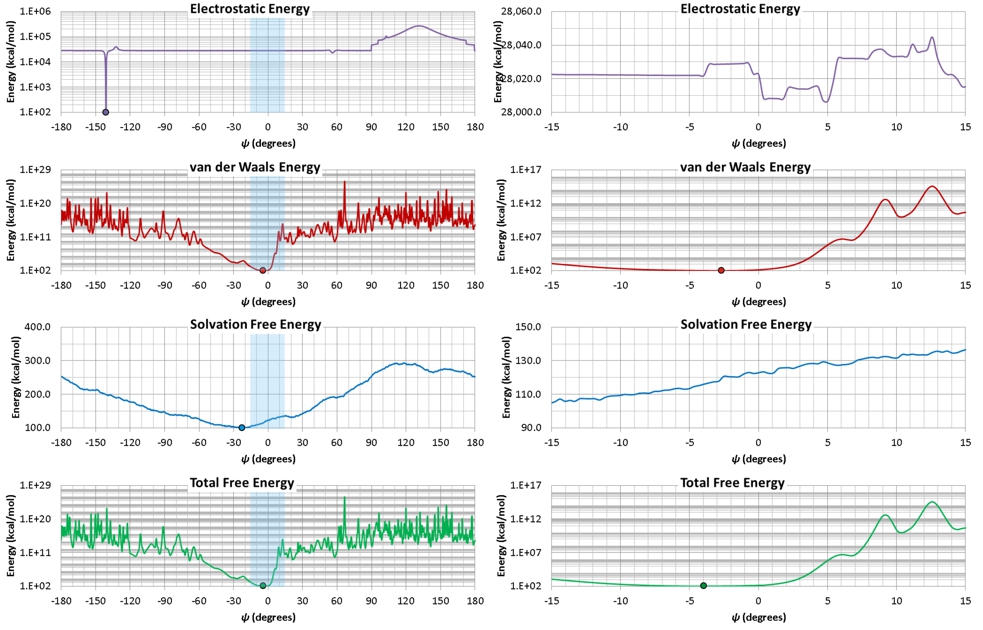}
    \caption{Energy variations for Gamma-B Crystallin (PDB: 1GCS) versus changes in $\psi_{81}$ for fixed native $\phi_{81}$.}
    \label{figure24}
\end{figure*}

Figures \ref{figure23} and \ref{figure24} show the variations of the different energy terms.\footnote{To enabel logarithmic plots, each energy ordinate is offset by a constant value to shift its minimum to (the arbitrary positive value of) 100 kcal per mol.}
It appears that the van der Waals effects are dominant in determining the profile of the energy well near the native conformation, which can be attributed to the extensive contact interface between the two domains in Fig. \ref{figure22}. The electrostatic and solvation effects substantially change the energy landscape thus can dramatically affect the folding pathway. However, they do not cause a significant change in the location of the energy minimum.
Once again, the solvation energy is noisier due to the discrete nature of the enumeration algorithm in Section \ref{sec_enum}. Another interesting observation is that the electrostatic energy has discontinuities due to the cut-off approximation when pairs of atoms are farther than $9.0~\angstrom$. Although it does not seem to affect the minimum location in this example, larger cut-off distances might be necessary for analyzing large proteins, since the accumulation of the pairwise errors grows quadratically with the number of atoms.


\section{Conclusion}

The KCM approach to protein folding \cite{Kazerounian2004,Kazerounian2005} originally implemented into {\sf Protofold I} \cite{Kazerounian2004a,Kazerounian2004b,Kazerounian2005a} provides a promising fast alternative to the popular MD simulation and MC sampling methods by
\begin{enumerate}
    \item modeling the protein chain as a kinematic linkage with restricted DOF to which the well-studied principles from mechanism synthesis and robotics can be readily applied; and
    \item replacing the 2nd-order dynamic response with 1st-order kinetostatic integration of the equations of motion to facilitate convergence to the free energy minima.
\end{enumerate}
In the present work, we introduced major model and implementation improvements in {\sf Protofold II} by
\begin{enumerate}
    \item incorporating the solvation effects that characterize the hydrophobic effect, i.e., the entropic changes due to cavity formation in the aqueous solvent;
    \item taking advantage from efficient auxiliary algorithms and data structures to improve the computational complexity from $O(n^2)$ to expected $O(n)$; and
    \item implementing fast and relatively accurate evaluation of the SASA and its gradient for solvation energy- and force-field computation, respectively, in parallel on both CPU and GPU.
\end{enumerate}

The presented enumeration algorithm for the latter provides a fast approximate method, in which the degree of accuracy is traded off with the performance by a proper choice of the sample size. We argued that the inclusion of the solvation free energy into the mix can significantly affect the folding pathway for water-soluble proteins whose folding {\it in vivo} is dominated by such effects. We also demonstrated that the performance gain of the GPU-accelerated implementation scales properly with the number of atoms, achieving up to two orders of magnitude in speed-ups after memory optimization.

{\sf Protofold II} has been completely rearchitectured and is evolving into a versatile analysis toolbox for studying the kinematic and structural behavior of molecular chains in protein engineering applications such as the design of nano-manipulators \cite{Tavousi2015,Tavousi2016} among the ongoing projects.
A hybrid force-field model was used, composed of
\begin{enumerate}
    \item the AMBER model \cite{Weiner1984} (for noncovalent interactions); and
    \item supplemental terms similar to the CHARMM model \cite{Wesson1992} and the GROMOS model \cite{Fraternali1996} (for solvation effects) except for the probabilistic SASA estimation \cite{Wodak1980} replaced with our own surface sampling algorithm.
\end{enumerate}
This model is by no means versatile enough to enable addressing the ultimate goal of arriving from sequence to 3D structure at a click of a button. Even though predicting folding of real 3D structures requires further developments, this study represents a major step toward this goal.

\section{Acknowledgements}

The authors are thankful to Sanguthevar Rajasekaran from the Department of Computer Science and Engineering and to Andrei Alexandrescu and Victoria Robinson from the Department of Molecular and Cellular Biology at UConn for providing instructive insight in this interdisciplinary project.

This work was supported in part by the National Science Foundation grants CMMI-1200089, CNS-0927105, and CMMI-1462759. The responsibility for any errors and omissions lies solely with the authors.

\appendix

\section{Peptide Chains} \label{app_peptide}

This appendix overviews the structural biochemistry of peptide chains. Amino acids (AAs) are composed of a central carbon atom (denoted C$_\alpha$) attached to 4 chemical components; namely a carboxylate group ($-$COO$^-$), an amino group ($-$NH$_3^+$), and a hydrogen atom ($-$H), common among all types, and a variable side chain (denoted $-$R) \cite{Kuriyan2012}. The amino group of one AA reacts with the carboxyl group of another to form a `peptide bond,' eliminating a water molecule. This so-called `condensation reaction' repeats over and over again to form a `peptide chain' \cite{Kuriyan2012}.

As depicted in Fig. \ref{figure1}, the 3D structure of the peptide chain can be uniquely represented by a set of bond lengths, and two sets of angles, namely the angles between adjacent bonds that share one atom (referred to as `bond angles'), and those describing rotation around the bonds (referred to as `torsion angles' or `dihedral angles'). It is reasonable to assume that the bond lengths and bond angles are constant \cite{Kazerounian2005}, thus the dihedral angles exclusively specify the protein conformation. For a protein with $m$ AA residues denoted by $AA_i ~(1 \leq i \leq m)$ numbered in order from N-terminus to C-terminus, the 3 set of dihedral angles in the main chain are defined for $1 \leq i \leq m$ as

\begin{itemize}
    \item the rotation angle $\omega_i$ around the backbone C$-$N bond that connects the residues $AA_i$ and $AA_{i+1}$;
    \item the rotation angle $\phi_i$ around the backbone N$-$C$_\alpha$ bond in the residue $AA_i$; and
    \item the rotation angle $\psi_i$ around the backbone C$_\alpha-$C bond in the residue $AA_i$.
\end{itemize}

Based on high resolution X-ray crystallographic studies, the angle $\omega_i$ is very close to either $0^\circ$ (the `cis' conformation) or $180^\circ$ (the `trans' conformation) \cite{Kuriyan2012}, and the 6 atoms in the peptide group C$_\alpha-$CO$-$NH$-$C$_\alpha$ are approximately coplanar, forming the so-called `peptide plane' \cite{Kazerounian2005}. Due to the partial double-bond characteristic of the peptide bond C$-$N, the peptide groups are almost rigid, hence modeled as rigid links hinged to the preceding and following peptide groups along the main chain \cite{Kazerounian2005}. These planes rotate about the N$-$C$_\alpha$ and C$_\alpha-$C bonds, which can be thought of as {\it revolute joints}. Hence the `main chain dihedral angles' $\phi_i$ and $\psi_i$ completely specify the conformation of the backbone. In addition, each side chain can be treated as a shorter linkage which can add up to 4 extra links with their associated joint angles, called `side chain dihedral angles' ($\chi_{i,1}$ to $\chi_{i,4}$). Therefore, the whole protein chain can be modeled as an open kinematic linkage, conformation of which is fully specified by a set of main chain and side chain dihedral angles. The resulted model has a reduced number of DOF of $2m + \sum_{i=1}^{m} {\rm DOF}({\rm R}_i)$, where the DOF of the side chain ${\rm R}_i$ is determined by the number of its side chain dihedral angles.

\section{Prefix Computation} \label{app_pref}

The prefix sum problem is fundamental to numerous important algorithms, and is defined as follows. Given a finite ordered sequence of elements $X = (x_1, x_2, \cdots, x_n) \in \Sigma^n$ and an arbitrary binary operator $\oplus: \Sigma \times \Sigma \rightarrow \Sigma$ that is $O(1)-$time computable and associative (i.e., $(x \oplus y) \oplus z = x \oplus (y \oplus z)$), compute another sequence $Y = (y_1, y_2, \cdots, y_n) \in \Sigma^n$ where $y_1 = x_1$, $y_2 = x_1 \oplus x_2$, $\cdots$, $y_n = x_1 \oplus x_2 \oplus \cdots \oplus x_n$; in other words $y_i = y_{i-1} \oplus x_n$ for $1 \leq i \leq n$ where $y_0$ is the left-identity element (i.e., $y_0 \oplus x = x$ for all $x \in \Sigma$).

It is trivial to show that the prefix sums can be computed sequentially in $O(n)$, which is optimal. In addition, there are work-optimal parallel algorithms with a total computational work of $TP = O(n)$ that carry out the prefix computation in $T = O(\log n)$ time using $P = O(n/\log n)$ CREW PRAM or in $T = O(\log n/\log \log n)$ time using $P = O(n \log \log n/\log n)$ CRCW PRAM processors with common conflict resolution \cite{Cole1989}---see Appendix \ref{app_PRAM} for details regarding PRAM.

\section{Parallel Computing}

\subsection{Abstract Machines} \label{app_PRAM}

The parallel random-access machine (PRAM) is a shared-memory abstract parallel computation model, typically assigned with the exclusive/concurrent-read (ER/CR) and exclusive/concurrent-write (EW/CW) attributes \cite{Maggs1995}. The most common attributes are CREW and CRCW, noting that multiple processors can concurrently read a memory cell but only one can write at a time to prevent race conditions. To enable concurrent writing, one needs to resolve possible conflicts with typical mechanisms such as 1) `common' meaning that all processors attempt to write the same value); 2) `arbitrary' meaning that one processor's attempt succeeds at random; and 3) `priority' meaning that the processors are prioritized by a prespecified order. Note that a CREW algorithm can always run in the same (if not fewer) number of steps on a CRCW machine.

\subsection{GPU SIMT Model} \label{app_SIMT}

A typical CUDA GPU program proceeds as follows. The data is first transferred from the CPU (i.e., {\it host}) memory to the GPU (i.e., {\it device}) memory. The host application invokes the so-called kernels on the GPU with specified granularity, i.e., issuing a 1D, 2D, or 3D grid of blocks, each block being a 1D, 2D, or 3D array of threads that are sent in groups of 16 or 32 (called `warps') to one of the streaming multiprocessors (SM). The threads within the same block can access the fast shared memory banks on the SM, and communication across blocks is done using the global memory. The computed results are transferred from the device memory back to the host memory.

There are different types of GPU memory locations, classified into two groups: 1) device (i.e., off-chip) memory including global and local memories; and 2) on-chip memory including shared memory, cache, and registers. The access latencies to the on-chip are much less (around $100\times$ faster) than those of the off-chip memory.

\bibliographystyle{asmems4}
{\scriptsize \bibliography{CDL-TR-16-03}}

\end{document}